\documentstyle [12pt,a4,epsf] {article}

\newcommand{\rr   }   { {\bf r} }
\newcommand{\dr  }   { {\rm d}\rr\ }
\newcommand{\asix}   {\frac{a^2}{6}}
\newcommand{\half} {\frac{1}{2}}
\newcommand{\e} { {\rm e} }
\newcommand{\ie} {{\it i.e.}}
\newcommand {\s}{{\rm salt}}
\newcommand{\eg} {{\it e.g.}}
\newcommand{\be} {\begin{equation}}
\newcommand{\ee} {\end{equation}}
\newcommand{\bea} {\begin{eqnarray}}
\newcommand{\eea} {\end{eqnarray}}
\newcommand{\dd} {{\rm d}}

\begin{document}


\title {Neutral and Charged  \\
            Polymers at Interfaces\thanks{to be published, {\em Physics Reports}, 2003}}
\author {Roland R. Netz \\
         Max-Planck Institute for Colloids and Interfaces\\
         D-14424 Potsdam, Germany \\
         Sektion Physik, Ludwig-Maximilians-Universit\"at\\
         Theresienstr. 37, 80333 M\"unchen, Germany\\
\and
David Andelman \\
         School of Physics and Astronomy   \\
         Raymond and Beverly Sackler Faculty of Exact Sciences \\
         Tel Aviv University, Ramat Aviv, Tel Aviv 69978, Israel \\
         \\}
\date{February 2003}
\maketitle
\begin{abstract}
\setlength {\baselineskip} {16pt}

Chain-like macromolecules (polymers) show characteristic
adsorption properties due to their flexibility and internal
degrees of freedom, when attracted to surfaces
and interfaces. In this review we discuss concepts and features
that are relevant to the adsorption of neutral and charged
polymers at equilibrium, including the type of polymer/surface
interaction, the solvent quality, the characteristics of the
surface, and the polymer structure. We pay special attention
to the case of charged polymers
(polyelectrolytes) that have a special importance due to their water
solubility. We present a  summary of recent progress in this
rapidly evolving field. Because many experimental studies are
performed with rather stiff biopolymers, we discuss in detail the
case of semi-flexible polymers in addition to flexible ones.
We first review the behavior of
neutral and charged chains in solution. Then, the adsorption of a
single polymer chain is considered. Next, the adsorption and
depletion processes in the many-chain case are reviewed. Profiles,
changes in the surface tension and polymer surface excess are
presented. Mean-field and corrections due to fluctuations and
lateral correlations are discussed. The force of interaction
between two  adsorbed layers, which is important in understanding
colloidal stability, is characterized. The behavior of grafted
polymers is also reviewed, both for neutral and charged polymer brushes.

\end{abstract}
\newpage
\tableofcontents
\newpage

\setlength {\baselineskip} {18pt}

\pagebreak
\section*{Legend of Symbols}

\begin{itemize}

\item $a$: Kuhn length or effective  monomer size
\item $b$: monomer size
\item $c_m$: monomer concentration (per unit volume)
\item $c_m(x)$: monomer density profile at distance $x$ from the surface
\item $c_m^b$: bulk monomer concentration in semi-dilute solutions.
\item $c_m^*$: overlap concentration of bulk polymer solution
\item $c_\s$: salt concentration in the solution
\item $c^\pm(x)$: profiles of $\pm$ ions
 \item $d$: polymer diameter (or cross-section)
 \item $D$: adsorption layer thickness, height of brush
\item $e$: electronic unit charge
\item $f$: fractional charge of the chain $0<f<1$
\item $\hat{f}$: force  in units of $k_BT$
\item $ F$: intensive free energy  in units of $k_BT$ (per unit area or unit volume)
\item $ {\cal F}$: extensive free energy in units of $k_BT$
 \item $g$: number of monomers per blob
\item $h(x)$: dimensionless PE adsorption profile
 \item $H$: height of counterion cloud (PE brush case)
\item $k_B T$: thermal energy
 \item $L$: contour length of a chain
\item $L_{\rm el}$: chain length inside one electrostatic blob
\item $L_{\rm sw}$: chain length inside one swollen blob
\item $\ell_B=e^2/(\varepsilon k_B T)$:  Bjerrum length
\item $\ell_0$: bare (mechanical) persistence length
\item $\ell_{\rm OSF}$: electrostatic contribution to persistence length (Odijk, Skolnick, and
Fixman length)
 \item $\ell_{\rm eff}$: effective persistence length
 \item $N$: polymerization index
\item  $R$: end-to-end polymer chain radius
\item $R_{\rm el}$: radius of one electrostatic blob
\item $S(q)$: structure factor (or scattering function) of a PE solution
 \item $S_0(q)$: form factor of a single chain
 \item $U(x)$: electrostatic potential at point $x$ from the surface
\item $U_s$: surface potential at $x=0$
 \item $u(x)=eU(x)/k_BT$: dimensionless  potential profile
\item $u_s=u(0)$: rescaled surface potential \item $v_2$:  2nd
virial coefficient of monomers in solution. $v_2>0$ for good
solvents \item $\tilde{v}_2=v_2/a^3$:  dimensionless 2nd virial
coefficient of monomers in solution
 \item $v(r)=e^2/k_BT \varepsilon r$: Coulomb interaction between two ions
in units of $k_BT$
 \item $v_{\rm DH}(r)=v(r)\exp(-\kappa r)$: Debye-H\"uckel
 interaction
 \item $z=\pm 1$, $\pm 2, \dots$ valency of the ions
\item $\Gamma$: polymer surface excess per unit area
\item $\varepsilon$: dielectric constant of the medium. $\varepsilon=80$ for water.
\item $\kappa^{-1}$: Debye-H\"uckel screening length
\item  $\kappa_{\rm salt}$: salt contribution to $\kappa$
\item $\mu^\pm$: chemical potential of $\pm$ ions
\item $\mu_p$: chemical potential of polymer
\item $\nu$: Flory exponent for the polymer size
\item $\xi_b, \xi_s$: correlation length (mesh size) of semi-dilute polymer solution
in bulk and at surface
\item  $\Pi$: osmotic pressure in units of $k_BT$
\item $\rho$: grafting density of a polymer  brush
\item $\sigma$: surface charge density (in units of $e$) at $x=0$
\item $\Delta \sigma=f\Gamma- \sigma$: overcharging parameter
 \item $\tau=f/b$: linear charge density on the chain
\item $\phi(x)=a^3c_m(x)$: monomer volume fraction (dimensionless) at distance $x$
from the surface
\item $\phi_b=a^3c_m^b$: bulk value of $\phi$
\item $\phi_s$: surface value of $\phi$

\item $\psi(x)=\sqrt{\phi(x)}$: polymer order parameter
\item $\psi_b$: bulk value of polymer order parameter

\end{itemize}

\newpage
\section{Introduction}
\setcounter{equation}{0}

Polymers are
long chain molecules which play important roles in industrial
applications and in biological processes.
On a more fundamental level, polymers exhibit
interesting behavior which can be derived
from the knowledge of their statistical mechanics properties.
We review the basic mechanisms underlying
the equilibrium properties of these macromolecules in solution
and, in particular, their behavior at surfaces and interfaces.
The understanding of polymer systems progressed tremendously from
the late 1960's because
of innovation in experimental techniques such as
X-ray and neutron diffraction and light scattering.
Some techniques
like ellipsometry, second harmonics generation (SHG), Brewster angle microscopy,
surface force apparatus, atomic force microscopy (AFM) and
X-ray or neutron reflectivity are especially
appropriate to study polymers at interfaces.
Of equal merit was the advancement in theoretical
methods ranging from field theoretical methods and scaling arguments
to numerical simulations.

The major progress in the field of polymer adsorption at liquid interfaces
and solid surfaces is even more recent. Even though several
excellent books~\cite{fleer,erich} and review articles
~\cite{cscv86}-\cite{fl97} exist, we feel that the present review
 is timely because we  address recent progress in
the field of chains at interfaces, paying particular attention
to charged chains. Charged polymers
are interesting from the application point of view,
since they allow for a number of water-based formulations which
are advantageous for economical and ecological reasons. Recent years
have seen a tremendous research activity on charged polymers in
bulk and at interfaces. Likewise, adsorption of biopolymers such
as DNA at planar or spherical substrates is an intermediate step
in the fabrication of gene-technology related structures, and
therefore of great current interest. In addition to being
charged, DNA is rather stiff  on the nanoscopic length scale.
On intermediate length scales, it can
be well described as a semi-flexible polymer, in contrast to
most synthetic polymers,   which are well represented by
flexible polymer models. Accordingly, we discuss the complexity of
charged and semi-flexible chains in addition to neutral and
flexible ones. We also contrast the situation of physical
adsorption of chains with that of terminally attached chains
(neutral or charged) to surfaces.

This review is focused on  physical aspects of polymer adsorption
at thermodynamical equilibrium and
summarizes the main theoretical and recent progress. We only outline
theoretical calculations and do not explain in detail theoretical and experimental
techniques. Whenever possible we try to
explain principal concepts in simple terms.
Experimental results are mentioned when they are of direct relevance but this review should
not be considered as an exhaustive review of various experimental
techniques and data.

The review  starts by explaining well known facts about
conformations of a single ideal chain as well as  self-avoiding chain
and their behavior in solution (Sec.~2). We then examine the
effect of charges on the statistics of an isolated chain and of
multi-chains in
solution (Sec.~3). The rest of the paper deals with adsorption in
several distinct situations: a general introduction to adsorption processes (Sec.~4),
 adsorption of a single neutral chain
(Sec.~5) and of a single polyelectrolyte chain (Sec.~6), mean
field theories for adsorption of neutral (Sec.~7) and charged (Sec.~8)
chains. Corrections to mean-field theories are considered in
Secs.~7 and 9. In Sec.~10 the interaction between two
adsorption layers is presented,
while adsorption on more complicated substrates
such as heterogeneous and curved interfaces are briefly discussed
in Secs.~11-12. Finally, chains that are terminally anchored to the
surface are mentioned in Sec.~13. These polymer brushes are
discussed both for neutral and charged chains.

Although this review is written as one coherent manuscript, expert readers
can skip the first three sections and concentrate on adsorption of neutral chains
(Secs.~4, 5, 7, 10, 12), adsorption of charged chains (Secs.~6, 8-12)
and grafted polymer layers (brushes) (Sec. 13).

\subsection{Types of Polymers}

\begin{figure}
 \epsfxsize=8cm
 \centerline{\vbox{\epsffile{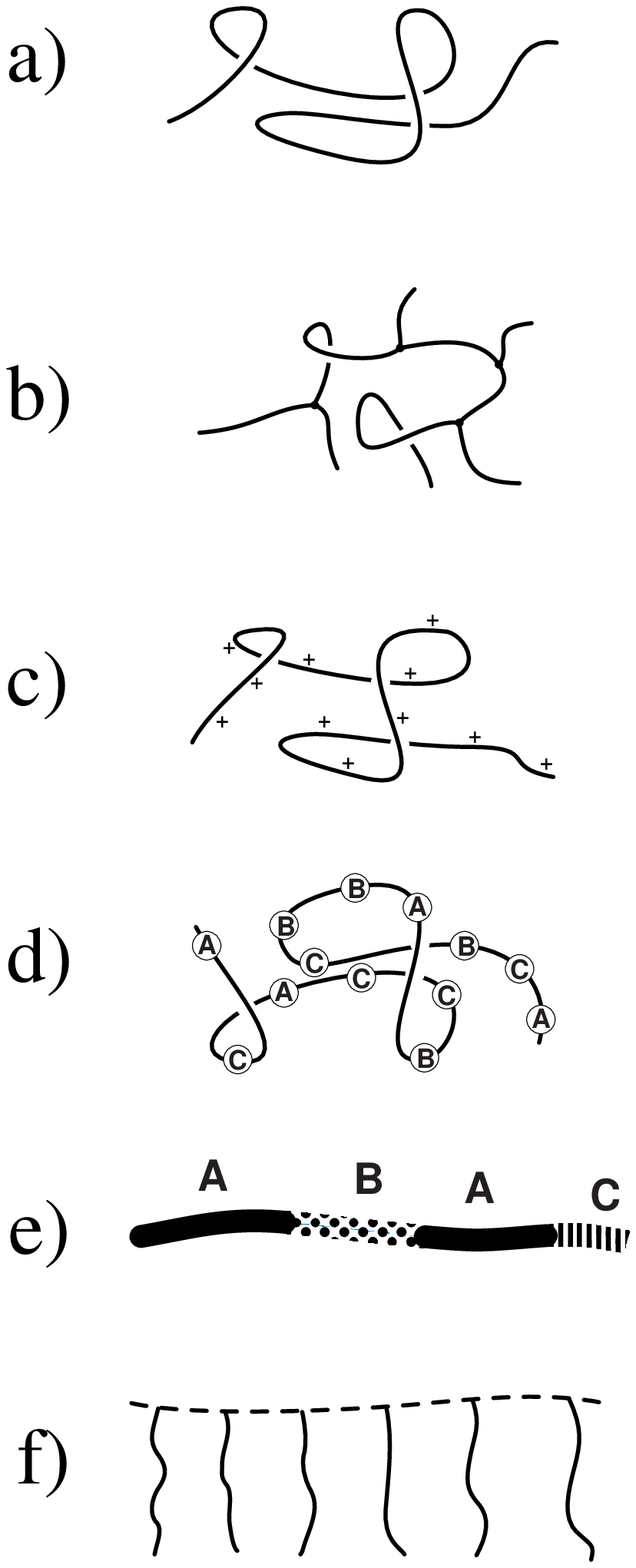} } }
\caption{ \label{fig1o}
 Schematic view of different polymer types.
a) Linear homopolymers that are the main subject of this
review. b) Branched polymers. c) Charged polymers or polyelectrolytes
(PE's), with a certain fraction of charged monomers;
d) A disordered (hetero) copolymer with no specific order
of the different monomers: A, B, C, etc.; e) A block co-polymer.
For example, a quatro-block A-B-A-C is drawn, where each of the blocks is
a homopolymer by itself; f) A copolymer composed
of a backbone (dashed line) and side chains (solid line) of different
chemical nature. The backbone could for example be hydrophilic and
make the polymer water-soluble as a whole,
while the side chain might be hydrophobic and attract other hydrophobic solutes
in the solution.}
\end{figure}

The polymers considered here are taken as linear and long chains,
 as is schematically depicted in Fig.~\ref{fig1o}a. We briefly
mention other, more complex, chain architectures. For example,
branched chains~\cite{Grosberg}, Fig.~\ref{fig1o}b,  appear in
many applications. One special type of branched structures,
Fig.~\ref{fig1o}f, is a chain having a backbone (main chain) with
repeated side branches. The chemical nature of the side and main
chain can be different. This demonstrates the difference between
homopolymers, formed from a single repeat unit (monomer) and
heteropolymers, formed from several chemical different monomers.
The heteropolymer can be statistical, e.g. DNA, where the
different units repeat in a non-periodic or random  fashion,
Fig.~\ref{fig1o}d. Another case is that of block copolymers
built from several blocks each being a homopolymer by itself.
For example, an A-B-A-C block copolymer is a chain composed of an
A, a B, an A and a C block linked serially to form a quarto-block
chain, Fig.~\ref{fig1o}e.

Synthetic polymers such as polystyrene and polyethylene are composed of flexible
chains which can be solubilized in a variety of organic solvents
like toluene, cyclohexane, etc. These polymers are highly insoluble
in water. Another class of polymers are water soluble ones. They either
have strong dipolar groups  which
are compatible with the strong polarizability of the aqueous media
(e.g., polyethylene oxide) or they
carry charged groups.

Charged polymers, also known as polyelectrolytes (PE),
are shown schematically in Fig.~\ref{fig1o}c.
They are extensively studied not only because of their numerous industrial
applications, but also from a pure scientific
interest~\cite{Oosawa}-\cite{barrat1}.
One of the most important properties of PE's is their water solubility giving rise
to a wide range of non-toxic, environmentally friendly and cheap
formulations. On the theoretical side, the physics of PE's combines the  field of
statistical mechanics of charged systems with the field of polymer
science and offers quite a number of surprises and challenges.

Two other concepts associated with PE's and water soluble polymers
are {\it associating polymers} (not discussed in this review)
and the flexibility of the polymer
chain. In cases when the copolymers have both hydrophobic and hydrophilic groups
(similarly to short-chain amphiphiles), they will self-assemble in solution to
form meso-structures such as lamellae, cylinders and spheres dispersed in solution.
The inside of these structures is where the hydrophobic chain
sections are packed, away from the
water environment.
In other cases, association of hydrophobic groups may lead to inter-chain
networking and drastically modify the visco-elasticity of the solution.
Another concept discussed at large in this review is
the chain flexibility. The chains considered here are either flexible or
semi-flexible. Flexible chains are chains where it does not cost energy to bend them, while
the stiffness of {\it semi-flexible} chains is an important property.
For PE's the charge groups contribute substantially to the
chain stiffness, and the chain conformational degrees of freedom
are coupled with the electrostatic ones.

\section{Neutral Polymer Chains}
\setcounter{equation}{0}

\subsection{Flexible Chains}

 The statistical thermodynamics of flexible chains
is well developed and the theoretical concepts can be applied
with a considerable degree of confidence
\cite{Grosberg,Flory1,Yamakawa,degennes,Cloizeaux}.
In contrast to other molecules or particles, polymer chains
contain not only translational and rotational degrees of
freedom, but also a vast number of conformational degrees of
freedom. This fact
plays a crucial role in determining their behavior in solution and
at surfaces.
When flexible chains adsorb on surfaces they form {\em diffusive}
adsorption layers extending away from the surface into the
solution. This is in contrast to semi-flexible or rigid chains,
which can  form dense and compact adsorption layers.

From the experimental point of view,
the main parameters used to describe a polymer chain are
the polymerization index $N$, which counts the number of repeat
units or monomers along the chain, and the monomer size $b$, being
the  size of one monomer or the distance between two neighboring
monomers. The  monomer size ranges from  a few Angstroms for synthetic
polymers to a few nanometers for biopolymers ~\cite{Flory1}.

\begin{figure}
 \epsfxsize=12cm
 \centerline{\vbox{\epsffile{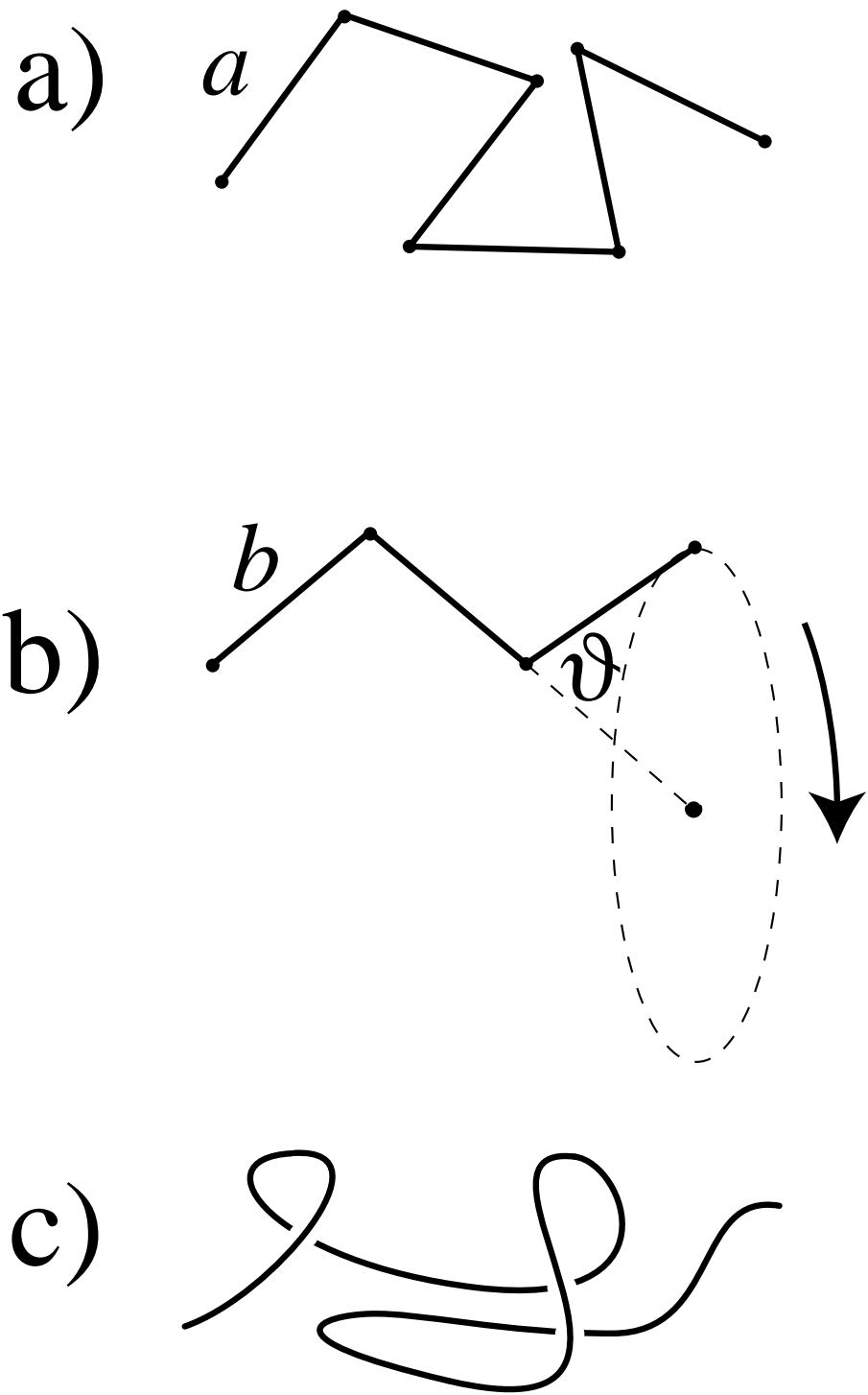} } }
\caption{ \label{fig5o}
a) Freely-jointed chain (FJC) model, where $N$ bonds of length $a$
are connected to form a flexible chain.
b) Freely-rotating chain (FRC) model, which describes a
polymer chain with a saturated carbon backbone. It consists of
a chain of $N$ bonds of length
$b$, with fixed bond  angles $\vartheta$, reflecting the chemical
bond structure, but with  freely rotating  torsional
angles.
c) The simplified model, appropriate for more advanced theoretical
calculations, consists of a structureless line, governed by some
bending rigidity or line tension. This continuous model can be used when
the relevant length scales are much larger than the monomer size.
}
\end{figure}

The simplest theoretical description
of flexible chain conformations is achieved
with the so-called freely-jointed chain (FJC) model,
where a polymer consisting
of $N+1$ monomers is represented by $N$ bonds defined by bond
vectors ${\bf r}_j$ with $j = 1, \ldots N$. Each bond vector has
a fixed length $|{\bf r}_j| = a$ corresponding to the Kuhn
length, but otherwise is allowed to rotate freely, as is
schematically shown in Fig.~\ref{fig5o}a. This model of course
only gives a coarse-grained description of real polymer chains,
but we will later see that by a careful adjustment of the Kuhn length $a$
(which is related but not identical to
the monomer size $b$), an accurate description
of the large-scale properties of real polymer chains is possible.
The main advantage is that due to the simplicity of the FJC model,
all interesting
observables (such as chain size or distribution functions) can
be calculated with relative ease.
Fixing one of the chain
ends at the origin, the position of the $(k+1)$-th monomer is
given by the vectorial sum
 \be
  {\bf R}_k = \sum_{j=1}^k {\bf r}_j ~.
 \ee
Because  two arbitrary bond vectors are uncorrelated in this
simple model, the thermal average over the scalar product of two
different bond vectors vanishes, $\langle {\bf r}_j\cdot {\bf r}_k
\rangle = 0 $ for $j \neq k$, while the mean squared bond vector
length is simply given by $\langle {\bf r}_j^2 \rangle = a^2 $.
It follows that the mean squared end-to-end radius $R^2$  is
proportional to the number of monomers,
 \be \label{idealscaling}
 \label{fjc} R^2 \equiv \langle {\bf R}_N^2 \rangle = N a^2 = La ,
 \ee
where the contour length of the chain is given by $L= Na$.
The same result is obtained for the mean quadratic displacement
of a freely diffusing particle  and alludes to the same underlying
physical principle, namely the statistics of Markov processes.

\begin{figure}
 \epsfxsize=12cm
 \centerline{\vbox{\epsffile{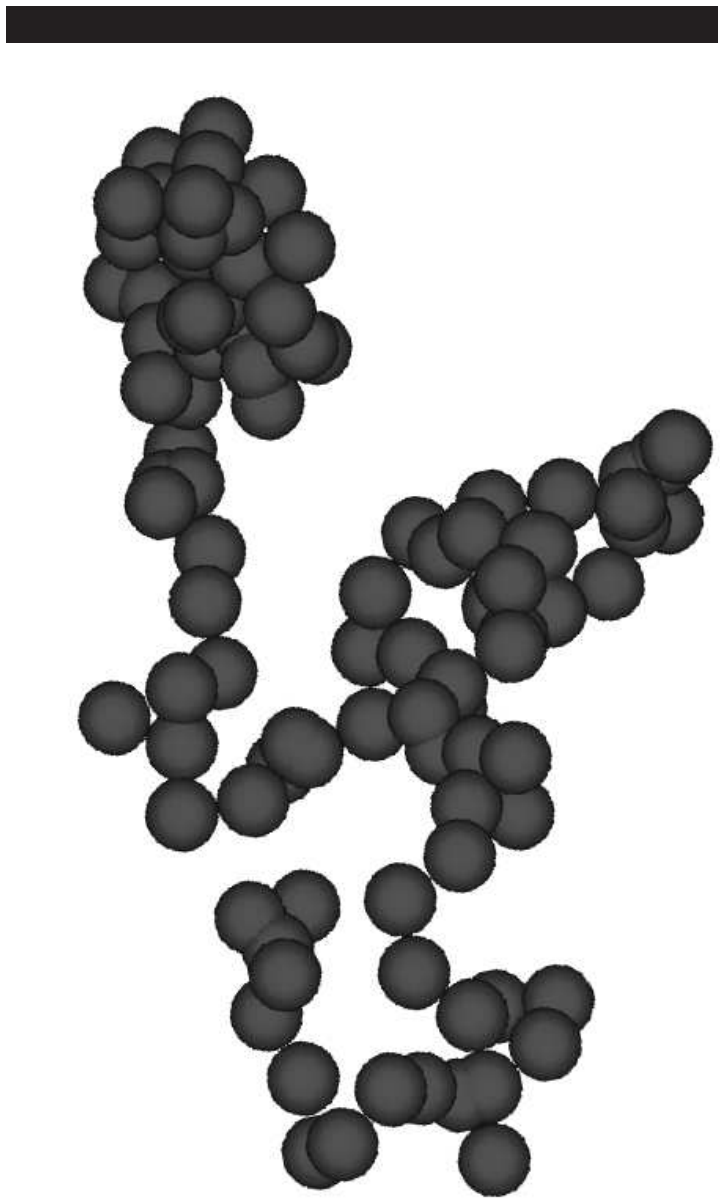} } }
\caption{ \label{fig1a} Snapshot of  a Monte-Carlo simulation of a
neutral freely-jointed chain (FJC) consisting of $N=100$ monomers with
a diameter corresponding to the Kuhn length $a$. The theoretical
end-to-end radius,  $R = 10 a$, is indicated by a straight bar. }
\end{figure}

In Fig.~\ref{fig1a} we show a snapshot of a Monte-Carlo
simulation of a freely-jointed chain consisting of 100
non-interacting monomers, each being represented by a sphere of
diameter $a$. The bar has a length of $10 a$,
which according to Eq.~(\ref{fjc}) is the average distance between the
chain ends. As can be seen in the figure,
the end-to-end radius gives a good idea of
the typical chain size.

In the so-called {\em freely rotating chain} (FRC) model,
different chain conformations are produced by torsional rotations of the
polymer backbone bonds of length $b$ at fixed bond angle $\vartheta$, as
shown schematically in Fig.~\ref{fig5o}b. This model is
closer to real synthetic polymers than the FJC model, but is also
more complicated to calculate. In contrast to the FJC model,
the correlation between two neighboring bond vectors does not
vanish and is given by $\langle {\bf r}_j\cdot {\bf r}_{j+1}
\rangle = b^2 \cos \vartheta $. Correlations between further-nearest
neighbors are transmitted through the backbone and one thus obtains
for the bond-vector correlation function\cite{Grosberg}
\be \label{FRCcorr}
\langle {\bf r}_j\cdot {\bf r}_{k}
\rangle = b^2 (\cos \vartheta)^{|j-k|}.
\ee
The mean-squared end-to-end radius is for this model in the limit
of long chains ($N \rightarrow \infty$) given by\cite{Grosberg}
 \be \label{FRC}
 R^2 \simeq N b^2 \frac{1+\cos \vartheta}{1- \cos \vartheta}.
 \ee
We will now demonstrate that the simple result for the FJC model,
Eq.~(\ref{idealscaling}), applies  on length scales which are large
compared with the microscopic chain details also to the more complicated
FRC model (which takes the detailed microscopic chain structure into
account). To make the connection between the two models, we observe that
the FRC contour length is $L = N b \cos (\vartheta/2)$. Using the scaling relation
$R^2 = a L $ (which we established for the FJC model) as a definition for the
Kuhn length $a$, we obtain for the FRC model
\be
a = b \frac{1+\cos \vartheta}{\cos (\vartheta /2) (1- \cos \vartheta)},
\ee
where the Kuhn length $a$ is now interpreted as an
effective monomer size. For a typical
saturated carbon backbone one finds a bond angle
$\vartheta \approx 70^o$ and thus obtains for the relation
between the  Kuhn length and the monomer size $ a \approx 2.5 b$.
With a typical bond length of $b \approx 0.15 nm$ this results
in a Kuhn length of $a \approx 0.38 nm$.
Clearly, the Kuhn length $a$ is always larger than the monomer size $b$.
We have thus shown that it is possible to use the simple FJC model
also for more detailed chain models if one interprets the Kuhn length
$a$ as an effective length which takes correlations between chemical
bonds into account. In the remainder of this review, we will in most cases
use a flexible chain model characterized by the Kuhn length $a$.
Only in cases where the microscopic structure of the polymer chains matters
will we use more detailed models (and then have to distinguish between the
Kuhn length $a$, characterizing the large-scale properties of a chain,
and the monomer size $b$).

In many theoretical calculations aimed at elucidating
large-scale properties, the simplification is carried even
a step further and a continuous model is used, as schematically
shown in Fig.~\ref{fig5o}c. In such models the polymer backbone is
replaced by a continuous line and all microscopic details are neglected.

The models discussed so far describe ideal Gaussian chains
and do not account for interactions between monomers which are
not necessarily close neighbors along the backbone. Including
these interactions will give a different scaling behavior for
long polymer chains. The end-to-end radius, $R=\sqrt{\langle
R^2_N\rangle}$, can be written more generally for $N\gg 1$ as
\be R \simeq a N^{\nu}.
\ee
For an ideal polymer chain (no interactions between monomers),
 Eq.~(\ref{idealscaling})
implies $\nu = 1/2$. This holds only for
polymers where the attraction between monomers (as compared with
the monomer-solvent interaction) cancels the steric repulsion
(which is due to the fact that the monomers cannot penetrate each
other). This situation can be achieved in the
condition of ``theta" solvents.
More generally, polymers in solution can experience three types of
solvent conditions, with theta solvent condition  being
intermediate between ``good" and ``bad" solvent conditions.
The solvent quality depends mainly on the specific chemistry
determining the interaction between the solvent molecules and
monomers. It can be changed by varying the temperature.

In good solvents the monomer-solvent
interaction is more favorable than the monomer-monomer one.
Single polymer chains in good solvents have ``swollen'' spatial
configurations dominated by the steric repulsion, characterized by
an exponent $\nu \simeq 3/5$~~\cite{Flory1}. This spatial size of
a polymer coil is much smaller than the extended contour length
$L=aN$ but larger than the size of an ideal chain $aN^{1/2}$. The
reason for this peculiar behavior is entropy combined with the
favorable interaction between monomers and solvent molecules in
good solvents, as we will see in the following section.
Similarly, for adsorption of polymer chains
on solid substrates, the conformational degrees of freedom of polymer
coils  lead to salient differences between  the adsorption
of polymers and small molecules.

In the opposite case of ``bad'' (sometimes called ``poor") solvent
conditions, the effective interaction between monomers is
attractive, leading to collapse of the chains and to their
precipitation from solution (phase separation between the
polymer and the solvent). In this case, the
polymer size, like any space filling object embedded in
three-dimensional space, scales as $N \sim R^3$, yielding
$\nu=1/3$.

\subsection{Chain Swelling and Chain Collapse: Flory Theory and Blob Formation}

The standard way  of taking into account interactions between
monomers is the Flory theory, which treats these interactions on
a mean-field level
\cite{Grosberg,Flory1,Yamakawa,degennes,Cloizeaux}. Let us first
consider the case of repulsive interactions between monomers,
which can be described by a positive second-virial
coefficient $v_2$.
 This corresponds to the aforementioned good-solvent
condition. For pure hard-core interactions and with no additional
attractions between monomers, the second virial coefficient
(which corresponds to the excluded volume) is of the order of
$a^3$,  the monomer volume. The repulsive interaction between monomers, which
tends to swell the chain, is counteracted and balanced by the
ideal chain elasticity, which is brought about by the entropy
loss associated with  stretching the chain.
The analogy with an external stretching force is helpful:
For a freely-jointed
chain, the stretching response due to an external  force $\hat{f}$
(measured in units of the thermal energy $k_BT$)
is $R
\simeq a^2 N \hat{f}/ 3 $ for weak forces
$\hat{f} \ll 1/a$ \cite{degennes}. Hence, a freely-jointed chain
acts like an ideal spring with a spring constant  (in units
of $k_BT$ ) of $3
/(2 a^2 N)$. The temperature dependence of the spring constant
tells us that the chain elasticity is purely entropic.
The origin is that the number
of polymer configurations having an end-to-end radius of the
order of the unperturbed end-to-end radius is large. These
configurations are entropically favored over
configurations characterized by a large end-to-end radius, for
which the number of possible polymer conformations is drastically
reduced. The standard Flory theory~\cite{Flory1} for a flexible
chain of radius $R$ is based on writing the free energy ${\cal
F}$ (in units of the thermal energy $k_BT$)
as a sum of two terms (omitting numerical prefactors)
\be {\cal F} \simeq  \frac{ R^2}{ a^2 N } + v_2 R^3 \left(
\frac{N }{R^3} \right)^2 ~,
\ee
where the first term is the entropic elastic energy associated
with swelling a polymer chain to a radius $R$, proportional to the
effective spring constant of an ideal chain, and the second term
is the second-virial repulsive energy  proportional to the
coefficient $v_2$,  and the segment density squared. It is
integrated over the volume $R^3$. The optimal radius $R$ is
calculated by minimizing  this free energy and gives the swollen
radius
\be \label{swollenflex} R \sim a (v_2/a^3)^{1/5} N^\nu ~,
\ee
with $\nu = 3/5$. For purely steric interactions with $v_2 \simeq
a^3$ we obtain $R \sim a N^\nu$. For $v_2  <a^3$ one finds that
the swollen radius Eq.~(\ref{swollenflex}) is only realized above
a minimal monomer number $N_{\rm sw} \simeq (v_2/a^3)^{-2}$ below
which the chain statistics is unperturbed by the interaction and
the scaling of the chain radius is Gaussian and given by
Eq.~(\ref{idealscaling}). A different way of looking at this
crossover from Gaussian to swollen behavior is to denote a
Gaussian coil of monomer number $N_{\rm sw}$ as a blob  with size
$R_{\rm sw} = a N_{\rm sw}^{1/2} \simeq a^4/v_2$, after which the
swollen radius Eq.~(\ref{swollenflex}) can be rewritten as
\be \label{swollenblob}
R \sim R_{\rm sw}  (N/N_{\rm sw})^\nu~.
\ee
The swollen chain can be viewed as chain of $N/N_{\rm sw}$
impenetrable blobs, each with a spatial size $R_{\rm sw}$
\cite{degennes}.

In the opposite limit of negative second virial coefficient,
corresponding to the bad or poor solvent regime,
the polymer coil will be collapsed due to the attractions between monomers.
In this case, the attraction term in the free energy
is balanced by the third-virial term in a
low-density expansion (where we assume that $v_3 >0$),
\be
{\cal F} \simeq v_2 R^3 \left( \frac{N }{R^3} \right)^2
+v_3 R^3 \left( \frac{N }{R^3} \right)^3~.
\ee
Minimizing this free energy with respect
to the chain radius one obtains
\be
\label{collapsed} R \simeq (v_3 / |v_2| )^{1/3} N^\nu~.
\ee
with $\nu = 1/3$. This indicates the formation of a compact
globule, since the monomer density  inside the globule, $c_m \sim
N / R^3 $, is independent of the chain length. The minimal  chain
length  to observe a collapse behavior is $N_{\rm col} \sim (v_3 /
a^3 v_2)^2$, and the chain radius Eq.~(\ref{collapsed}) can be
rewritten as $R \sim R_{\rm col} (N/N_{\rm col})^{1/3}$, where the
size of a Gaussian blob is $R_{\rm col} \sim a N_{\rm col}^{1/2}$.
For not too long chains and a second virial coefficient not too
much differing from zero, the interaction is irrelevant and one
obtains effective Gaussian or ideal behavior. It should be noted,
however, that even small deviations from the exact theta
conditions (defined by strictly $v_2 =0$) will lead to chain
collapse  or swelling for very long chains.

\subsection{Semi-Flexible Chains}

The freely-rotating  chain model exhibits orientational
correlations between bonds that are not too far from each other,
see Eq.~(\ref{FRCcorr}). These correlations give rise to a
certain chain stiffness, which plays an important role for the
local structure of polymers, and leads to more rigid structures.
For synthetic polymers with bond torsional degrees of freedom,
this stiffness is due to fixed bond angles and is further enhanced
by the hindered rotations around individual back-bone bonds
\cite{Flory1}, as schematically shown in Fig.~\ref{fig5o}b. This
effect is even more pronounced for polymers with bulky side
chains, where, because of steric constraints, the persistence
length can be of the order of a few nanometers \cite{Flory1}. This
stiffness can be conveniently characterized by the persistence
length $\ell_0$, defined as the length over which the normalized
bond (tangent) vectors at different locations on the chain are
correlated. In other words, the persistence length gives an
estimate for the typical radius of curvature, while taking into
account thermal fluctuations. For the FRC model, the persistence
length $\ell_0$ is defined by
\be  \label{FRCpers} \langle {\bf r}_j\cdot {\bf r}_k \rangle =
b^2 {\rm e}^{-|j-k| b \cos(\vartheta/2) / \ell_0}. \ee With the
result Eq.~(\ref{FRCcorr}), one obtains for the FRC model the
persistence length
\be
\ell_0 = \frac{ b \cos (\vartheta/2)}{| \ln \cos \vartheta |}.
\ee
For typical  saturated carbon backbones  with
$\vartheta \approx 70^o$ one obtains a persistence length
of $\ell_0 \approx 0.8 b$ which is thus of the order of the bond length.
Clearly, as the bond angle goes down, the persistence length
increases dramatically.

Biopolymers with a more complex structure on the molecular level
tend to be stiffer than simple synthetic polymers. Some typical
persistence lengths encountered in biological systems are $\ell_0
\approx 5$\,mm for tubulin \cite{tubulin},
 $\ell_0 \approx 20$\,$\mu$m for actin \cite{actin1,actin2}, and
$\ell_0 \approx 50$\,nm for double-stranded DNA \cite{DNA}.
Because some of these biopolymer are charged, we will discuss in
Sec.~3.2 at length the dependence of the persistence length on
the electrostatic conditions. In some cases the main contribution
to the persistence length comes from the repulsion between
charged monomers. In these cases, it is important  to include the
effect of stiffness into the theoretical description, even if the
bare or mechanical stiffness is only slightly larger than the
monomer size.

\begin{figure}
 \epsfxsize=12cm
 \centerline{\vbox{\epsffile{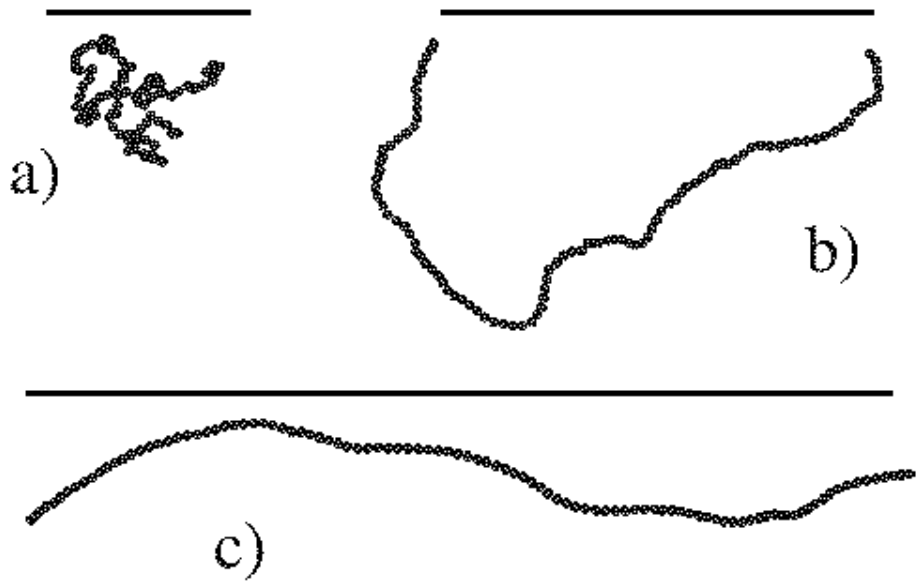} } }
\caption{ \label{fig1} Snapshots of  Monte-Carlo simulations of a
neutral and semi-flexible chain consisting of $N=100$ monomers
with a diameter $b$. The theoretical end-to-end radius $R$ is
indicated by a straight bar. The persistence lengths used in the
simulations are:  a) $\ell_0/b = 2 $, leading according to
Eq.~(\ref{Re2}) to $R/b = 19.8 $, b) $\ell_0/b = 10 $, leading to
$R/b = 42.4 $, c) $\ell_0/b = 100 $, leading to $R/b = 85.8 $.}
\end{figure}

To describe the bending rigidity of neutral polymers, it is easier
to use a continuum model, where one neglects the discrete nature
of monomers, as shown in Fig.~\ref{fig5o}c. In this approach the
bending energy (rescaled by the thermal energy, $k_B T$) of a
stiff or semi-flexible polymer of contour length $L$,
which is parameterized by the space curve ${\bf r}(s)$, is given by
\cite{Grosberg}
 \be \label{bendingenergy}
 \frac{\ell_0}{2} \int_0^L {\rm d} s \; \left( \frac{ {\rm
d}^2 {\bf r}(s)}{{\rm d} s^2} \right)^2~,
 \ee
where ${\rm d}^2 {\bf r}(s)/{\rm d} s^2$ is the local curvature
of the polymer. We assume here that the polymer segments are
non-expendable, \ie\ the tangent vectors
$\dot{\bf r} (s) = {\rm d} {\bf r}(s)/{\rm d} s $
are always normalized,
$| \dot{\bf r} (s) |=1$. Clearly, this continuum
description will only be good if the persistence length is larger
than the monomer size $b$. For the semi-flexible polymer
model, the correlations between tangent vectors exhibit
a purely exponential decay,
\be
\langle \dot{\bf r} (s) \cdot \dot{\bf r} (s') \rangle =
{\rm e}^{-|s-s'|/\ell_0 }.
\ee
From this result, the mean-squared end-to-end radius of a
semi-flexible chain, described by the bending energy
Eq.~(\ref{bendingenergy}), can be calculated  and
reads~\cite{Grosberg}
\begin{equation} \label{Re2}
R^2
=2  \ell_0 L +2 \ell_0^2\left( {\rm e}^{-L/\ell_0}-1\right)~,
\end{equation}
where the persistence length  is $\ell_0$ and the total contour
length of a chain is $L$. Two limiting behaviors  are obtained
for $R$ from Eq.~(\ref{Re2}): for long chains  $L \gg \ell_0$, the
chain behaves as  a flexible one, $R^2 \simeq 2 \ell_0 L $; while
for rather short chains, $ L \ll \ell_0$, the chain behaves as
a rigid rod, $R \simeq L$. Comparison with the scaling of the
freely-jointed chain model (having no persistence length, $\ell_0=0$),
Eq.~(\ref{fjc}), shows that a
semi-flexible chain can, for $L \gg \ell_0$, be described by a
freely-jointed chain model with an effective Kuhn length of
 \be
  a= 2 \ell_0~,
\ee
 and an effective number of segments
 \be
  N = \frac{L}{2 \ell_0}~,
 \ee
In this case the Kuhn length takes into account the chain
stiffness.
In Fig.~\ref{fig1} we
show snapshots taken from a Monte-Carlo simulation of a
semi-flexible chain consisting of 100 polymer beads of diameter
$b$. The persistence length is varied from $\ell_0=2b$
(Fig.~\ref{fig1}a), over $\ell_0=10b$ (Fig.~\ref{fig1}b), to
$\ell_0=100b$ (Fig.~\ref{fig1}c). Comparison with the
freely-jointed chain model is given in Fig.~\ref{fig1a} ($a=b$,
$\ell_0=0$). It is seen that as the persistence length is
increased, the chain structure becomes more expanded. The average
end-to-end radius $R$, Eq.~(\ref{Re2}), is shown as the bar on
the figure and gives a  good estimate on typical sizes of
semi-flexible polymers.

The main point here is that even though the semi-flexible polymer
model describes biopolymers much better than the freely-rotating
model does, on large length scales both models coincide if the
Kuhn length $a$ of the freely-jointed chain model is the effective
length which is extracted from the scaling of the end-to-end
radius in the semi-flexible model, Eq.~(\ref{Re2}). When the
small-scale behavior is probed, as for example in the case of
polymer adsorption with  short-ranged potentials, see Sec.~6, the
difference between the models matters and one has to use the
semi-flexible model. On the other hand, it should be kept in mind
that the semi-flexible polymer model is an idealization, which
neglects the detailed architecture of the polymer at the molecular
level. For synthetic polymers, a freely-rotating chain model with
a bond length $b$ and a bond angle $\vartheta$ as shown in
Fig.~\ref{fig5o}b is closer to reality but is more complicated to
handle theoretically \cite{Grosberg}.

\subsection{Dilute, Semi-Dilute and Concentrated Solutions}

It is natural to generalize the discussion of single chain
behavior to that of many chains for  dilute monomer
concentrations. The dilute regime is defined by $c_m< c_m^{*}$,
for which $c_m$ denotes the monomer concentration (per unit volume)
and $c_m^{*}$ is the concentration where individual chains start
to overlap. Clearly, the overlap concentration is reached when
the average bulk monomer concentration exceeds the monomer
concentration inside a polymer coil. To estimate the overlap
concentration $c_m^*$, we simply note that the average monomer
concentration inside a coil with radius $R \sim a N^\nu$ is
given by
\be  \label{overlap}
c_m^* \simeq \frac{N}{ R^3} \sim N^{1-3\nu} a^{-3}~.
\ee
For ideal chains with $\nu=1/2$ the overlap concentration scales
as $a^3 c_m^* \sim N^{-1/2}$ and thus decreases slowly as the
polymerization index $N$ increases. For swollen chains with $\nu =
3/5$, on the other hand, the overlap concentration scales as $a^3
c_m^* \sim N^{-4/5}$ and thus decreases more rapidly with
increasing chain length. The crossover to the concentrated or
melt-like regime occurs when the monomer concentration in the
solution reaches the local monomer concentration  inside a
Gaussian blob, which is for good solvent conditions given by (see
the discussion before Eq.~(\ref{swollenblob})) \be c_m^{**} \simeq
N_{\rm sw}/R_{\rm sw}^3 \simeq v_2 /a^6~. \ee It is seen that the
semi-dilute regime, obtained for concentrations $c_m^* < c_m <
c_m^{**}$, spans for long chains and under good solvent conditions
a rather wide range of concentrations and is thus important for
typical applications.

For chains characterized by a negative second virial coefficient,
attractions between collapsed single-chain globules lead to phase
separation between a very dilute solution of single-polymer
globules and a dense melt-like phase of entangled polymer coils
\cite{degennes}.

\section{Charged Polymer Chains}
\setcounter{equation}{0}

\subsection{Interactions Between Charged Objects}

A polyelectrolyte (PE) is a polymer where a fraction $f$ of its
monomers are charged. When this fraction is small, $f\ll 1$, the
PE is weakly charged, whereas when $f$ is close to
unity, the polyelectrolyte is strongly charged. There are two
common ways to control $f$~\cite{barrat1}. One
way is to polymerize a
heteropolymer using strongly acidic and neutral monomers as building
blocks. Upon contact
with water, the acidic groups dissociate into positively charged
protons (H$^+$) that bind immediately to water molecules, and negatively
charged monomers. Although this process effectively charges the
polymer molecules, the counterions make the PE solution
electro-neutral on larger length scales.
The charge distribution along the chain is quenched
(``frozen'') during the polymerization stage, and it is
characterized by the fraction of charged monomers on the chain,
$f$. In the second way, the PE is a weak polyacid or
polybase. The effective charge of each monomer is controlled by
the pH of the solution. Moreover, this annealed fraction depends
on the local electric potential. This is in particular important
for adsorption processes since the local electric field close to a
strongly charged surface can be very different from its value in
the bulk solution.

The counterions are attracted to the charged polymers via
long-ranged Coulomb interactions, but this physical association
typically only leads to a rather loosely bound counterion cloud
around the PE chain.
Because PE's are present in a background of a polarizable and
diffusive counterion cloud, there is a strong influence of the
counterion distribution on the PE structure, as will be discussed
at length in this section. Counterions contribute significantly
towards bulk properties, such as the osmotic pressure, and their
translational entropy is responsible for the generally good water
solubility of charged polymers. In addition, the statistics of PE
chain conformation is governed by intra-chain Coulombic repulsion
between charged monomers, resulting in a more extended and
swollen conformation of PE's as compared to neutral polymers.

For polyelectrolytes, electrostatic interactions provide the driving force
for their salient features and have to be included in
any theoretical description. The reduced electrostatic interaction
between two point-like charges
 can be written as $ z_1 z_2 v(r)$ where
\begin{equation} \label{intro1}
v (r) = \ell_B / r
\end{equation}
is the Coulomb interaction between two elementary charges
in units of $k_BT$ and $z_1$
and $z_2$ are the valencies (or the reduced charges in units of
the elementary charge $e$).
The Bjerrum length $\ell_B$ is defined as
\begin{equation}
\ell_B = \frac{e^2}{ \varepsilon k_BT},
\end{equation}
where $\varepsilon$ is the medium dielectric constant.
It denotes the distance at which the Coulombic interaction
between two unit charges in a dielectric medium is equal to
thermal energy ($k_BT$). It is a measure of the distance below
which the Coulomb energy is strong enough to compete with the
thermal fluctuations; in water at room temperatures, one finds
$\ell_B \approx 0.7$\,nm.

The electrostatic interaction in a homogeneous medium
depends only on the distance $r$ between the
charges. The total electrostatic energy of a given distribution
of charges is obtained from adding up all pairwise interactions
between charges according to Eq.~(\ref{intro1}). In principle,
the equilibrium behavior of an ensemble of charged particles
(\eg\  a salt solution) follows from the partition function, \ie\
the weighted sum over all different microscopic configurations,
which
--- via the Boltzmann factor --- depends on the electrostatic
energy of each configuration. In practice, however, this route is
very complicated for several reasons:

 i) The Coulomb interaction, Eq.~(\ref{intro1}), is long-ranged and
 couples
 many charged particles. Electrostatic problems are
typically {\em many-body problems}, even for low densities.

ii) Charged objects in most cases are dissolved in water. Like any
material, water is polarizable and  reacts to the presence of a
charge with polarization charges. In addition, and this is by far
a more important effect, water molecules carry a permanent dipole
moment that partially orients in the vicinity of charged objects.
Within linearized response theory, these  polarization effects
can be incorporated by the dielectric constant of water,
a procedure which of course neglects non-local and non-linear effects.
Note that for water, $\varepsilon \approx 80$, so that
electrostatic interactions and self energies are much weaker in
water than in air (where $\varepsilon \approx 1$) or some other
low-dielectric solvents. Still, the electrostatic interactions
are especially important in polar solvents because in these
solvents, charges dissociate more easily than in apolar solvents.

iii) In  biological systems and most industrial applications, the
aqueous solution contains mobile salt ions. Salt ions of opposite
charge are drawn to the charged object and form a loosely bound
counterion cloud around it. They effectively reduce or {\em
screen} the charge of the object. The effective (screened)
electrostatic interaction between two charges $z_1 e$ and $z_2 e$
in the presence of salt ions and a polarizable solvent can be
written as $z_1 z_2 v_{\rm DH}(r)$, with the Debye-H\"uckel (DH)
potential $v_{\rm DH}(r)$ given (in units of $k_BT$) by
\begin{equation}
 \label{introDH}
v_{\rm DH} (r) = \frac{\ell_B}{r} {\rm e}^{-\kappa r}~.
\end{equation}
 The exponential decay is characterized
by the screening length $\kappa^{-1}$, which is related to the
salt concentration $c_\s$ by
\begin{equation}
\kappa^2 = 8 \pi z^2 \ell_B c_\s~,
\end{equation}
where $z$ denotes the valency of $z:z$ salt. At
physiological conditions the salt concentration is  $c_\s \approx
0.1$\,M and for monovalent ions ($z=1$) this leads to $\kappa^{-1}
\approx 1$\,nm. This means that although the Coulombic interactions
are long-ranged, in physiological conditions they are highly screened
above length scales of a few nanometers, which results from multi-body
correlations between ions in a salt solution.

The Debye-H\"uckel potential in Eq.~(\ref{introDH}) results from a
linearized mean-field procedure, and becomes inaccurate when i)
the number of correlated ions is small and ii) when the typical
interaction between ions exceeds the thermal energy. In the
following we estimate the validity of the DH approximation using
simple scaling arguments: The number of ions which are correlated
in a salt solution with concentration $c_\s$ is of the order of $n
\sim \kappa^{-3} c_\s$, where one employs the screening length
$\kappa^{-1}$ as the scale over which ions are correlated. Using
the definition $\kappa^2 = 8 \pi z^2 \ell_B c_\s$, one obtains $n
\sim (z^2 \ell_B c_\s^{1/3})^{-3/2}$. The average distance between
ions is roughly $r_\s \sim c_\s^{-1/3}$. The typical electrostatic
interaction between two ions in the solution, rescaled by the
thermal energy, thus is $W_{\rm el}  \sim z^2 \ell_B /r_\s \sim
z^2 \ell_B c_\s^{1/3}$ and we obtain $W_{\rm el} \sim n^{-2/3}$.
Using these scaling arguments one obtains that either i) many ions
are weakly coupled together (\ie\ $n\gg 1$ and $W_{\rm el} \ll
1$), or ii) a few ions interact strongly with each other ( $n
\simeq W_{\rm el} \simeq 1$). In the first case, and in the
absence of external fields, the approximations leading to the
Debye-H\"uckel approximation, Eq.~(\ref{introDH}), are valid. In
the second case, correlation effects and nonlinear effects become
important, as will be  discussed at various points in this review.

\subsection{Isolated Polyelectrolyte Chains}

  We discuss now the
scaling behavior of a single semi-flexible PE  in the bulk,
including chain stiffness and electrostatic repulsion between
monomers. For charged polymers, the effective persistence length
is increased due to electrostatic repulsion between monomers. This
effect modifies considerably not only the PE behavior in solution
but also their adsorption characteristics.

The scaling analysis is a simple extension of previous
calculations for flexible (Gaussian)
PE's~\cite{gpv,Khokhlov,barrat2,Netz2}. The semi-flexible polymer
chain is characterized by a bare persistence length $\ell_0$ and
a linear charge density $\tau$. Using the monomer length $b$ and
the fraction of charged monomers $f$ as parameters, the linear
charge density can be expressed as $\tau = f/b$. Note that in the
limit where the persistence length is small and comparable to a
monomer size, only a single length scale remains, $\ell_0 \simeq
a \simeq b$. Many interesting effects, however, are obtained in
the general case treating the persistence length $\ell_0$ and the
monomer size $b$ as two independent parameters. In the regime
where the electrostatic energy is weak, and for long enough
contour length $L$, $L \gg \ell_0$,  a polymer coil will be
formed with a radius $R$ unperturbed by the electrostatic
repulsion between monomers. According to Eq.~(\ref{Re2}) we get
$R^2 \simeq 2 \ell_0 L$. To estimate when the electrostatic
interaction will be sufficiently strong to swell the polymer coil
we recall that the electrostatic energy (rescaled by the thermal
energy $k_BT$) of a homogeneously charged sphere of total charge
$Z$ (in units of the elementary charge $e$)  and radius $R$ is
\be
W_{\rm el} = \frac{ 3 \ell_B Z^2}{5 R}~.
\ee
The exact charge distribution inside the sphere only changes the
prefactor of order unity and is not important for the scaling
arguments. For a polymer of length $L$ and line charge density
$\tau$ the total charge is $Z= \tau L$. The electrostatic energy
of a (roughly spherical) polymer coil  is then
\be
W_{\rm el} \simeq \ell_B \tau^2 L^{3/2} \ell_0^{-1/2}~.
\ee
The polymer length $L_{\rm el}$ at which the electrostatic self energy is of
order $k_B T$, \ie\ $W_{\rm el}  \simeq 1$, is then
\begin{equation}
 \label{elblob}
  L_{\rm el} \simeq \ell_0 \left( \ell_B \ell_0
\tau^2 \right)^{-2/3}~,
\end{equation}
and defines the electrostatic blob size or electrostatic polymer
length. We expect a locally crumpled polymer configuration if
$L_{\rm el} > \ell_0$, \ie\  if
\be
\tau \sqrt{\ell_B \ell_0} < 1 ~,
\ee
because the electrostatic repulsion between two segments of
length $\ell_0$ is smaller than the thermal energy and is not
sufficient to align the two segments. This is in accord with more
detailed calculations by  Joanny and Barrat~\cite{barrat2}. A
recent general Gaussian variational calculation confirms this
scaling result and in addition yields logarithmic
corrections~\cite{Netz2}. Conversely, for
\be \label{pers}
\tau \sqrt{\ell_B \ell_0} > 1 ~,
\ee
electrostatic chain-chain repulsion is already relevant on length
scales comparable to the persistence length. The chain is
expected to have a  conformation characterized by an effective
persistence length $\ell_{\rm eff}$, larger than the bare
persistence length $\ell_0$, \ie\  one expects $\ell_{\rm eff}
>\ell_0$.

\begin{figure}
 \epsfxsize=12cm
 \centerline{\vbox{\epsffile{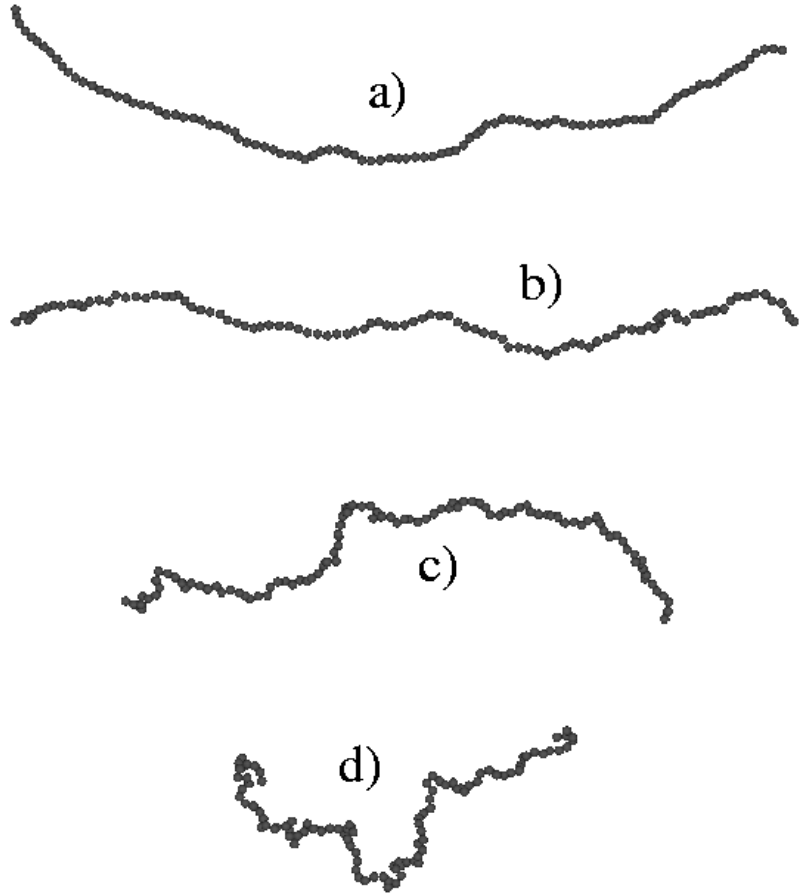} } }
\caption{  \label{fig2} Snapshots of  Monte-Carlo simulations of a
PE chain of $N=100$ monomers of size $b$. In all simulations the
bare persistence length is fixed at $\ell_0/b =1$, and the
screening length and the charge interactions are tuned such that
the electrostatic persistence length is constant and $\ell_{\rm
OSF}/b = 100 $ according to Eq.~(\ref{OSF}). The parameters used
are: a) $\kappa^{-1}/b =\protect \sqrt{50}  $ and $\tau^2 \ell_B
\ell_0 =8$, b)  $\kappa^{-1}/b =\protect \sqrt{200} $ and $\tau^2
\ell_B \ell_0 =2$, c)  $\kappa^{-1}/b =\protect \sqrt{800} $ and
$\tau^2 \ell_B \ell_0 =1/2$, and d) $\kappa^{-1}/b =\protect
\sqrt{3200} $ and $\tau^2 \ell_B \ell_0 =1/8$. Noticeably, the
weakly charged chains crumple at small length scales and show a
tendency to form electrostatic blobs.}
\end{figure}

This effect is visualized in Fig.~\ref{fig2}, where we show
snapshots of  Monte-Carlo  simulations for charged chains
consisting of  100 monomers of size $b$. The monomers are
interacting solely via screened DH potentials as defined in
Eq.~(\ref{introDH}). In all simulations the bare persistence
length equals the monomer size, $\ell_0 = b$. The screening length
$\kappa^{-1}$ and the linear charge density $\tau$ are varied such
that the ratio $\tau/\kappa$ is the same  for all four
simulations. The number of persistent segments in an electrostatic
blob can be written according to Eq.~(\ref{elblob}) as $ L_{\rm
el}/\ell_0 = (\tau^2\ell_B \ell_0 )^{-2/3}$ and yields for
Fig.~\ref{fig2}a) $ L_{\rm el}/\ell_0 = 0.25$, for \ref{fig2}b) $
L_{\rm el}/\ell_0 = 0.63$, for  \ref{fig2}c) $ L_{\rm el}/\ell_0 =
1.6$, and for \ref{fig2}d)  $ L_{\rm el}/\ell_0 = 4$. In other
words, in \ref{fig2}d) the electrostatic blobs consist of four
persistent segments, and indeed this weakly charged chain
crumples at small length scales. On the other hand, in
Fig.\ref{fig2}a) the persistence length is four times larger than
the electrostatic blob length and therefore the chain is straight
locally. A typical  linear charge density reached with synthetic
PE's is one charge per two carbon bonds (or, equivalently, one
charge per monomer), and it corresponds to
 $\tau \approx 4$\,nm$^{-1}$.
Since for these highly flexible synthetic PE's the bare persistence length is of
the order of the monomer size, $\ell_0 \approx b \approx $ 0.25 nm,
the typical value of $\tau^2 \ell_B \ell_0 $  is roughly
$\tau^2 \ell_B \ell_0  \approx 3$,
and thus intermediate between the values in Fig.~\ref{fig2}a) and b).
Smaller
linear charge densities can always be obtained by replacing some
of the charged monomers on the polymer backbone with neutral ones.
In this case the crumpling observed in Fig.~\ref{fig2}d) becomes
relevant. On the other hand, increasing the bare stiffness
$\ell_0$, for example by adding bulky side chains to a
synthetic PE backbone, increases the value of $\tau^2 \ell_B \ell_0 $
and, therefore, increases the electrostatic stiffening of the
backbone. This is an interesting illustration of the fact that
electrostatic interactions and chain architecture (embodied via the
persistence length) combine to control the polymer configurational behavior.

The question now arises as to what are the typical chain
conformations at much larger length scales. Clearly, they will be
influenced by the electrostatic repulsions between monomers.
Indeed, in the {\em persistent
regime}, obtained for $\tau \sqrt{\ell_B \ell_0} > 1 $, the
polymer remains locally stiff even for contour lengths larger than
 the bare persistence length $\ell_0$ and the
effective persistence length is given by
\begin{equation}
 \label{elleff} \ell_{\rm eff} \simeq \ell_0 + \ell_{\rm OSF}~.
\end{equation}
The electrostatic contribution to the effective
persistence length, first derived by Odijk and
independently by Skolnick and Fixman, reads \cite{Odijk0,Skolnick}
\begin{equation}
\label{OSF} \ell_{\rm OSF} = \frac{\ell_B \tau^2}{4 \kappa^2}~,
\end{equation}
and is calculated from the electrostatic energy of a slightly bent
polymer using the linearized Debye-H\"uckel approximation,
Eq.~(\ref{introDH}). It is valid  only for polymer conformations
which do not deviate too much from the rod-like reference state
and for weakly charged polymers
(two conditions that are often not simultaneously
satisfied in practice and therefore
have led to criticism of the OSF result, as will be detailed below).
 The electrostatic persistence
length  gives a sizable contribution to the effective
persistence length only for $\ell_{\rm OSF} > \ell_0$. This is
equivalent to the condition
\be
 \label{Gauss} \tau \sqrt{\ell_B \ell_0} > \ell_0 \kappa~.
 \ee
The {\em persistent regime} is obtained for parameters  satisfying both
conditions (\ref{pers}) and (\ref{Gauss})
and exhibits chains that do not crumple locally and are stiffened electrostatically.
Another regime called
the {\em Gaussian regime} is obtained in the opposite limit of
$\tau \sqrt{\ell_B \ell_0} < \ell_0\kappa$ and does not exhibit chain stiffening
due to electrostatic monomer-monomer repulsions.

\begin{figure}
 \epsfxsize=12cm
 \centerline{\vbox{\epsffile{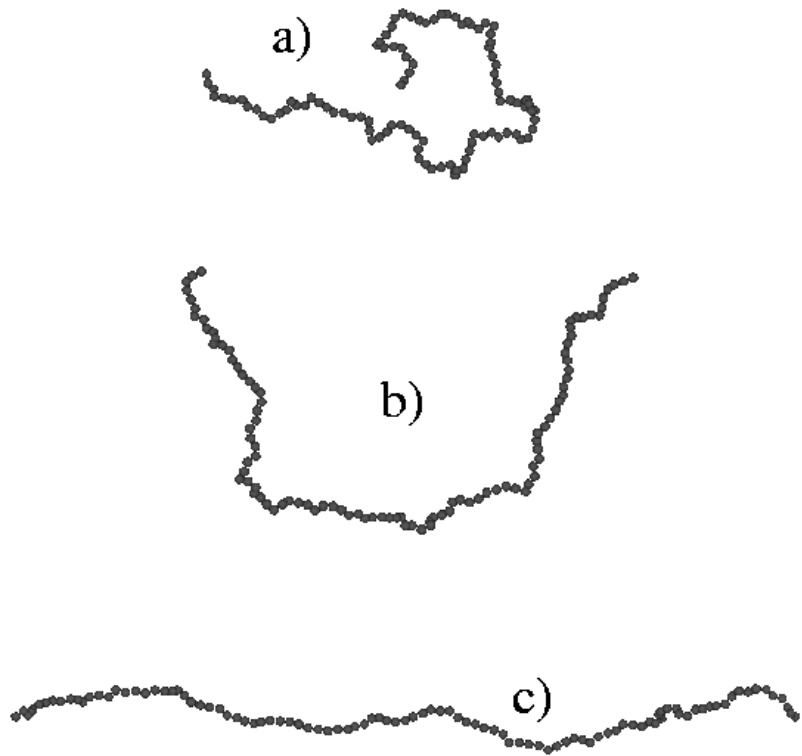} } }
\caption{  \label{fig3} Snapshots of Monte-Carlo simulations of a
PE chain consisting of $N=100$ monomers of size $b$. In all
simulations, the bare persistence length is fixed at $\ell_0/b
=1$, and the charge-interaction parameter is chosen to be $\tau^2
\ell_B \ell_0 =2$. The snapshots correspond to varying screening
length of: a) $\kappa^{-1}/b = \protect \sqrt{2}$, leading to an
electrostatic contribution to the persistence length of
$\ell_{\rm OSF}/b = 1$, b) $\kappa^{-1}/b =\protect \sqrt{18}$,
leading to  $\ell_{\rm OSF}/b= 9$, and c) $\kappa^{-1}/b
=\protect \sqrt{200}$, leading to $\ell_{\rm OSF}/b =100$.
According to the simple scaling principle, Eq. (\ref{elleff}), the
effective  persistence length in the snapshots a)-c) should be
similar to the bare persistence length in Fig.~\ref{fig1}a)-c).}
\end{figure}

The effects of the electrostatic persistence length are visualized in
Fig.~\ref{fig3}, where we present snapshots of a Monte-Carlo
simulation of a charged chain consisting of 100 monomers of size
$b$. The bare persistence length is  fixed at $\ell_0 =b$, and the
charge-interaction parameter is chosen to be $\tau^2 \ell_B b
=2$ for all three simulations, close to the typical charge density
obtained with fully charged synthetic PE's. In Fig.~\ref{fig3} we
show configurations for three different values of the screening length,
namely a) $\kappa^{-1}/b = \protect \sqrt{2} $, leading to an electrostatic
contribution to the persistence length of $\ell_{\rm OSF}/b = 1$;
b) $\kappa^{-1}/b =\protect \sqrt{18} $, or $\ell_{\rm OSF}/b = 9$;
and c) $\kappa^{-1}/b =\protect \sqrt{200}$, equivalent to
an electrostatic persistence length of
$\ell_{\rm OSF}/b = 100 $. According to the simple scaling
principle, Eq.~(\ref{elleff}), the effective persistence length
in the snapshots, Fig.~\ref{fig3}a-c, should be similar to the
bare persistence length in Fig.~\ref{fig1}a-c, and indeed, the
chain structures in ~\ref{fig3}c) and ~\ref{fig1}c) are very
similar. Figure~\ref{fig3}a and ~\ref{fig1}a) are clearly
different, although the effective persistence length is predicted to be
quite similar. This deviation is
mostly due to self-avoidance effects which are
present in charged chains and which will be discussed in detail
in Sec.~\ref{selfavoid}.

\begin{figure}
 \epsfxsize=8cm
 \centerline{\vbox{\epsffile{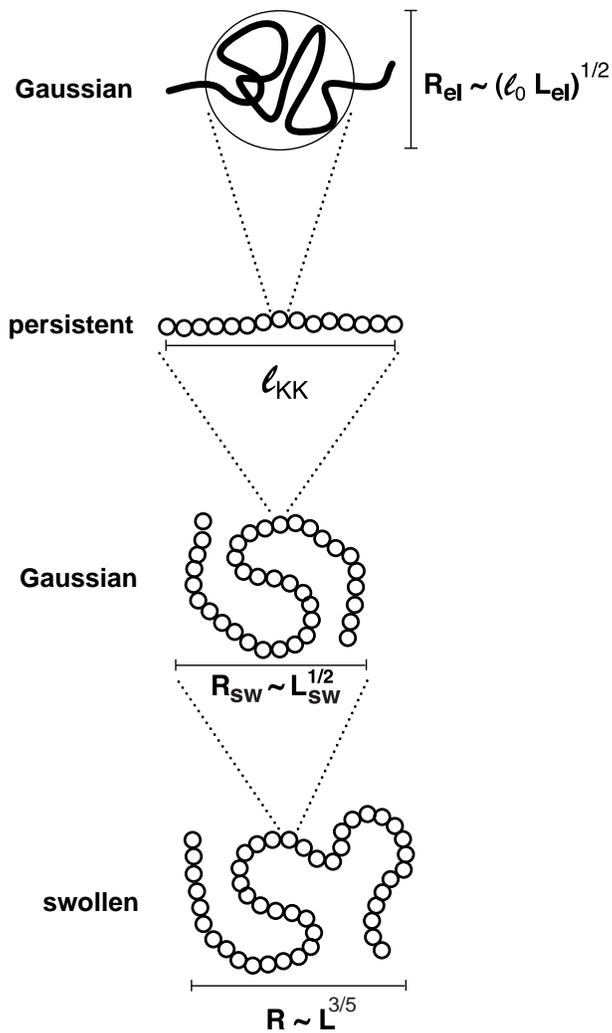} } }
\caption{  \label{fig4}
Schematic view of the four scaling
ranges in the Gaussian-persistent regime. On spatial scales
smaller than $R_{\rm el}$ the chain behavior is Gaussian; on
length scales larger than $R_{\rm el}$ but smaller than $\ell_{\rm
KK}$ the Gaussian blobs are aligned linearly. On larger length
scales the chain is isotropically swollen with an exponent
$\nu=1/2$, and on even larger length scales self-avoidance effects
become important and the exponent changes  to $\nu = 3/5$.}
\end{figure}

For the case where the polymer crumples on length scales larger
than the bare persistence length, \ie\  for $L_{\rm el} > \ell_0$ or
$\tau \sqrt{\ell_B \ell_0} < 1 $, the electrostatic repulsion
between polymer segments is not strong enough to prevent crumpling
on length scales comparable to $\ell_0$, but can give rise to a
chain stiffening on larger length scales, as explained by
Khokhlov and Khachaturian~\cite{Khokhlov} and confirmed by
Gaussian variational methods~\cite{Netz2}. Figure~\ref{fig4} schematically
shows the PE structure in this Gaussian-persistent
regime, where the chain on small
scales consists of Gaussian blobs of size $R_{\rm el}$, each
containing a chain segment of length $L_{\rm el}$.
Within these blobs electrostatic interactions are not important. On
larger length scales electrostatic repulsion leads to a chain
stiffening, so that the PE forms a linear array of electrostatic
blobs. To quantify this effect, one defines an effective line
charge density  $\tilde{\tau}$ of a linear array of electrostatic blobs with blob
size $R_{\rm el} \simeq \sqrt{\ell_0 L_{\rm el}}$,
\begin{equation}
\label{tautilde}
\tilde{\tau} \simeq \frac{\tau L_{\rm el}}{R_{\rm el}} \simeq
\tau \left( \frac{L_{\rm el}}{\ell_0} \right)^{1/2}~.
\end{equation}
Combining Eqs.~(\ref{tautilde}) and (\ref{OSF}) gives the
effective electrostatic persistence length for a string of
electrostatic blobs,
\begin{equation}
\label{OSFtilde}
\ell_{\rm KK} \simeq \frac{\ell_B^{1/3}
\tau^{2/3}}{ \ell_0^{2/3} \kappa^2}~.
\end{equation}
This electrostatic stiffening is only relevant for the so-called
{\em Gaussian--persistent regime} valid for $\ell_{\rm KK} >
R_{\rm el}$, or equivalently
\be \label{Gausspers} \tau \sqrt{\ell_B \ell_0} >( \ell_0
\kappa)^{3/2}~.
\ee
When this inequality is inverted the Gaussian persistence regime
crosses over to the Gaussian one.

\begin{figure}
 \epsfxsize=12cm
 \centerline{\vbox{\epsffile{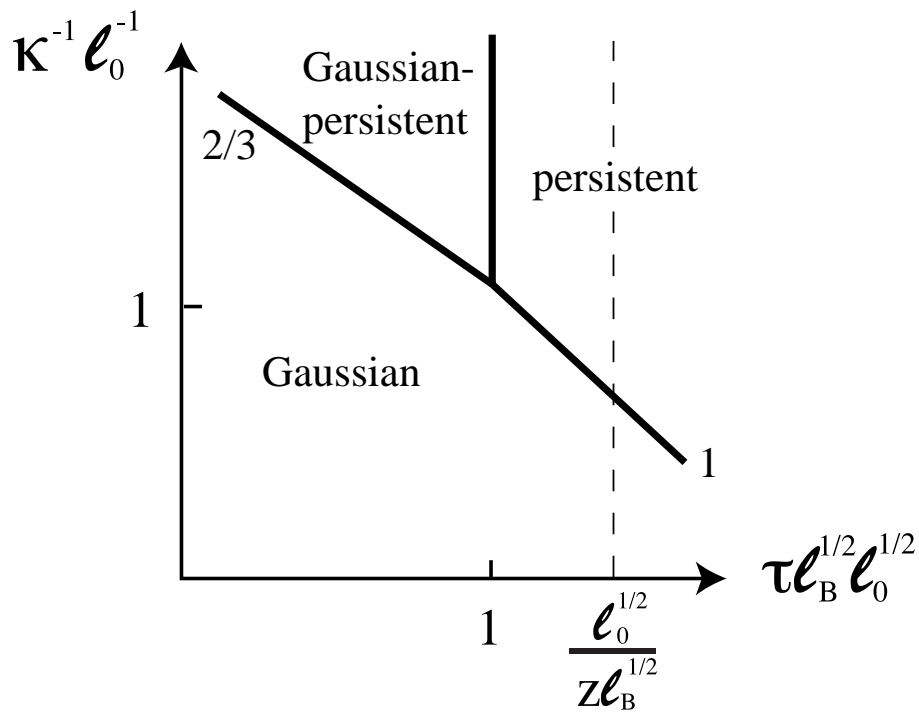} } }
\caption{  \label{fig5} Behavior diagram of a single
semi-flexible PE in bulk solution with bare persistence length
$\ell_0$ and line charge density $\tau$, exhibiting various
scaling regimes. High--salt concentration and small $\tau$
correspond to the Gaussian regime, where the electrostatic
interactions are irrelevant. In the persistent regime, the
polymer persistence length is increased, and in the
Gaussian-persistent regime the polymer forms a persistent chain
of Gaussian blobs as indicated in Fig.~\ref{fig4}. The broken
line indicates the Manning condensation, at which counterions
condense on the polymer and reduce the effective polymer line
charge density. We use a log-log plot, and the various power-law
exponents  for the crossover boundaries are denoted by numbers.}
\end{figure}

The crossover boundaries (\ref{pers}), (\ref{Gauss}),
(\ref{Gausspers}) between the various scaling regimes are
summarized in Fig.~\ref{fig5}. We obtain three distinct regimes.
In the persistent regime, for $\tau \sqrt{\ell_B \ell_0} > \ell_0
\kappa$ and $\tau \sqrt{\ell_B \ell_0} >1 $, the polymer takes on
a rod-like structure with an effective  persistence length
given by the OSF expression, and larger
than the bare persistence length
Eq.~(\ref{OSF}). In the Gaussian-persistent regime, for $\tau
\sqrt{\ell_B \ell_0} < 1 $ and $\tau \sqrt{\ell_B \ell_0} >
(\ell_0 \kappa)^{3/2}$, the polymer consists of a linear array of
Gaussian electrostatic blobs, as shown in Fig.~\ref{fig4}, with
an effective persistence length $\ell_{\rm KK}$ larger than the
electrostatic blob size and given by Eq.~(\ref{OSFtilde}).
Finally, in the Gaussian regime, for $\tau \sqrt{\ell_B \ell_0}
<(\ell_0 \kappa)^{3/2} $ and $\tau \sqrt{\ell_B \ell_0} < \ell_0
\kappa$, the electrostatic repulsion does not lead to stiffening
effects at any length scale (though the chain will be non-ideal).

The persistence length $\ell_{\rm KK}$ was also obtained from
Monte-Carlo simulations with parameters similar to the ones used
for the snapshot shown in Fig.~\ref{fig2}d), where chain crumpling
at small length scales and chain stiffening at large length scales
occur simultaneously \cite{Sim1,Sim2,Sim3,Sim4}. However,
extremely long chains are needed in order to obtain reliable
results for the persistence length, since the stiffening occurs
only at intermediate length scales and, therefore, fitting of the
tangent-tangent correlation function is nontrivial. Whereas
previous simulations for rather short chains point to a different
scaling than in Eq.~(\ref{OSFtilde}), with a dependence on the
screening length closer to a linear one, in qualitative agreement
with experimental results \cite{Foerster}, more recent simulations
for very long chains exhibit a persistence length in agreement
with Eq.~(\ref{OSFtilde})\cite{Ralf,Nguyen3}. The situation is
complicated by the fact that recent theories for the single PE
chain make conflicting predictions, some confirming the simple
scaling results described in Eqs.~(\ref{OSF}) and (\ref{OSFtilde})
\cite{Netz2,Li,Ha2}, while others confirming Eq.~(\ref{OSF}) but
disagreeing with  Eq.~(\ref{OSFtilde})
\cite{barrat2,Ha1,Liverpool}. This issue is not resolved and still
under intense current investigation. For multivalent counterions
fluctuation effects can even give rise to a PE collapse  purely
due to electrostatic interactions
\cite{PEcoll0,PEcoll1,PEcoll2,PEcoll3,PEcoll4,PEcoll5}, which is
accompanied by a negative contribution to the effective
persistence length \cite{PEcoll6,PEcoll7,PEcoll8,Ariel1,Ariel2}. A
related issue is the effective interaction between highly charged
parallel rods, which has been shown to become attractive in the
presence of multivalent
counterions\cite{cyl1,cyl2,cyl3,cyl4,cyl5}.

\subsection{Manning Condensation} \label{SecManning}

A peculiar phenomenon occurs for highly charged PE's and is known
as the Manning condensation of counterions~\cite{Man1,Man2,Tracy,Man3}.
Strictly speaking, this phenomenon constitutes a true
phase transition only in the absence
of any added salt ions.
For a single  rigid PE chain represented by an infinitely
long and straight cylinder
with a linear charge density larger than
\be \label{Manning}
\ell_B \tau z =1~,
\ee
where $z$ is the counterion valency, it was shown that
counterions condense on the oppositely charged cylinder in the
limit of infinite solvent dilution. Namely, in the limit where the
inter-chain distance tends to infinity. This is an effect which is
not captured by the linear Debye-H\"uckel theory used in the last
section to calculate the electrostatic persistence length
Eq.~(\ref{OSF}). A simple heuristic way to incorporate the
non-linear effect of Manning condensation is to replace the bare
linear charge density $\tau$ by the renormalized one $\tau_{\rm
renorm} = 1/(z \ell_B)$ whenever $\ell_B \tau z
>1$ holds. This procedure, however, is not totally
satisfactory at high--salt concentrations
\cite{Fixman2,LeBret}. Also, real polymers have a finite length,
and are neither completely straight nor in the infinite dilution
limit \cite{Man4,Deserno,Limbach}. Still, Manning condensation has an
experimental significance for polymer solutions\cite{Man5,Kuhn,Deshkovski}
because thermodynamic quantities, such as counterion
activities~\cite{Wandrey} and osmotic coefficients~\cite{Blaul},
show a pronounced signature of Manning condensation. Locally,
polymer segments can be considered as straight over length scales
comparable to the persistence length. The Manning condition
Eq.~(\ref{Manning}) usually denotes a region where the binding of
counterions to charged chain sections begins to deplete the
solution from free counterions. Within the scaling diagram of
Fig.~\ref{fig5}, the Manning threshold (denoted by a vertical
broken line) is reached typically for  charge densities larger
than the one needed to straighten out the chain. This holds for
monovalent ions provided $\ell_0 > \ell_B$, as is almost always
the case. The Manning condensation of counterions will therefore not have
a profound influence on the local chain structure since the chain
is rather straight already due to monomer-monomer repulsion. A
more complete description of various scaling regimes related to
Manning condensation, chain collapse and chain swelling has
recently been given in Ref.~\cite{Schiessel}.

\subsection{Self-Avoidance and Polyelectrolyte
Chain Conformations } \label{selfavoid}

Let us now consider how the self-avoidance of PE chains comes
into play, concentrating on the persistent regime defined by  $\tau
\sqrt{\ell_B \ell_0} >1 $. The end-to-end radius $R$ of a
strongly charged PE chain shows three distinct scaling ranges.
For a chain length $L$ smaller than the effective persistence
length $\ell_{\rm eff}$, which according to Eq.~(\ref{elleff}) is
the sum of the bare and electrostatic persistence lengths,
$R$ grows linearly
with the length, $R \sim L$. Self-avoidance plays no role in this
case, because the chain is too short to fold back on itself.

For much longer chains, $L \gg \ell_{\rm eff}$, we envision a
single polymer coil as a solution of separate polymer pieces of
length $\ell_{\rm eff}$, and treat their interactions using a
virial expansion.  The second virial coefficient $v_2$
of a rod of length $\ell_{\rm eff}$ and diameter $d$ scales as
$v_2 \sim \ell_{\rm eff}^2 d$~\cite{houwaart,Fixman}. The chain
connectivity is taken into account by adding the entropic chain
elasticity as a separate term. The standard Flory
theory~\cite{Flory1} (see Sec.~2.2) modified to apply to a
semi-flexible chain is based on writing the free energy ${\cal
F}$ (in units of $k_B T$) as a sum of two terms
\be
{\cal F} \simeq  \frac{R^2}{\ell_{\rm eff} L} +
v_2 R^3 \left( \frac{L/\ell_{\rm eff} }{R^3} \right)^2~,
\ee
where the first term is the entropic elastic energy associated
with swelling a semi-flexible polymer chain to a radius $R$ and the second term
is the second-virial repulsive energy  proportional to the
coefficient $v_2$ and the segment density squared. It is
integrated over the volume $R^3$. The optimal radius $R$ is
calculated by minimizing  this free energy and gives the swollen
radius
\be \label{swollen}
R \sim (v_2 / \ell_{\rm eff})^{1/5} L^\nu~,
\ee
with $\nu = 3/5$ which is the semi-flexible analogue of
Eq.~(\ref{swollenflex}). This radius is only realized above a
minimal chain length $L > L_{\rm sw} \simeq \ell_{\rm eff}^7/v_2^2
\sim \ell_{\rm eff}^3 / d^2$. For elongated segments with $\ell
_{\rm eff} \gg d$, or, equivalently, for a highly charged PE, we
obtain an intermediate range of chain lengths $\ell_{\rm eff} < L
< L_{\rm sw}$ for which the chain is predicted to be Gaussian and
the chain radius scales as
\be \label{Grange}
R \sim \ell_{\rm eff}^{1/2} L^{1/2}~.
\ee
For charged chains, the effective rod diameter $d$ is given in
low--salt concentrations by the screening length, \ie\   $d \sim
\kappa^{-1}$ plus logarithmic corrections~\cite{houwaart,Fixman}.
The condition to have a Gaussian scaling regime,
Eq.~(\ref{Grange}), thus becomes $\ell_{\rm eff} \gg \kappa^{-1}$.
For the case $\tau \sqrt{\ell_B \ell_0} < 1 $, where the chain
crumples and locally forms Gaussian blobs, a similar calculation
to the one outlined here leads to the condition $\ell_{\rm KK} >
\kappa^{-1}$ in order to see a Gaussian regime between the
persistent and the swollen one. Within the persistent and the
Gaussian-persistent scaling regimes depicted in Fig.~\ref{fig5}
the effective persistence length is dominated by the electrostatic
contribution and given by Eqs.~(\ref{OSF}) and (\ref{OSFtilde}),
respectively, which in turn are always larger than the screening
length $\kappa^{-1}$. It follows that a Gaussian scaling regime,
Eq.~(\ref{Grange}), always exists between the persistent regime
where $R \sim L$ and the asymptotically swollen scaling regime,
Eq.~(\ref{swollen}). This situation is depicted in
Fig.~\ref{fig4} for the Gaussian-persistent scaling regime, where
the chain shows two distinct Gaussian scaling regimes at the small
and large length scales. This multi-hierarchical scaling
structure is only one of the many problems one faces when trying
to understand the behavior of PE chains, be it experimentally,
theoretically, or by simulations.

A different situation occurs when the polymer backbone is under
bad-solvent conditions, in which case an intricate interplay
between electrostatic chain swelling and short-range collapse
occurs \cite{Khokhlov2}. Quite recently, this interplay was
theoretically shown to lead to a Rayleigh instability in the form
of a necklace structure consisting of compact globules connected by
stretched chain segments\cite{Kantor,Dobrynin2,Solis,Lyulin,Micka2}. Small-angle
X-ray scattering on solvophobic PE's in a series of
polar organic solvents of various solvent quality could
qualitatively confirm these theoretical predictions \cite{Waigh}.

\subsection{Dilute Polyelectrolyte Solutions}

In accordance with our discussion for neutral chains in Sec.~2.4,
the dilute regime is defined by $c_m< c_m^{*}$, where $c_m$
denotes the monomer concentration (per unit volume) and $c_m^{*}$
is the concentration where individual chains start to overlap.
Using Eq.~(\ref{overlap}), for rigid polymers with $\nu = 1$ the
overlap concentration scales as $c_m^* \sim a^{-3} N^{-2}$ and
decreases strongly as $N$ increases. This means that the dilute
regime for semi-flexible PE chains corresponds to extremely low
monomer concentrations. For example taking a Kuhn length $a
\approx 0.25$\,nm (corresponding to the projected length of two
carbon bonds) and a polymerization index of $N = 10^4$, the
overlap concentration becomes $c_m^* \approx 6 \times
10^{-7}$\,nm$^{-3} \approx 10^{-3}$\,mM, which is a very small
concentration.

The osmotic pressure (rescaled by $k_BT$ )
in the dilute regime in the limit $c_m \rightarrow 0$ is given by
\be
\Pi = \frac{f c_m}{z} + \frac{c_m}{N}~,
\ee
and consists of the ideal pressure of non-interacting counterions
(first term) and polymer coils (second term). Note that since the
second term scales as $N^{-1}$, it is quite small for large $N$
and can be neglected. Hence, the main contribution to the osmotic
pressure comes from the counterion entropy. This entropic term
explains also why charged polymers can be dissolved  in water
even when their backbone is quite hydrophobic. Precipitation of
the PE chains will also mean that the counterions are confined
within the precipitate. The entropy loss associated with this
confinement is too large and keeps the polymers dispersed  in
solution. In contrast, for neutral polymers there are no
counterions in solution. Only the second term in the osmotic
pressure exists and contributes to the low osmotic pressure of
these polymer solutions. In addition, this explains the trend
towards precipitation even for very small attractive interactions
between neutral polymers: The entropic pressure scale as $c_m/N$,
while the enthalpic pressure which favors precipitation scales as
$-c_m^2$ with no additional $N$ dependence, thus dominating the
entropic term for large $N$ \cite{degennes}.

\subsection{Semi-Dilute Polyelectrolyte Solutions}

\begin{figure}
 \epsfxsize=12cm
 \centerline{\vbox{\epsffile{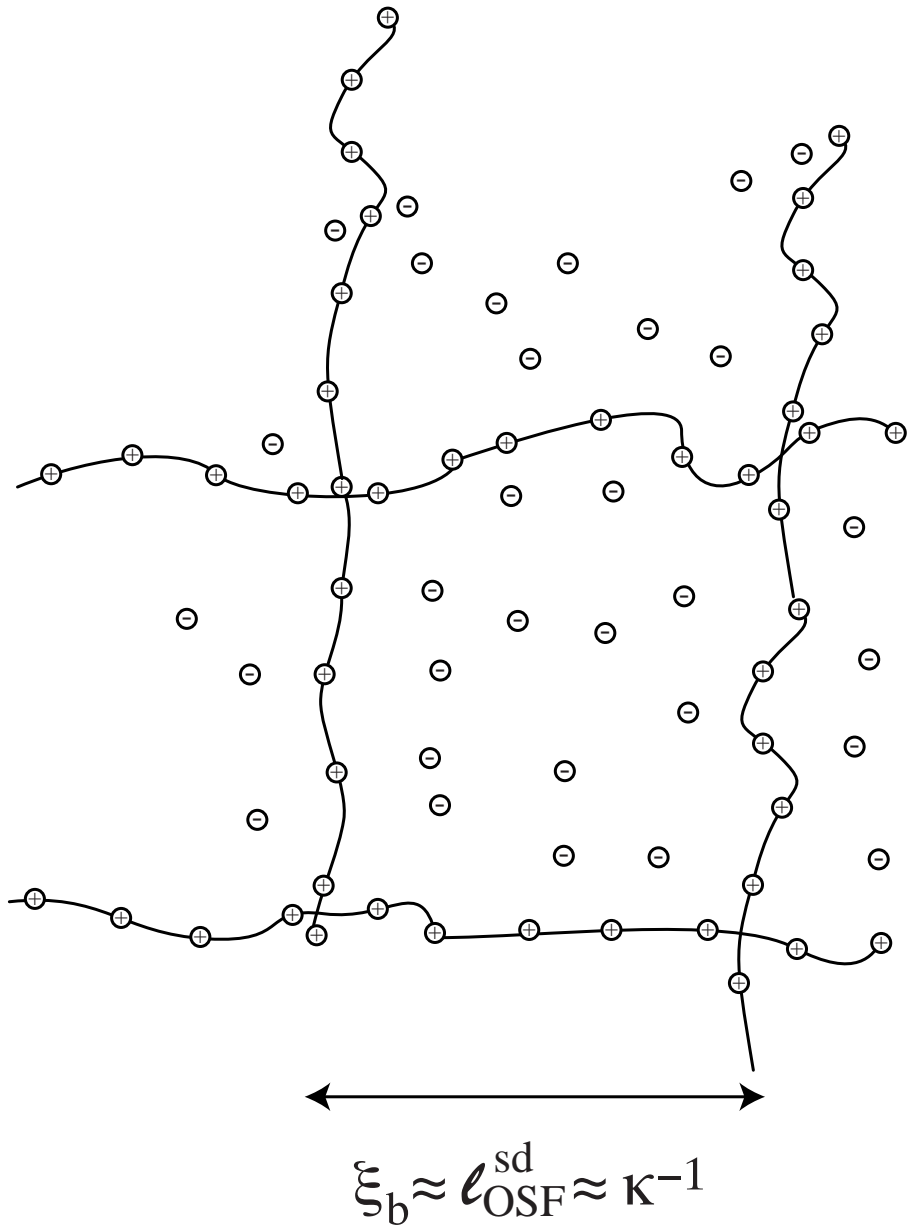} } }
\caption{  \label{fig6}
Schematic view of the PE chain structure in
the semi-dilute concentration range. The mesh size $\xi_b$ is about
equal to the persistence length $\ell^{\rm sd}_{\rm OSF}$
and to the screening length $\kappa^{-1}$ (if no salt is added to
the system).}
\end{figure}

In the semi-dilute concentration regime, $c_m > c_m^*$, different
polymer coils are strongly overlapping, but the polymer solution
is still far from being concentrated. This  means that the volume
fraction of the monomers in solution is much smaller than unity,
$a^3 c_m \ll 1$. In this concentration range, the statistics of
counterions and polymer fluctuations are intimately connected.
 One example where this
feature is particularly prominent is furnished by neutron and
X-ray scattering from semi-dilute PE
solutions~\cite{Nierlich1}-\cite{Essafi2}.
The structure factor $S(q)$  shows a pronounced peak, which
results from a competition between the connectivity of polymer
chains and the electrostatic repulsion between charged monomers,
as will be discussed below. An important length scale,
schematically indicated in Fig.~\ref{fig6}, is the mesh-size or
correlation length $\xi_b$, which measures the length below which
entanglement effects between different chains are unimportant.
The mesh size can be viewed as the polymer (blob) scale below
which single-chain statistics are valid. A semi-dilute solution
can be roughly thought of as being composed of a close-packed array
of polymer blobs of size $\xi_b$.

The starting point for the present discussion is the screened
interaction between two charges immersed in a semi-dilute PE
solution containing charged polymers, their counterions and,
possibly, additional salt ions. Screening in this case is
produced not only by the ions, but also by the charged chain
segments which can be easily polarized and shield any free
charges.

Using the random-phase approximation (RPA), the effective
Debye-H\"uckel (DH) interaction can be written in Fourier space
as~\cite{Borue,Joanny}
\begin{equation} \label{vRPA}
v_{\rm RPA}(q)  = \frac{1+ v_2 c_m S_0(q)}
{c_m f^2  S_0(q) + v_{\rm DH}^{-1}(q)+
v_2 c_m v_{\rm DH}^{-1}(q) S_0(q)}~,
\end{equation}
recalling that $c_m$ is the average density of monomers in
solution and $f$ is the fraction of charged monomers on the PE
chains. The second virial coefficient of non-electrostatic
monomer-monomer interactions is $v_2$ and the single-chain form
factor (discussed  below) is denoted by $S_0(q)$. In the case
where no chains are present, $c_m = 0$, the RPA expression
reduces to $v_{\rm RPA}(q) =v_{\rm DH}(q)$, the Fourier-transform
of the Debye-H\"uckel potential of Eq.~(\ref{introDH}), given by
\begin{equation} \label{DH}
v_{\rm DH}(q) = \frac{ 4 \pi \ell_B}{q^2 + \kappa^2}~.
\end{equation}
As before, $\kappa^{-1}$ is the DH screening length, which is due
to all mobile ions. We can write $ \kappa^2 = \kappa_\s^2 +4 \pi
\ell_B f c_m $, where $\kappa^2_\s = 8 \pi z^2\ell_B c_\s$
describes the screening due to added salt of concentration
$c_\s$, and the second term describes the screening due to the
counterions of the PE monomers. Within the same RPA
approximation the monomer-monomer structure factor $S(q)$ of a
polymer solution with monomer density $c_m$ is given
by~\cite{Borue,Joanny}
\begin{equation} \label{Sinv}
S^{-1}(q) =
f^2 v_{\rm DH}(q) + S_0^{-1}(q) /c_m + v_2 ~.
\end{equation}
 The structure factor (or scattering function)
 depends only on the form factor of an isolated, non-interacting
polymer chain, $S_0(q)$, the second virial coefficient  $v_2$,
the fraction $f$ of charged monomers, and the interaction between
monomers, which in the present case is taken to be the
Debye-H\"uckel potential $v_{\rm DH}(q)$. The  structure factor
of a non-interacting semi-flexible polymer is characterized, in
addition to the monomer length $b$, by its persistence length
$\ell_{\rm eff}$. In general, this form factor is a complicated
function which cannot be written down in closed
form~\cite{Cloizeaux,Yoshizaki}. However, one can separate between
three different ranges of wavenumbers $q$, and within each range
the form factor shows a rather simple scaling behavior, namely
\begin{equation}
\label{Sasym}
S^{-1}_0(q) \simeq
     \left\{
\begin{array}{llll}
     & N^{-1}
         & {\rm for} &  q^2 < 6/N b \ell _{\rm eff}  \\
     & q^2 b \ell_{\rm eff}/6
         & {\rm for} &  6/N b \ell _{\rm eff}  < q^2 < 36 /
                      \pi^2 \ell_{\rm eff}^2  \\
     & q b / \pi
         & {\rm for }     &  36 / \pi^2 \ell_{\rm eff}^2  < q^2~.   \\
\end{array} \right.
\end{equation}
For small wavenumbers the polymer acts like a point scatterer,
while in the intermediate wavenumber regime the polymer behaves
like a flexible, Gaussian polymer, and for the largest
wavenumbers the polymer can be viewed as a stiff rod.

One of the most interesting features of semi-dilute PE solutions
is the fact that the  structure factor $S(q)$  shows a pronounced peak.
 For weakly charged PE's, the
peak position scales as $q \sim c_m^{1/4}$ with the monomer
density~\cite{Moussaid}, in agreement with the above random-phase
approximation (RPA) \cite{Borue,Joanny} and other theoretical approaches
\cite{Dymitrowska,Yethi}. We now discuss the scaling of the characteristic
scattering peak within the present formalism. The position of the
peak follows from the inverse structure factor, Eq.~(\ref{Sinv}),
via $\partial S^{-1}(q) / \partial q =0$. which leads to the
equation
\begin{equation} \label{qmax}
q^2 + \kappa_\s^2 + 4 \pi \ell_B f c_m  = \left(\frac{8 \pi  q
\ell_B f^2  c_m } { \partial S_0^{-1}(q) /\partial q
}\right)^{1/2}~.
\end{equation}
In principle, there are two distinct scaling behaviors possible
for the peak, depending on whether the chain form factor of
Eq.~(\ref{Sasym}) exhibits flexible-like or rigid-like scaling
\cite{NetzRPA}. We concentrate now on the flexible case, i.e. the
intermediate $q$-range in Eq.~(\ref{Sasym}). A peak is only
obtained if the left-hand side of Eq.~(\ref{qmax}) is dominated
by the $q$-dependent part, \ie\  if $q^2 > \kappa_\s^2 + 4 \pi
\ell_B f c_m$. In this case, the peak of $S(q)$ scales as
\begin{equation} \label{qflex}
q^* \simeq  \left(
 \frac{24 \pi \ell_B f^2 c_m}{b \ell_{\rm eff}} \right)^{1/4}~,
\end{equation}
in agreement with experimental results.

\begin{figure}
 \epsfxsize=12cm
 \centerline{\vbox{\epsffile{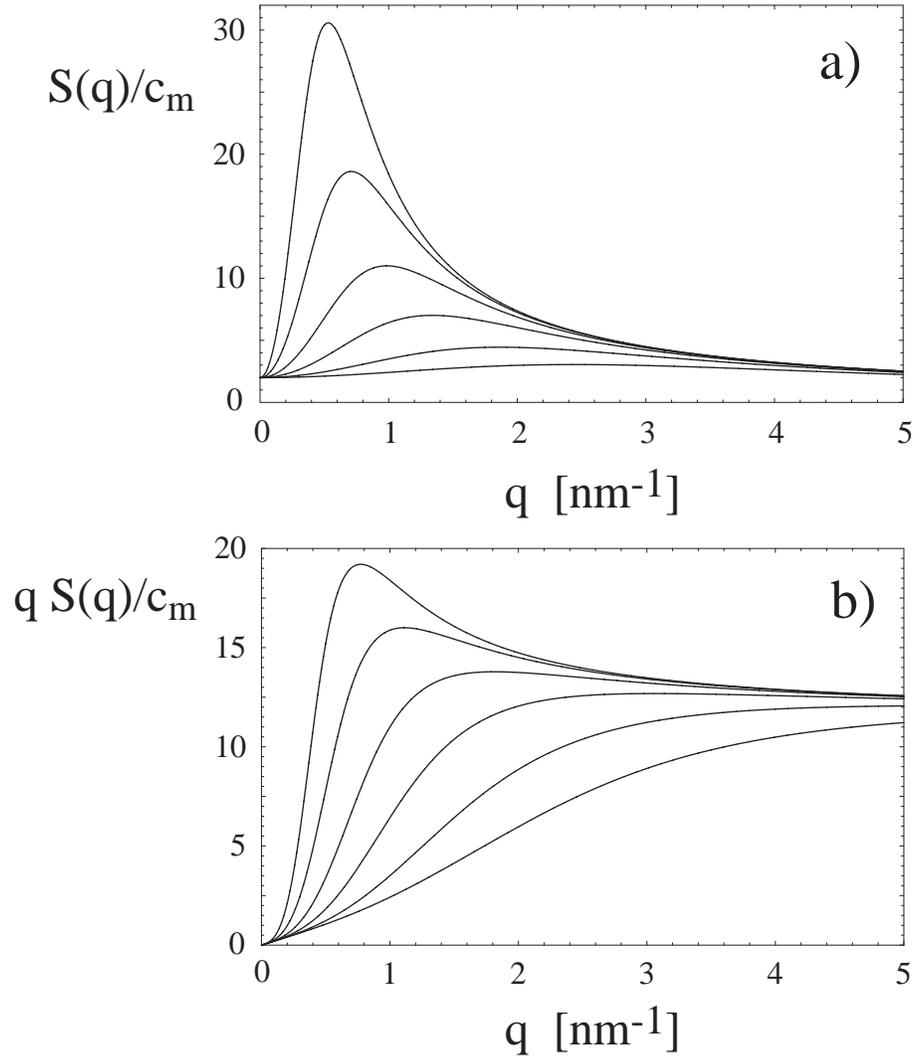} } }
\caption{  \label{fig7}
a) RPA prediction for the rescaled
structure factor $S(q)/c_m$ of a semi-dilute PE solution with
persistence length $\ell_{\rm eff}=1$\,nm, monomer length
$b=0.38$\,nm, polymerization index $N=500$
 and charge fraction $f=0.5$ in the salt-free case.
The monomer densities are (from bottom to top) $c_m = 1$\,M, 0.3\,M,
10\,mM, 3\,mM, 1\,mM, 0.3\,mM. b) For the same series of $c_m$ values as
in a) the structure factor is multiplied by the wavenumber $q$.
The semi-flexibility becomes more apparent because for large $q$
the curves tend towards a constant.}
\end{figure}

In Fig.~\ref{fig7}a we show density-normalized scattering curves
for a PE solution characterized by the persistence length
$\ell_{\rm eff}=1$\,nm (taken to be constant and thus independent
of the monomer concentration), with monomer length $b=0.38$\,nm
(as appropriate for Poly-DADMAC),
polymerization index  $N=500$, charge fraction $f=0.5$ and
with no added salt.
As the monomer density decreases (top to bottom in the figure),
the peak moves to smaller wavenumbers and sharpens, in agreement
with previous implementations of the RPA. In Fig.~\ref{fig7}b we
show the same data in a different representation. Here we clearly
demonstrate that the large-$q$ region already is dominated by the
$1/q$ behavior of the single-chain structure factor, $S_0(q)$.
Since neutron scattering data easily extend to wavenumbers as high
as $q \sim 5$\,nm$^{-1}$, the stiff-rod like behavior in the high
$q$-limit, exhibited on such a plot, will be important in
interpreting and fitting experimental data even at lower
$q$-values.

In a semi-dilute solution there are three different, and in
principle, independent length scales: The mesh size $\xi_b$, the
screening length $\kappa^{-1}$, and the persistence length
$\ell_{\rm eff}$. In the absence of added salt, the screening
length scales as
\be
\kappa^{-1} \sim \left( \ell_B f c_m \right)^{-1/2}~.
\ee
Assuming that the persistence length is larger or of the same
order of magnitude as the mesh size, as is depicted in
Fig.~\ref{fig6}, the polymer chains can be thought of as straight
segments between different cross-links. Denoting the number of
monomers inside a correlation blob as $g$, this means that $\xi_b
\sim b g$. The average monomer concentration scales as $c_m \sim
g / \xi_b^3$ , from which we conclude that
\be
\xi_b \sim \left( b c_m \right)^{-1/2}~.
\ee
Finally, the persistence length within a semi-dilute PE solution
can be calculated by considering the electrostatic energy cost
for slightly bending a charged rod. In PE solutions, it is
important to include in addition to the screening by salt ions
also the screening due to charged chain segments. This can be
calculated by using the RPA interaction, Eq.~(\ref{vRPA}). Since
the screening due to polymer chains is scale dependent and
increases for large separations, a $q$-dependent instability is
encountered and leads to  a persistence length~\cite{NetzRPA}
\be
\ell_{\rm OSF}^{\rm sd} \sim \left( b c_m \right)^{-1/2}~,
\ee
where the `$sd$' superscript stands for `semi-dilute'. This result
is a  generalization of the OSF result for a single chain,
Eq.~(\ref{OSF}), and applies to semi-dilute solutions. Comparing
the three lengths, we see that
\be
\xi_b \sim \ell_{\rm OSF}^{\rm sd} \sim \sqrt{\frac{\ell_B f
}{b}} \kappa^{-1}~.
\ee
Since the prefactor $ \sqrt{\ell_B f / b}$ for synthetic fully
charged polymers is  roughly of order unity, one finds that for
salt-free semi-dilute PE solutions, all three length-scales scale
in the same way with $c_m$, namely as $\sim c_m^{-1/2}$. This scaling
relation has been found first
in experiments~\cite{Nierlich1,Nierlich2,Spiteri}
and was later confirmed by theoretical calculations \cite{PEsemi1,PEsemi2}.
The screening effects due to neighboring PE chains, which form the basis
for the reduction of the electrostatic PE stiffness in a semi-dilute
solution, have also been observed in computer
simulations\cite{Sim6,Sim7,Sim8,Sim9}.

\section{General Considerations on Adsorption}
\setcounter{equation}{0}

\subsection{Adsorption and Depletion}

\begin{figure}
 \epsfxsize=12cm
 \centerline{\vbox{\epsffile{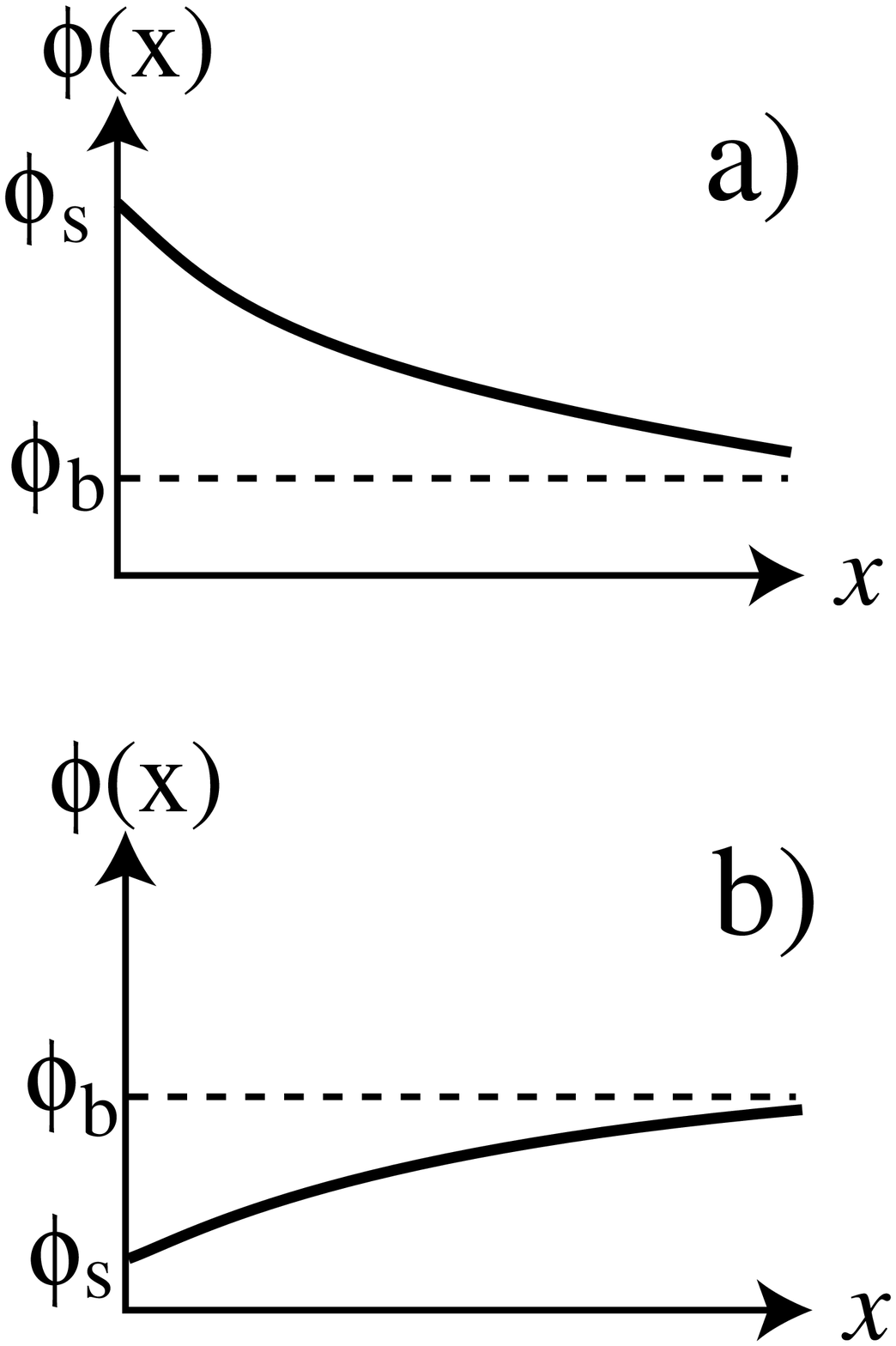} } }
\caption{ \label{fig2o}
Schematic profile of the monomer volume fraction $\phi(x)$ as a
function of the distance $x$ from a flat substrate as appropriate
a) for the case of adsorption, where the substrate attracts
monomers, leading to an increase of the polymer concentration
close to the surface; and, b) for the case of depletion, where the
substrate repels the monomers  leading to a depression of the
polymer concentration close to the surface. The bulk volume fraction, \ie,
the monomer volume fraction
infinitely far away from the surface is denoted by $\phi_b$, and $\phi_s$ denotes the
surface volume fraction right at the substrate surface.}
\end{figure}

Polymers can adsorb spontaneously from solution onto surfaces if
the interaction between the polymer and the surface is more
favorable than that of the solvent with the surface. For example,
a charged polymer like poly-styrene-sulfonate (PSS) is soluble in water
but will adsorb on various hydrophobic surfaces and on the
water/air interface \cite{KlitzingPSS}. This is the case of
equilibrium adsorption where the concentration of the polymer
monomers  increases close to the surface with respect to their
concentration in the bulk solution. We discuss this phenomenon at
length both on the level of a single polymer chain (valid only
for extremely dilute polymer solutions), Secs. 5 and 6, and
for polymers adsorbing from (semi-dilute) solutions,  Secs. 7
and 8. In Fig.~\ref{fig2o}a we show schematically the volume
fraction profile $\phi(x)$ of monomers as a function of the
distance $x$ from the adsorbing substrate. In the bulk, namely
far away from the substrate surface, the volume fraction of the
monomers is $\phi_b$, whereas at the surface, the corresponding
value is $\phi_s > \phi_b$. The theoretical models address
questions in relation to the polymer conformations at the
interface, the local concentration of polymer in the vicinity of
the surface and the total amount of adsorbing polymer chains. In
turn, the knowledge of the polymer interfacial behavior is used
to calculate thermodynamical properties like the surface tension
in the presence of polymer adsorption.

 The opposite case of {\em
depletion} can occur when the monomer-surface interaction is less
favorable than the solvent-surface interaction, as entropy of
mixing will always disfavor adsorption. This is, \eg, the case
for polystyrene in toluene which is depleted from a mica
substrate \cite{lk85}. The depletion layer is defined as the
layer adjacent to the surface from which the polymer is
depleted. The concentration in the vicinity of the surface is
lower than the bulk value, as shown schematically in
Fig.~\ref{fig2o}b.

\subsection{Surface Characteristics}

\begin{figure}
 \epsfxsize=6cm
 \centerline{\vbox{\epsffile{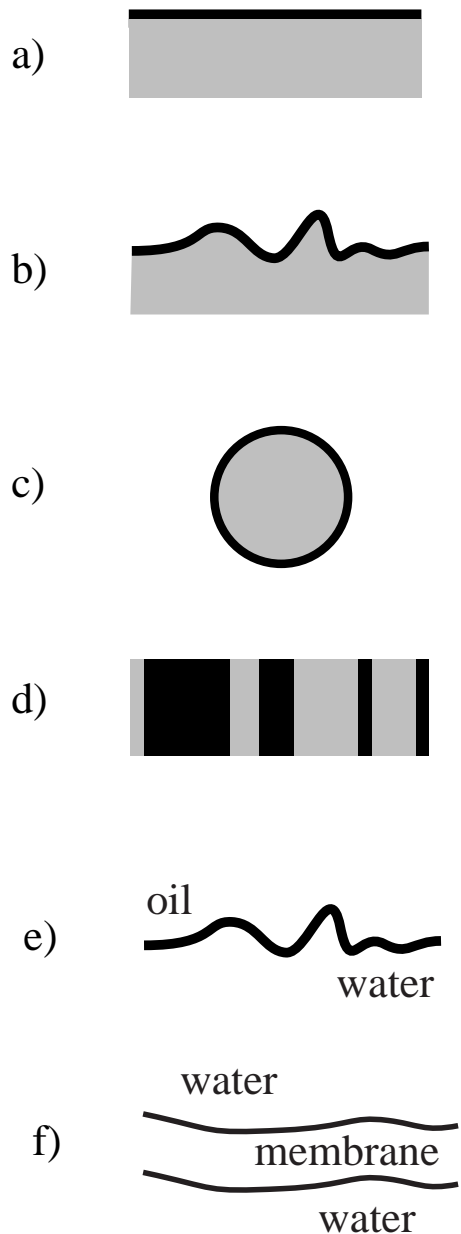} } }
\caption{ \label{fig4o}
Different possibilities of substrates: a) the prototype, a flat,
homogeneous substrate; b) a corrugated, rough substrate. Note that
experimentally, every substrate exhibits a certain degree of roughness
on some length scale; c) a spherical adsorption substrate, such as
a colloidal particle. If the colloidal radius is much larger than
the polymer size, curvature effects (which means the deviation
from the planar geometry) can be neglected; d) a flat but
chemically heterogeneous substrate; e) a liquid/liquid ``soft'' interface.
For example between water and oil; f) A lipid bilayer (membrane) which can
have both shape undulations and lateral composition variations.}
\end{figure}

 Clearly, any adsorption process will be sensitive
to the type of surface and its internal structure. As a starting
point for adsorption problems we assume that the solid surface is
atomically smooth, flat, and homogeneous, as shown in
Fig.~\ref{fig4o}a. This ideal solid surface is impenetrable to
the chains and imposes on them a surface interaction. The surface
potential can be either short-ranged, affecting only monomers
which are in direct contact with the substrate or in close
vicinity of the surface. The surface can also have a longer range
effect, like van der Waals, or electrostatic interactions  if it
is charged. Interesting extensions beyond ideal surface
conditions are expected in several cases: (i)~rough or corrugated
surfaces, such as depicted in Fig.~\ref{fig4o}b;   (ii)~surfaces
that are curved, \eg, adsorption on spherical colloidal
particles, see Fig.~\ref{fig4o}c; (iii)~substrates which are
chemically inhomogeneous, \ie, which show some lateral
organization, as shown schematically in Fig.~\ref{fig4o}d;
(iv)~polymer adsorbing on ``soft" and ``flexible" interfaces
between two immiscible fluids or at the liquid/air surface,
Fig.~\ref{fig4o}e; and, (v)~surfaces that have internal degrees of
freedom like surfactant monolayers or amphiphilic bilayer
(membrane), Fig.~\ref{fig4o}f. We briefly mention those situations
in Secs.~11-12.

\subsection{Surface--Polymer Interactions}

\begin{figure}
 \epsfxsize=12cm
 \centerline{\vbox{\epsffile{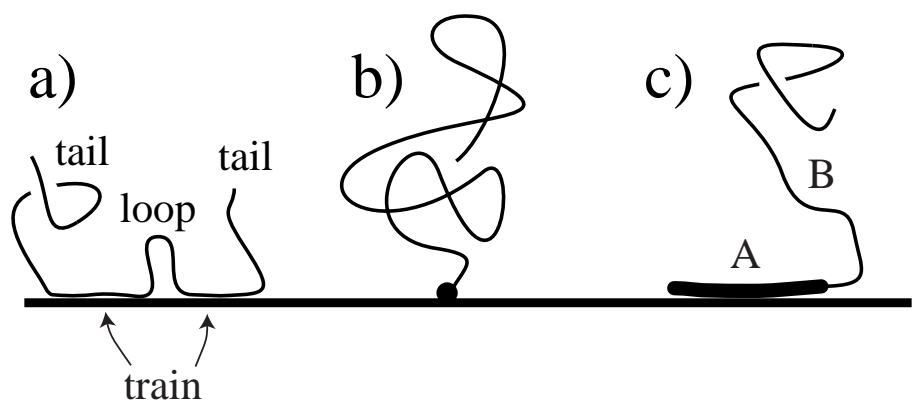} } }
\caption{ \label{fig3o}
The different adsorption mechanisms discussed in this chapter: a)
adsorption of a homopolymer, where each monomer has the same
interaction with the substrate. The `tail', `train' and `loop'
sections of the adsorbing chain are shown; b) grafting of an
end-functionalized polymer via a chemical or a physical bond, and;
c) adsorption of a diblock copolymer where  one of the two block
is attached to the substrate surface, while the other is not.}
\end{figure}

Equilibrium adsorption of polymers is only one of the methods used
to create a change in the polymer concentration close to a
surface. For an adsorbed polymer, it is interesting to look at the
detailed conformation of a single polymer chain at the substrate.
One distinguishes polymer sections that are bound to the
surface (trains), sections that form loops, and end
sections  that can form dangling tails. This
is schematically depicted in Fig.~\ref{fig3o}a.

We mention two other methods to produce
polymer layers at surfaces for polymers which do
not adsorb spontaneously on a given surface.

(i) In the first method, the polymer is chemically attached
(grafted) to the surface by one of the chain ends, as shown in
Fig.~\ref{fig3o}b. In good solvent conditions the polymer chains
look like ``mushrooms'' on the surface when the distance between
grafting points is larger than the typical size of the chains. In
some cases, it is possible to induce a much higher grafting density,
resulting in a polymer ``brush'' extending in the
perpendicular direction from the surface, as is discussed in
detail in Sec.~\ref{sectionbrush}.

(ii) A variant on the grafting method is to use a diblock
copolymer made out of two distinct blocks, as shown in Fig.~\ref{fig3o}c.
The first block is insoluble and is attracted to the substrate.
 Thus, it acts as an ``anchor'' fixing
the chain to the surface; it is drawn as a thick line in
Fig.~\ref{fig3o}c. It should be long enough to cause irreversible
fixation on the surface. The other block is a soluble one (the
``buoy''), forming the brush layer (or ``mushroom"   ). For example, fixation on
hydrophobic surfaces from a water solution can be made using a
polystyrene-polyethylene oxide (PS-PEO) diblock copolymer. The PS
block is insoluble in water and attracted towards the substrate,
whereas the PEO forms the brush layer. The process of diblock
copolymer fixation has a complex dynamics during the formation
stage but is very useful in applications~\cite{napper}. A related
application is to employ a polyethylene glycol (PEG) polymer
connected to a lipid (PEG-lipid) chain and use the lipid to
anchor the PEG chain onto a lipid membrane \cite{lipopol1}.

There are a variety of other adsorption phenomena not discussed
in this review. For example the influence of different polymer
topologies on the adsorption characteristics. In Ref.~\cite{star}
the adsorption of star polymers, where a number of polymer chains
are connected to one center, is discussed. The adsorption of ring
polymers has also received considerable attention
\cite{ring1,ring2}. Another important class of polymers is made
up of random copolymers, which are used to manipulate the
interfacial properties of a variety of systems. The adsorption of
such random copolymers has been studied at solid substrates
\cite{ran1,ran2,ran3} and at penetrable interfaces
\cite{ran4,ran5,ran6,ran6b,ran7}.

\section{Adsorption of a Single Neutral Chain}
\setcounter{equation}{0}

\begin{figure}
 \epsfxsize=12cm
 \centerline{\vbox{\epsffile{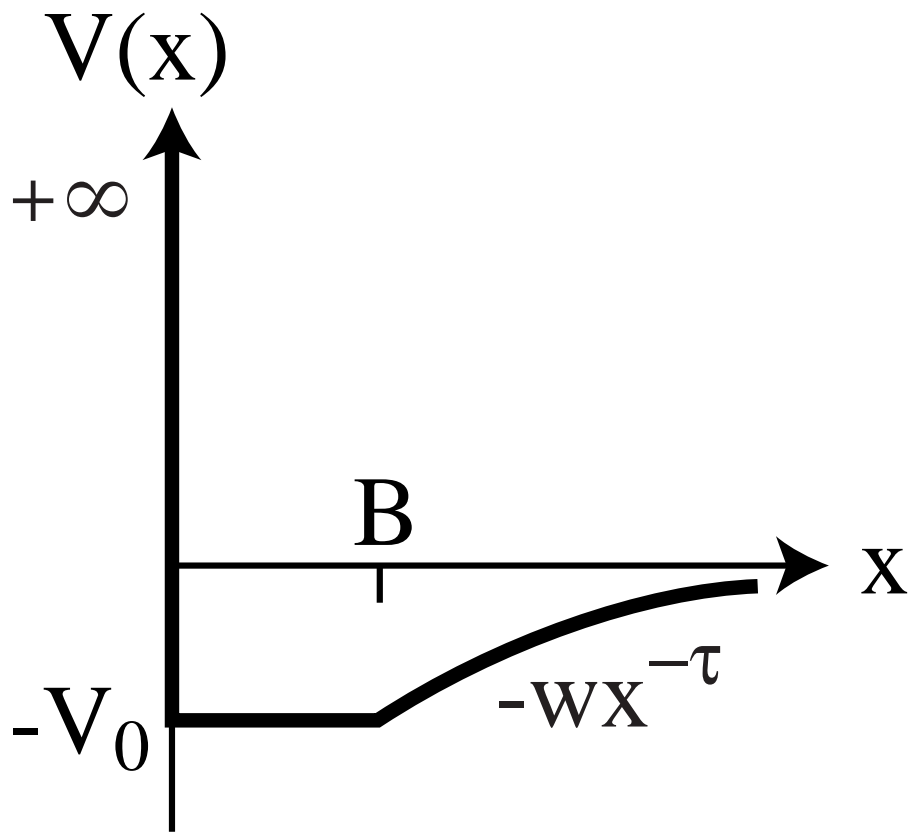} } }
\caption{ \label{fig6o} A typical surface potential felt by a
monomer as a function of the distance $x$ from an adsorbing
surface. First the surface is impenetrable. Then, the attraction
is of strength $V_0$ and range $B$. For separations larger than
$B$, typically a long-ranged tail exists and is modelled by $-w
x^{-\tau}$. }
\end{figure}

Let us consider now the interaction of a single  polymer chain
with a solid substrate. The main effects particular to the
adsorption of polymers (as opposed to the adsorption of simple
molecules) are due to the reduction of conformational states of
the polymer at the substrate, which is caused by the impenetrability
of the substrate for monomers~\cite{fse53}-\cite{jr77}.
The second factor determining the adsorption behavior is the
substrate-monomer interaction. Typically, for the case of an
adsorbing substrate, the interaction potential $V(x)$
(measured in units of $k_BT$)
between the
substrate and a single monomer has a form similar to the one shown
in Fig.~\ref{fig6o}, where $x$ measures the distance of the monomer from the
substrate surface,

\begin{equation}
\label{freepot}
V(x) \simeq
     \left\{ \begin{array}{llll}
     &  \infty
         & {\rm for} &  x<0  \\
     &  -V_0
         & {\rm for} &  0<x< B \\
     & -w x^{-\tau}
         & {\rm for }     & x>B ~.\\
                \end{array} \right.
\end{equation}

\noindent The separation of $V(x)$ into three parts is done for
convenience. It consists of a hard wall at $x=0$, which embodies
the impenetrability of the substrate, \ie, $V(x) = \infty$ for
$x<0$. For positive $x$ we assume the potential to be given by an
attractive well of depth $V_0$ and width $B$. At large distances,
$x>B$, the potential can be modelled by a long-ranged attractive
tail decaying as $V(x) \sim -w x^{-\tau}$.

For the important case of (non-retarded)
van-der-Waals interactions between the substrate and the polymer
monomers, the potential shows a decay governed by the exponent
$\tau=3$ and can be attractive or repulsive, depending on the
solvent, the chemical nature of the monomers and the substrate
material. The decay power $\tau=3$ follows from  the van-der-Waals
pair interaction, which decays as the inverse sixth power with
distance, by integrating over the three spatial dimensions of the
substrate, which is supposed to be a semi-infinite half
space~\cite{Israel}.

\begin{figure}
 \epsfxsize=12cm
 \centerline{\vbox{\epsffile{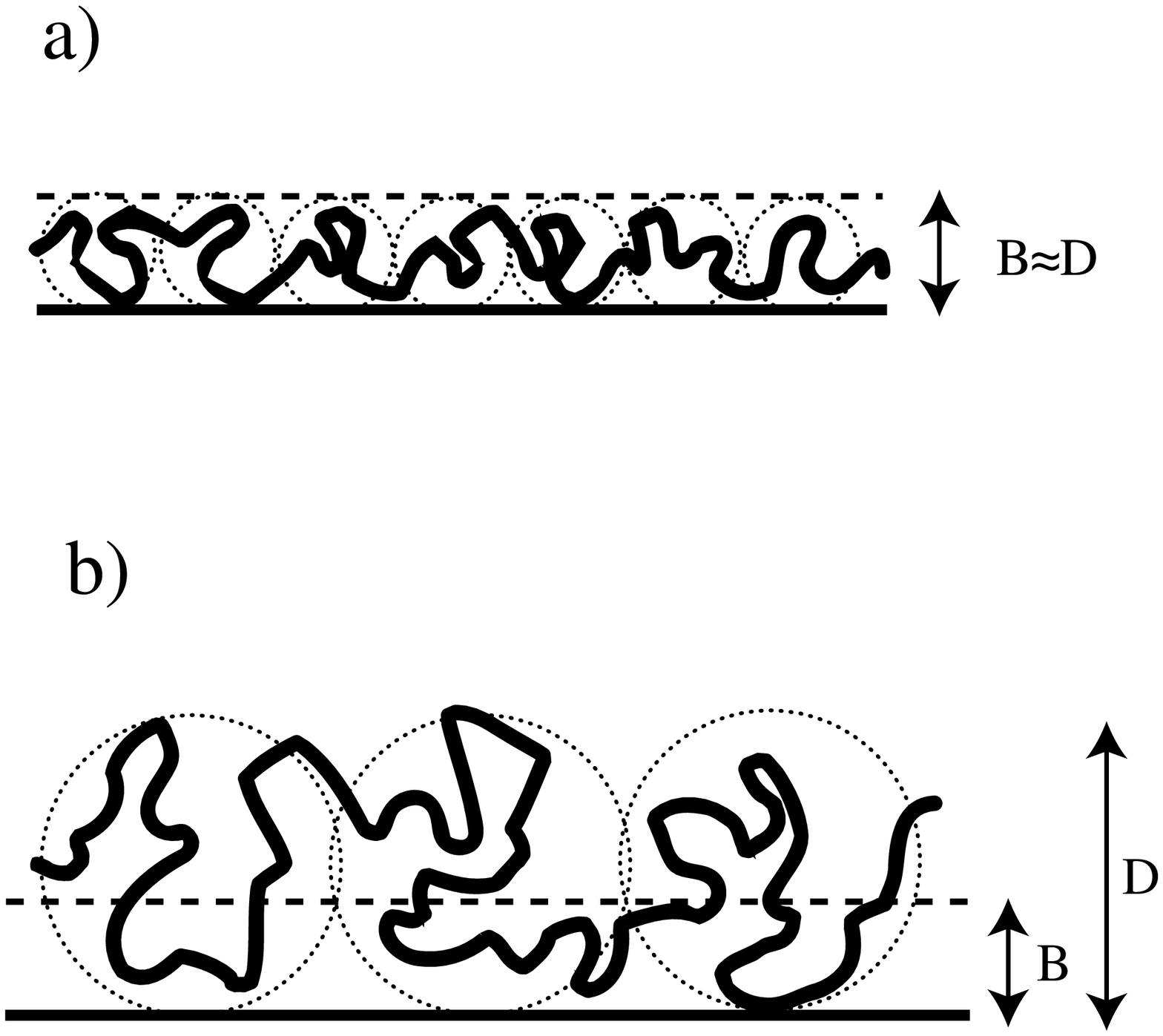} } }
\caption{ \label{fig7o}
Schematic drawing of single-chain adsorption. a) In the limit of
strong coupling, the polymer decorrelates into a  number of
blobs
(shown as dotted circles) and the chain is confined to a
layer thickness $D$, of the same order of magnitude as the
potential range $B$; b) in the case of weak coupling, the width of
the polymer layer $D$ is much larger than the interaction range
$B$ and the polymer forms large blobs, within which the polymer
is not perturbed by the surface.
}
\end{figure}

The strength of the potential well is measured by $V_0$, \ie,
by comparing the potential depth with the thermal energy
$k_BT$. For strongly attractive  potentials, \ie, for $V_0$ large
or, equivalently, for low temperatures, the polymer is strongly
adsorbed and the thickness of the adsorbed layer, $D$,
approximately equals the potential range $B$. The resulting
polymer structure is shown in Fig.~\ref{fig7o}a, where the width of the
potential well, $B$, is denoted by a broken line.

For weakly attractive potentials, or for high temperatures, we anticipate a
weakly adsorbed polymer layer, with a diffuse layer thickness $D$ much
larger than the potential range $B$. This structure is depicted in Fig.~\ref{fig7o}b.
For both cases shown in Fig.~\ref{fig7o}, the polymer conformations are unperturbed
on a spatial scale of the order of $D$; on larger length scales, the
polymer is broken up into decorrelated {\em polymer
blobs}~\cite{degennes,Cloizeaux}, which are denoted by dotted circles in
Fig.~\ref{fig7o}. The idea of introducing polymer blobs is related to the fact that
very long and flexible chains have different spatial arrangement at small
and large length scales. Within each blob the short range interaction is
irrelevant, and the polymer structure inside the blob is similar to the
structure of an unperturbed polymer far from the surface. Since all
monomers are connected, the blobs themselves are linearly connected and
their spatial arrangement represents the behavior on large length scales.
In the adsorbed state, the formation of each blob leads to an entropy loss
of the order of one $k_BT$ (with a numerical prefactor of order unity that
is neglected in this scaling argument), so the total entropy loss of a
chain of $N$ monomers is ${\cal F}_{\rm rep} \sim  N/g$ in units of
$k_BT$, where $g$
denotes the number of monomers inside each blob.

Using the scaling relation $D \simeq a g^\nu$ for the blob size
dependence on the number of monomers $g$, Eq.~(\ref{swollenflex}),
the entropy penalty for the confinement of a polymer chain to a
width $D$ above the surface can be written as~\cite{daoud77} :

\be
\label{freerep}
{\cal F}_{\rm rep} \simeq N \left(\frac{a}{D} \right)^{1/\nu}.
\ee

\noindent The adsorption behavior of a  polymer chain results from
a competition between the attractive potential $V(x)$, which tries
to bind the monomers to the substrate, and the entropic repulsion
${\cal F}_{\rm rep}$, which tries to maximize entropy, and
 favors a delocalized state where a large fraction of
 monomers is located farther away from the surface.

It is of interest to compare the adsorption of long--chain
polymers with the adsorption of small molecular solutes. Small
molecules adsorb onto a surface only if there is a bulk reservoir
with non-zero concentration in equilibrium with the surface. An
infinite polymer chain $N\rightarrow \infty$ behaves differently
as it remains adsorbed also in the limit of zero bulk
concentration. This corresponds to a true thermodynamic phase
transition in the limit
 $N\rightarrow \infty$ \cite{dg69}.
 For finite polymer length, however, the equilibrium
adsorption resembles that of small
molecules. Only a non-zero bulk polymer concentration will lead to
adsorption of finite-length polymer chains on the substrate.
Indeed, as all real
polymers are of finite length, the adsorption of single polymers
is never observed in practice. However, for fairly long polymers,
the desorption of a single polymer is almost a `true' phase
transition, and corrections due to finite (but long) polymer
length are often below experimental resolution.

\subsection{Mean--Field Regime}

Fluctuations of the local monomer concentration are of importance
for polymers at surfaces because of the large number of
possible chain conformations. These fluctuations are treated
theoretically using field-theoretic or transfer-matrix techniques.
In a field-theoretic formalism, the problem of accounting for
different polymer conformations is converted into a functional
integral over different monomer-concentration
profiles~\cite{Cloizeaux}. Within transfer-matrix techniques, the
Markov-chain property of ideal polymers is exploited to re-express
the conformational polymer fluctuations as a product of
matrices~\cite{Flory2}.

However, there are cases where fluctuations in the local monomer
concentration become unimportant. Then, the adsorption behavior of
a single polymer chain is obtained using simple {\em mean--field
theory} arguments. Mean--field theory is a very useful
approximation applicable in many branches of physics, including
polymer physics. In a nutshell, each monomer is placed in a
``field", generated by the external potential plus the
averaged interaction with all the other monomers.

The mean--field theory can be justified for two cases: (i) a {\em
strongly adsorbed polymer chain}, \ie, a polymer chain which is
entirely confined inside the potential  well; and, (ii) the case
of {\em long-ranged attractive surface potentials}. To proceed, we
assume that the  adsorbed polymer layer is confined with an
average thickness $D$, as depicted in Fig.~\ref{fig7o}a or b. Within
mean--field theory, the polymer chain feels an average of the
surface potential, $\langle V(x) \rangle$, which is replaced by
the potential evaluated at the average distance from  the surface,
$\langle x \rangle \simeq D/2$. Therefore, $\langle V(x) \rangle
\simeq V(D/2)$. Further stringent conditions when such a
mean--field theory is valid are detailed below. The full free
energy of one chain, ${\cal F}$, of polymerization index $N$, can
be expressed as the sum of the repulsive entropic term,
Eq.~(\ref{freerep}), and the average potential

\be
\label{free1}
{\cal F} \simeq N \left(\frac{a}{D} \right)^{1/\nu}
+N V(D/2) ~.
\ee

\noindent
Let us consider first the case of a strongly adsorbed polymer, confined
to a potential well of depth $\sim V_0$. In this case the potential energy
per monomer becomes $V(D/2) \simeq - V_0$.
Comparing  the repulsive entropic term
with the potential term, we find
the two terms to be of equal strength for
a well depth $V_0^* \simeq (a/D)^{1/\nu}$.
Hence, the strongly adsorbed state, which is depicted in Fig.~\ref{fig7o}a,
should be realized for a high  attraction strength $V_0> V_0^* $.
For intermediate attraction strength, $V_0 \approx  V_0^*$,
the adsorbed chain will actually be adsorbed
in a layer of width $D$  larger
than the potential width $B$, as shown in Fig.~\ref{fig7o}b, which  will
be discussed further below.
Since the potential depth  $V_0$ is measured in units of $k_BT$,
it follows that at high temperatures
it becomes increasingly difficult to confine
the chain. This can be seen from expressing the bare potential depth as
$\tilde{V}_0 = k_BT V_0$, so that the critical potential depth becomes
$\tilde{V}_0^* \simeq k_BT (a/D)^{1/\nu}$ and thus increases linearly
with temperature.
In fact, for an ideal chain,
with $\nu=1/2$, the resulting scaling relation for the critical well depth,
$V_0^* \sim (a/D)^2$, agrees with exact transfer-matrix predictions
for the adsorption threshold in  a square-well potential~\cite{square}.

We turn now to the case of a weakly adsorbed polymer layer. The
potential depth is smaller than the threshold, \ie, $V_0< V_0^*$, and
the stability of the weakly adsorbed polymer chain, depicted in
Fig.~\ref{fig7o}b,  has to be examined. The thickness $D$ of this polymer
layer follows from the minimization of the free energy,
Eq.~(\ref{free1}),  with respect to $D$, where we use the
asymptotic form of the surface potential, Eq.~(\ref{freepot}), for
large separations.  The result is
\be
\label{Dmean}
D \simeq \left( \frac{a^{1/\nu}}{w} \right)^{\nu/(1-\nu \tau)}~.
\ee
Under which circumstances is the prediction Eq. (\ref{Dmean})
correct, at least on a qualitative level? It turns out that the
prediction for $D$, Eq.~(\ref{Dmean}), obtained within the simple
mean--field theory, is correct if the attractive tail of the
substrate potential in Eq.~(\ref{freepot}) decays for large values
of $x$ slower than the entropic repulsion in Eq.~(\ref{freerep})
~\cite{lip89}. In other words, the mean--field theory is valid for
weakly-adsorbed polymers only for $\tau < 1/\nu$. This can already
be guessed from the functional form of  the layer thickness, Eq.
(\ref{Dmean}), because for $\tau > 1/ \nu$ the layer thickness $D$
goes to zero as $w$ diminishes. Clearly an unphysical result. For
ideal polymers (theta solvent, $\nu=1/2$), the validity condition
is $\tau<2$, whereas for swollen polymers (good solvent
conditions, $\nu = 3/5$), it is $\tau < 5/3$. For most
interactions (including van der Waals interactions with $\tau =
3$) this condition on $\tau$ is not satisfied, and fluctuations
are in fact important, as is discussed in the next section.

There are two notable exceptions. The first is for charged
polymers close to an oppositely charged surface, in the {\it
absence} of salt ions. Since the attraction of the polymer to an
infinite, planar and charged surface is  linear in $x$, the
interaction is described by Eq.~(\ref{freepot}) with  an exponent
$\tau=-1$, and the inequality $\tau<1/\nu$ is satisfied. For
charged surfaces, Eq.~(\ref{Dmean}) predicts the thickness $D$ to
increase to infinity as the temperature increases or as the
attraction strength $w$ (proportional to the surface charge
density) decreases. The resultant exponents for the scaling of $D$
follow from Eq. (\ref{Dmean}) and are $D\sim w^{-1/3}$ for
ideal chains, and $D\sim w^{-3/8}$ for swollen
chains~\cite{Borisov,netz95}. This case will be considered in more
detail in Sec. \ref{PEsingle}.

A second example where the mean--field theory can be used is the
adsorption of polyampholytes on charged surfaces~\cite{pa1,pa2}.
Polyampholytes are polymers consisting of negatively and
positively charged monomers. In cases where the total charge on
such a polymer adds up to zero, it might seem that the interaction
with a charged surface should vanish. However, it turns out that
local charge fluctuations (\ie, local spontaneous dipole moments)
lead to a strong attraction of polyampholytes to charged
substrates. In the absence of salt this attractive interaction has an
algebraic decay with an exponent $\tau = 2$ \cite{pa1}. On the
other hand, in the presence of salt, the effective interaction is
exponentially screened, yielding a decay faster than the
fluctuation repulsion, Eq.~(\ref{freerep}). Nevertheless, the
mean--field theory, embodied in the free energy expression
Eq.~(\ref{free1}), can be used to predict the adsorption phase
behavior within the strongly adsorbed case (\ie, far from any
desorption transition)~\cite{pa3,pa4,pa5,pa6}.

\subsection{Fluctuation Dominated Regime}

Here we consider  the weakly adsorbed case for substrate
potentials which decay (for large separations from the surface)
faster than the entropic repulsion Eq.~(\ref{freerep}), \ie, $\tau
> 1/\nu$. This applies, \eg, to van-der-Waals attractive
interaction between the substrate and monomers, screened
electrostatic interactions,
 or any other short-ranged potential. In this case,
fluctuations play a decisive role. In fact, for {\em ideal
chains}, it can be rigorously proven (using transfer-matrix
techniques) that all potentials decaying faster than $x^{-2}$ for
large $x$ have a  continuous adsorption transition at a finite
critical temperature $T^*$ \cite{lip89}. This means that the
thickness of the adsorbed polymer layer diverges as

\be
\label{Dideal}
D \sim (T^*-T)^{-1}~.
\ee
for $T \rightarrow T^*$ \cite{Poland}.
The power law divergence of $D$ is universal. Namely, it
does not depend on the specific functional form and strength of
the potential as long as they satisfy the above condition.

The case of {\em non-ideal chains} is much more complicated
\cite{Fisher}. First progress has been made by de Gennes who
recognized the analogy between the partition function of a
self-avoiding chain and the correlation function of an
$n$-component spin model in the zero-component ($n\rightarrow 0$)
limit \cite{gennes72}. The adsorption behavior of non-ideal
chains has been treated by field-theoretic methods  using the
analogy to surface critical behavior of magnets (again in the
$n\rightarrow 0$ limit) \cite{erich,ekb82}. The resulting
behavior is similar to the ideal-chain case and shows an
adsorption transition at a finite temperature, and a continuous
increase towards infinite layer thickness characterized by a
power law divergence as function of $T-T^*$ \cite{ekb82}.

The complete behavior for ideal and swollen chains can be
described using scaling ideas in the following way. The entropic
loss due to the confinement of the chain to a region of thickness
$D$ close to the surface is  again given by Eq.~~(\ref{freerep}).
Assuming that the adsorption layer is much thicker than  the range
of the attractive potential  $V(x)$, the attractive potential can
be assumed to be localized  at the substrate surface $V(x)\simeq
V(0)$. The attractive free energy of the chain due to the substrate
surface can then be written as~\cite{dg76}

\be
\label{freeatt1} {\cal F}_{\rm att}\simeq - \tilde{\gamma}
\frac{(T^*-T)}{T} N f_1 = - \gamma_1 a^2 N f_1 \ee

\noindent where $f_1$ is the probability to find a monomer at the
substrate surface and $\tilde{\gamma}$ is a dimensionless
interaction parameter. Two surface excess energies are typically
being used: $\gamma_1=\tilde{\gamma}(T^*-T)/ Ta^2$ is the excess
energy per unit area, while  $\gamma_1 a^2$ is the
(dimensionless) excess energy per monomer at the surface. Both are
positive for the attractive case (adsorption) and negative for
the depletion case. The dependence of $\gamma_1$ on $T$ in
Eq.~~(\ref{freeatt1}) causes the attraction to vanish at a
critical temperature, $T=T^*$, in accord with our expectations.

The contact probability for a swollen chain with the surface,
$f_1$, can be calculated as follows~\cite{dgp83}. In order to force
the chain of polymerization index $N$
to be in contact with the surface, one of the chain ends is pinned to the
substrate.
The number of monomers which are in contact with the surface can
be calculated using field-theoretic methods and is given by
$N^{\varphi}$, where $\varphi$ is called the {\em surface
crossover exponent}~\cite{erich,ekb82}. The fraction of bound
monomers follows to be $f_1 \sim N^{\varphi-1} $, and thus goes to
zero as the polymer length increases, for $\varphi<1$. Now instead
of speaking of the entire chain, we refer to a `chain of blobs'
(see Fig.~\ref{fig7o}) adsorbing on the surface,
each blob consisting of $g$ monomers.  We proceed by assuming that
the size of an adsorbed blob $D$ scales with the number of monomers
per blob $g$ similarly as in the bulk, $D \sim a g^{\nu}$, as is
indeed confirmed by field theoretic calculations. The fraction of
bound monomers can be expressed in terms of $D$ and is given by

\be
\label{fprob1}
f_1 \sim \left(\frac{D}{a}\right)^{(\varphi-1)/\nu}~.
\ee

\noindent
Combining the entropic repulsion, Eq.~(\ref{freerep}),
and the substrate attraction, Eqs.~(\ref{freeatt1}-\ref{fprob1}),
the total free energy is given by

\be
\label{free2}
{\cal F} \simeq N \left(\frac{a}{D}
\right)^{1/\nu}- N\frac{\tilde{\gamma}(T^* - T)}{T}
\left(\frac{D}{a}\right)^{(\varphi-1)/\nu}~.
\ee

\noindent
Minimization with respect to $D$ leads to the final result

\be \label{Dnonideal} D \simeq a \left[\frac{\tilde{\gamma}
(T^*-T)}{T}\right]^{-\nu / \varphi} \simeq a \left(a^2\gamma_1
\right)^{-\nu / \varphi} ~.
\ee

\noindent For ideal chains, one has $\varphi=\nu=1/2$, and thus
we recover the prediction from the transfer-matrix calculations,
Eq.~(\ref{Dideal}). For non-ideal chains, the crossover exponent
$\varphi$ is in general different from the swelling exponent
$\nu$. However, extensive Monte Carlo computer
simulations~\cite{ekb82} and recent field-theoretic calculations
\cite{Diehl}  point to a value for $\varphi$ close to $\nu$,
such that the adsorption exponent $\nu/\varphi$ appearing in
Eq.~(\ref{Dnonideal}) is close to unity, for polymers embedded in
three dimensional space.

A further point which has been calculated using field theory is
the behavior of the monomer volume fraction $\phi(x)$ close to the
substrate. Rather general arguments borrowed from the theory
of critical phenomena suggest a power-law behavior for
$\phi(x)$ at sufficiently small distances from the substrate
~\cite{ekb82,dgp83,bd87}

\be
\label{proxscale} \phi(x) \simeq   (x/a)^{-m}~\phi_s ,
\ee
\noindent
recalling that the monomer density is related to $\phi(x)$ by
$c_m(x)=\phi(x)/a^3$.

In the following, we  relate the so-called {\em proximal exponent}
$m$ with the two other exponents introduced above, $\nu$ and
$\varphi$. First note that the surface value of the monomer volume
fraction, $\phi_s = \phi(x \approx a)$, for one adsorbed blob
follows from the number of monomers at the surface per blob, which
is given by $f_1 g$, and the cross-section area of a blob, which
is of the order of $D^2$. The surface volume fraction is given by

\be
\label{surf}
\phi_s \sim \frac{ f_1 g a^2}{D^2} \sim g^{\varphi-2 \nu} ~.
\ee

\noindent
Using the scaling prediction Eq.~(\ref{proxscale}), we see that
the monomer volume fraction at  the blob center, $x \simeq D/2$,
is given by $\phi(D/2) \sim  g^{\varphi-2 \nu} (D/a)^{-m}$,
which (again using $D  \sim a g^\nu$) can be rewritten as
$\phi(D/2) \simeq g^{\varphi-2 \nu-m\nu}$.

On the other hand, at a distance $D/2$ from the surface, the
monomer volume fraction should have decayed to the average monomer
volume fraction $a^3 g/D^3 \sim  g^{1-3\nu}$ inside the blob since
the statistics of the chain inside the blob is like for a chain in the bulk.
By direct comparison of the two volume fractions, we see that the
exponents $\varphi-2 \nu-m\nu$ and $1-3\nu$ have to match in order
to have a consistent result, yielding

\be
m = \frac{\varphi+\nu-1}{\nu}~.
\label{proxm}
\ee

\noindent
 For ideal chain (theta solvents), one has
$\varphi=\nu=1/2$. Hence, the proximal exponent vanishes, $m=0$.
This means that the proximal exponent has no mean--field analog,
explaining  why it was discovered only  within field-theoretic
calculations~\cite{erich,ekb82}. In the presence of correlations
(good solvent conditions) one has $ \varphi \simeq \nu \simeq 3/5$
and thus $m \simeq 1/3$.

Using $D \simeq a g^\nu$ and Eq.~(\ref{Dnonideal}),
the surface volume fraction, Eq.~(\ref{surf}),
can be rewritten as

\be
\phi_s \sim \left( \frac{D}{a} \right)^{(\varphi-2\nu)/\nu}
\sim \left(a^2 \gamma_1\right)^{(2\nu-\varphi)/\varphi} \simeq
a^2 \gamma_1 ~,\label{phi213}
\ee

\noindent where in the last approximation appearing in
Eq.~(\ref{phi213}) we used the fact that $\varphi \simeq \nu$. The
last result shows that the surface volume fraction within one blob
can become large if the adsorption energy per monomer, $a^2 \gamma_1$,
measured in units of $k_BT$, is of order unity.
Experimentally, this is
often the case, and additional interactions (such as multi-body
interactions) between monomers at the surface have
to be taken into account. Note that the polymer concentration in
the adsorbed layer can become so high that a transition into a
glassy state is induced. This glassy state depends on the details
of the molecular interaction, which are not considered here. It
should be kept in mind that such high-concentration effects can
slow down considerably the adsorption dynamics   while
prolonging equilibration times \cite{glassy}.

After having discussed the adsorption behavior of a single chain,
a word of caution is in order. Experimentally, one never looks at
single chains adsorbed to a surface. First, this is due to the
fact that one always works with polymer solutions, where there is
a large number of polymer chains contained in the bulk reservoir,
even when the bulk monomer (or polymer) concentration is quite
low. Second, even if the bulk polymer concentration is very low,
and in fact so low that polymers in solution barely interact with
each other, the surface concentration of polymer is enhanced
relative to that in the bulk. Hence, adsorbed polymers at the
surface usually do interact with neighboring chains, due to the
higher polymer  concentration at the surface~\cite{bd87}.

Nevertheless, the adsorption behavior of a single chain serves as
a basis and guideline for the more complicated adsorption
scenarios involving many-chain effects. It will turn out that the
scaling of the adsorption layer thickness $D$ and the proximal
volume fraction profile, Eqs.~(\ref{Dnonideal}) and
(\ref{proxscale}), are not affected by the presence of other
chains. This finding as well as other many-chain effects on
polymer adsorption is the subject of Sec.~7.

\section{Adsorption of a Single Polyelectrolyte Chain} \label{PEsingle}
\setcounter{equation}{0}

  After reviewing bulk properties of PE solutions we address
the complete adsorption diagram of a single semi-flexible
PE on an oppositely charged substrate. In contrast
to the adsorption of neutral polymers, the resulting phase
diagram shows a large region where the adsorbed polymer is
flattened out on the substrate and creates a dense adsorption
layer.

The results on single PE adsorption summarized in
this section are most relevant to the adsorption of highly charged
synthetic PE's from dilute solutions
\cite{Kerstin,Kerstin2,Berlepsch,Stubenrauch,Asnacios2,Baltes} or the
adsorption of rather stiff charged biopolymers such as
DNA\cite{Fang,Raedler,Maier}. In all these experiments, the
adsorbed phases can be quite dilute, and the
description of a single adsorbing polymer is a good starting
point.  Repeated adsorption of anionic and cationic PE's can lead
to well characterized multilayers on
planar~\cite{hong}-\cite{caruso2}
and spherical substrates~\cite{donath,caruso}.
The adsorption of a single PE chain has been
treated theoretically employing a variety of
methods~\cite{Wiegel,Muthukumar,Borisov,xav98}.
The adsorption process results from a subtle balance
between electrostatic repulsion between charged monomers, leading
to chain stiffening, and electrostatic attraction between the
substrate and the polymer chain.  It poses a much more
complicated problem than the corresponding adsorption of neutral
polymers.

The adsorption of a single semi-flexible and charged chain on an
oppositely charged plane~\cite{Netz4} can be treated as a
generalization of the adsorption of flexible
polymers~\cite{Borisov}. A PE characterized by a
linear charge density  $\tau$, is subject to an electrostatic
potential created by $\sigma$, the homogeneous surface charge
density (per unit area). Because this potential is attractive for
an oppositely charged substrate, we consider it as the driving
force for the adsorption.
Effects due to bad solvent\cite{Borisov2} and
more complex interactions  are neglected.
One example for the latter are  interactions
due to the dielectric discontinuity at the
substrate surface and to the impenetrability of the substrate for
salt ions.
\footnote{An ion in solution has a repulsive interaction from the
surface when the solution dielectric constant is higher than that
of the substrate. This  effect can lead to desorption for highly
charged PE chains. On the contrary, when the substrate is a metal
there is a possibility to induce PE adsorption on non-charged
substrates or on substrates bearing charges of the same sign as
the PE. See Ref.~\cite{Netz4} for more details.}

Within the linearized DH theory, the electrostatic potential of a
homogeneously charged plane is in units of $k_BT$
\be
V_{\rm plane} (x) =  4\pi \ell_B \sigma \kappa^{-1} {\rm
e}^{-\kappa x}~.
\ee
Assuming that the polymer is adsorbed over a layer of width
$D$ smaller than the screening length $\kappa^{-1}$, the
electrostatic attraction force per monomer unit length can be
written as
\begin{equation}
\label{fatt} \hat{f}_{\rm att}  = -4 \pi \ell_B \sigma
\tau ~.
\end{equation}
For simplicity, we neglect non-linear effects due to counterion condensation
on the PE (as obtained by the Manning theory, see Sec. \ref{SecManning})
and on the surface (as obtained within the Gouy-Chapman theory).
Although these effects
are important for highly charged system\cite{Dobrynin4},
most of the important features of single PE adsorption already appear on
the linearized Debye-H\"uckel level.

\begin{figure}
 \epsfxsize=12cm
 \centerline{\vbox{\epsffile{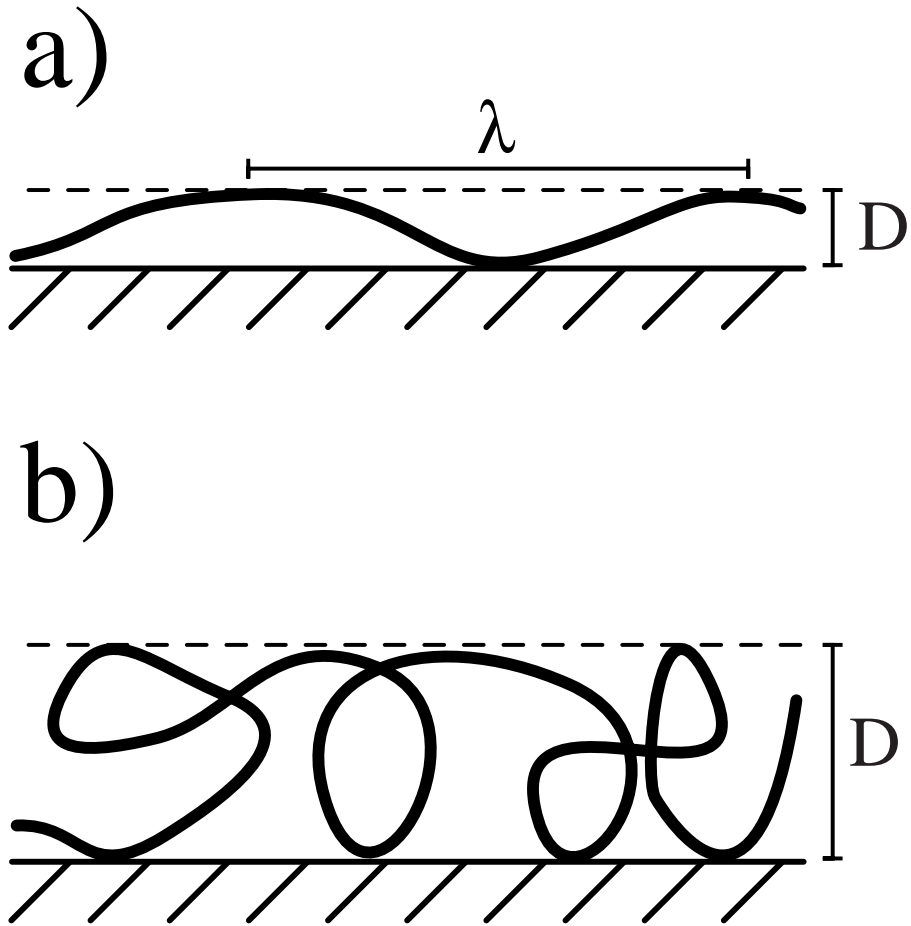} } }
\caption{  \label{fig8}
a) Schematic picture of the adsorbed
polymer layer when  the effective persistence length is larger
than the layer thickness, $ \ell_{\rm eff} > D$. The distance
between two contacts of the polymer with the substrate, the
so-called deflection length, scales as $\lambda \sim D^{2/3}
\ell_{\rm eff}^{1/3}$.
 b) Adsorbed layer for the case when the persistence length is smaller
than the layer thickness, $\ell_{\rm eff} < D$. In this case
the polymer forms a random coil with many loops and a description
in terms of a flexible polymer model becomes appropriate.}
\end{figure}

Because of the confinement in the adsorbed layer, the polymer
feels an entropic repulsion. If the layer thickness $D$ is much
smaller than the effective persistence length of the polymer,
$\ell_{\rm eff}$, as depicted in Fig.~\ref{fig8}a, a new length
scale, the so-called deflection length $\lambda$, enters the
description of the polymer statistics. The deflection length
$\lambda$ measures the average distance between two contact
points of the polymer chain with the substrate. As shown by
Odijk, the deflection length scales as $\lambda \sim D^{2/3}
\ell_{\rm eff}^{1/3}$ and is larger than the layer thickness $D $
but smaller than the persistence length $ \ell_{\rm
eff}$~\cite{Odijk1}. The entropic repulsion follows in a simple
manner from the deflection length by assuming that the polymer
loses roughly an energy of one $k_B T$ per deflection length.

On the other hand, if $D > \ell_{\rm eff}$, as shown in
Fig.~\ref{fig8}b, the polymer forms a random coil with many loops
within the adsorbed layer. The chain can be viewed as an assembly
of decorrelated blobs, each of a chain length of
$L \sim D^2/\ell_{\rm eff}$, within which
the polymer obeys Gaussian statistics.
The decorrelation into blobs has an entropic cost of roughly one $k_B T $
per blob. The entropic repulsion force per polymer unit length is
thus ~\cite{Odijk1}
\begin{equation}
\label{frep} \hat{f}_{\rm rep}  \sim
     \left\{ \begin{array}{llll}
     & D^{-5/3} \ell_{\rm eff}^{-1/3}
         & {\rm for} &  D \ll \ell_{\rm eff}  \\
     & \ell_{\rm eff} D^{-3}
         & {\rm for }     & D \gg \ell_{\rm eff}~,   \\
                \end{array} \right.
\end{equation}
where we neglected a logarithmic correction factor which is not
important for our scaling arguments. As shown in the preceding
section, the effective persistence length $\ell_{\rm eff}$
depends on the screening length and the line charge density; in
essence, one has to keep in mind that $\ell_{\rm eff}$ is  larger
than $\ell_0$ for a wide range of parameters because of
electrostatic stiffening effects.
\footnote{The situation is complicated by the fact that the
electrostatic contribution to the persistence length is scale
dependent and decreases as the chain is bent at length scales
smaller than the screening length. This leads to modifications of
the entropic confinement force, Eq.~(\ref{frep}), if the
deflection length is smaller than the screening length. As can be
checked explicitly, all results reported here are not changed by
these modifications.}

The equilibrium layer thickness follows from equating the
attractive and repulsive forces, Eqs.~(\ref{fatt}) and
(\ref{frep}). For rather stiff polymers and   small layer
thickness, $D < \kappa^{-1} < \ell_{\rm eff}$, we obtain
\begin{equation}
\label{delta}
D \sim \left(
\ell_B \sigma \tau \ell_{\rm eff}^{1/3} \right)^{-3/5} ~.
\end{equation}
For a layer thickness corresponding to the screening length, $D
\approx \kappa^{-1}$, scaling arguments predict a rather abrupt
desorption transition~\cite{Netz4}. This is in accord with
previous calculations~\cite{Maggs,Gompper,Bundschuh,semi1,semi2}
and simulations \cite{semi3} for a semi-flexible  polymer bound
by short-ranged (square-well) potentials. Setting $D \sim
\kappa^{-1}$ in Eq.~(\ref{delta}), we obtain an expression for the
adsorption threshold (for $\kappa \ell_{\rm eff} > 1$)
\begin{equation}
\label{strong} \sigma^* \sim \frac{\kappa^{5/3} } {\tau \ell_B
\ell_{\rm eff}^{1/3}  } ~.
\end{equation}
For $\sigma   > \sigma^*$ the polymer is adsorbed and localized
over a layer with a width smaller than the screening length (and
with the condition $\ell_{\rm eff}> \kappa^{-1}$ also satisfying
$D < \ell_{\rm eff}$, indicative of a flat layer).
As $\sigma$ is decreased, the polymer
abruptly desorbs at the threshold $\sigma=\sigma^*$ . In the
Gaussian regime, the effective persistence length $\ell_{\rm eff}
$ is given by the bare persistence length $\ell_0$ and the
desorption threshold is obtained by replacing $\ell_{\rm eff} $
by $\ell_0$ in Eq.~(\ref{strong}), \ie\
\begin{equation}
\sigma^* \sim \frac{\kappa^{5/3} } {\tau \ell_B \ell_0^{1/3}  }~.
\end{equation}
In the persistent regime, we have $\ell_{\rm eff} \sim \ell_{\rm
OSF}$ with $\ell_{\rm OSF}$ given by Eq.~(\ref{OSF}). The
adsorption threshold follows from Eq.~(\ref{strong}) as
\begin{equation}
\sigma^* \sim \frac{\kappa^{7/3} } {\tau^{5/3} \ell_B^{4/3}  }~.
\end{equation}
Finally, in the Gaussian-persistent regime, we have an effective
line charge density from Eq.~(\ref{tautilde}) and a modified
persistence length, Eq.~(\ref{OSFtilde}). For the adsorption
threshold we obtain from Eq.~(\ref{strong})
\begin{equation}
\sigma^* \sim \frac{\kappa^{7/3} \ell_0^{5/9} } {\tau^{5/9}
\ell_B^{7/9}  } ~.
\end{equation}

Let us now consider the opposite limit, $\ell_{\rm eff} <
\kappa^{-1}$.
\footnote{From Eq.~(\ref{delta}) we see that the layer thickness
$D$ is of the same order as $\ell_{\rm eff}$ for $ \ell_B
\sigma \tau \ell_{\rm eff}^2 \sim 1$, at which point the
condition $D \ll \ell_{\rm eff} $ used in deriving
Eq.~(\ref{delta}) breaks down. }
If the layer thickness is larger than the persistence length but
smaller than the screening length, $\ell_{\rm eff} < D <
\kappa^{-1} $, the prediction for $D$ obtained from
balancing Eqs.~(\ref{fatt}) and (\ref{frep}) becomes
\begin{equation}
\label{delta2}
D \sim \left( \frac{\ell_{\rm eff}}
{\ell_B \sigma \tau} \right)^{1/3}~,
\end{equation}
which is in accord with our mean-field result in Eq.~(\ref{Dmean})
for a linear potential characterized by $\tau =-1$ and an ideal
polymer chain with $\nu =1/2$. From the expression
Eq.~(\ref{delta2}) we see that $D$ has the same size as
 the screening length $\kappa^{-1}$ for
\begin{equation}
 \label{weak}
\sigma^*  \sim \frac{\ell_{\rm eff} \kappa^3} {\tau \ell_B  } ~.
\end{equation}
This in fact denotes the location of a continuous adsorption
transition at which the layer grows to infinity \cite{Netz4}. The
scaling results for the adsorption behavior of a flexible polymer,
Eqs.~(\ref{delta2})-(\ref{weak}), are in agreement  with previous
results~\cite{Muthukumar}.

\begin{figure}
 \epsfxsize=12cm
 \centerline{\vbox{\epsffile{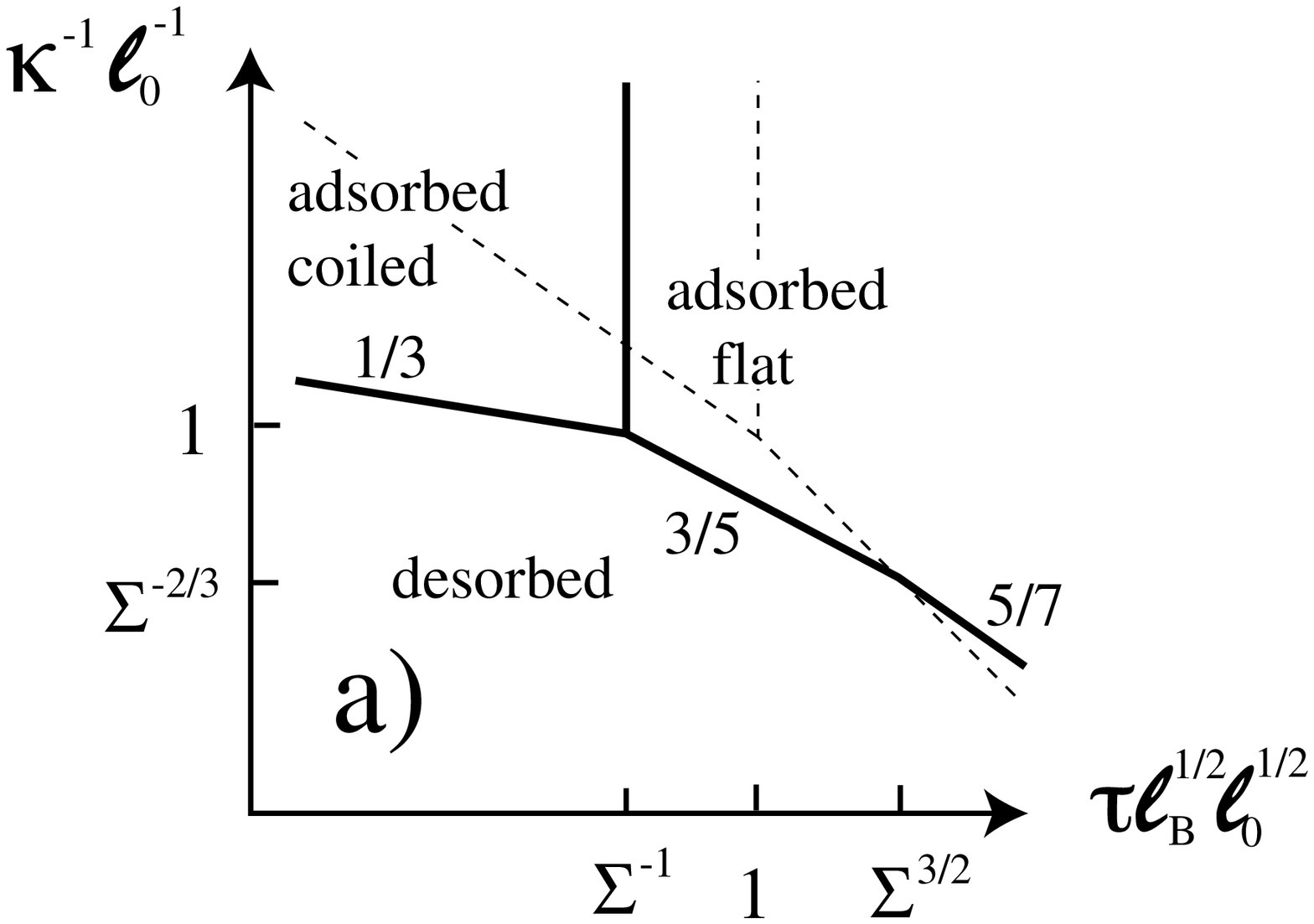} } }
 \epsfxsize=12cm
 \centerline{\vbox{\epsffile{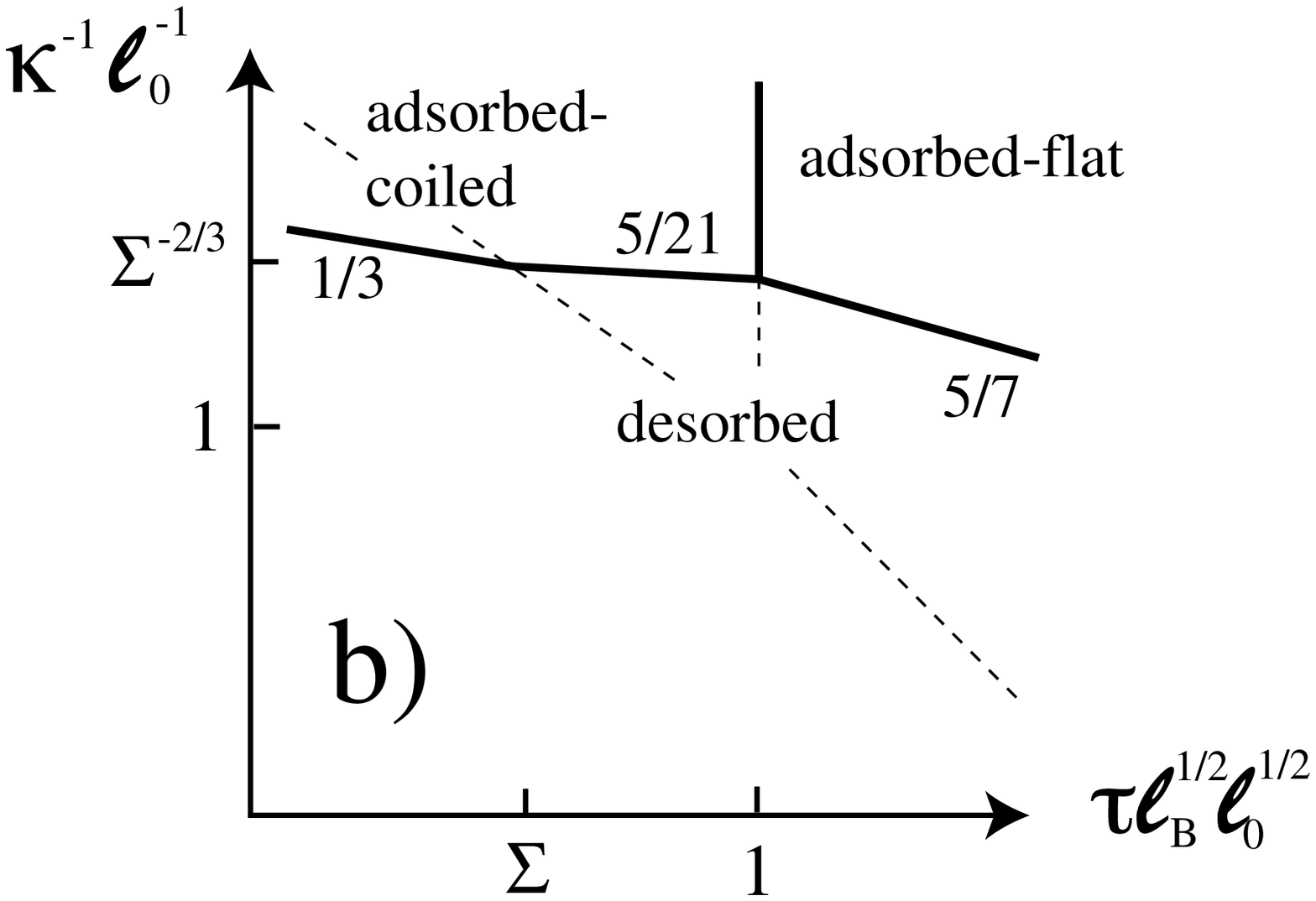} } }
\caption{  \label{fig9} Adsorption scaling diagram shown on a
log-log plot for a) strongly charged surfaces, $\Sigma = \sigma
\ell_0^{3/2} \ell_B^{1/2} >1$ and for b) weakly charged surfaces
$\Sigma <1$. We find a desorbed regime, an adsorbed phase where
the polymer is flat and dense, and an adsorbed phase where the
polymer shows loops. It is seen that a fully charged PE is
expected to adsorb as a flat layer, whereas charge-diluted PE's
can form coiled layers with loops and dangling ends. The broken
lines denote the scaling boundaries of PE chains in the bulk as
shown in Fig.~\ref{fig5}. The numbers on the lines indicate the
power law exponents of the crossover boundaries between the
regimes.}
\end{figure}

In Fig.~\ref{fig9} we show the desorption transitions and the
line at which the adsorbed layer crosses over from being flat, $D
< \ell_{\rm eff}$, to being crumpled or coiled, $D > \ell_{\rm
eff}$. The underlying PE behavior in the bulk, as shown in Fig.
\ref{fig5}, is denoted by broken lines. We obtain two different
phase diagrams, depending on the value of the parameter
\begin{equation}
\Sigma = \sigma \ell_0^{3/2} \ell_B^{1/2} ~.
\end{equation}
For strongly charged surfaces, $\Sigma > 1$, we obtain the phase
diagram as in Fig.~\ref{fig9}a, and for weakly charged surfaces,
$\Sigma < 1$, as in Fig.~\ref{fig9}b. We see that strongly
charged PE's, obeying $\tau \sqrt{\ell_0 \ell_B} > 1$, always
adsorb in flat layers. The scaling of the desorption transitions
is in general agreement with recent computer simulations of
charged PE's~\cite{Yamakov}. Assuming an image-charge repulsion
between the charged monomers and the substrate, as relevant for
low-dielectric substrates, some of the phase boundaries in Fig.
\ref{fig9} are eliminated, as explained in Ref.~\cite{Netz4}.
However, note that not all substrates are low-dielectric
materials, so that the full phase structure in Fig.~\ref{fig9}
might be relevant to some experiments.


\section{Neutral Polymer Adsorption from Solution}
\setcounter{equation}{0}


\subsection{The Mean--Field Approach: Ground State Dominance}

In this section we look at the equilibrium behavior of many chains
adsorbing on (or equivalently depleting from) a surface in contact
with a bulk reservoir of chains at equilibrium. The polymer chains
in the reservoir are assumed to be in a semi-dilute concentration
regime defined by  $c_m> c_m^{*}$, where $c_m$ denotes the monomer
concentration (per unit volume) and $c_m^{*}$ is the
overlap-concentration Eq.~(\ref{overlap}).

As in the previous section, the adsorbing surface is taken as an
ideal and smooth plane. Neglecting lateral concentration
fluctuations (which will be considered in Sec. 9),
one can reduce the problem to an effective
one-dimensional problem, where the monomer concentration depends
only on the distance $x$ from the surface, $c_m=c_m(x)$. The two
boundary values are: $c_m^b=c_m(x\rightarrow \infty)$ in the bulk,
while  $c_m^s=c_m(x=0)$ on the surface.

In addition to the monomer concentration $c_m$, it is more
convenient to work with the monomer volume fraction: $\phi(x)=a^3
c_m(x)$ where $a$ is the Kuhn length which characterizes the
effective monomer size. While the bulk value (far
away from the surface) is fixed by the concentration in the
reservoir, the value on the surface at $x=0$ is self-adjusting in
response  to a given surface interaction. The simplest
phenomenological surface interaction is linear  in the surface
polymer concentration. The resulting contribution to the surface
free energy (per unit area) is

\be
{F}_s=-\gamma_1\phi_s \label{d1} ~,\ee

\noindent where $\phi_s=a^3 c_m^s$ and a positive (negative)
value of $\gamma_1=\tilde{\gamma} (T-T^*)/Ta^2$,
 defined in Eq.~(\ref{freeatt1}),
enhances adsorption (depletion) of the chains on (from) the
surface. However, ${F}_s$ represents only the local reduction in
the interfacial free energy due to the adsorption. In order to
calculate the full interfacial free energy, it is important to
note that monomers adsorbing on the surface are connected to other
monomers belonging to the same polymer chain. The latter
accumulate in the vicinity of the surface. Hence, the interfacial
free energy does not only depend on the surface concentration of
the monomers but also on their concentration in the {\it vicinity}
of the surface. Due to the polymer flexibility and connectivity,
the entire adsorbing layer can have a considerable width. The {\it
total} interfacial free energy of the polymer chains will depend
on this width and is quite different from the interfacial free
energy for simple molecular liquids.

There are several theoretical frameworks
to treat this polymer adsorption. One of the simplest methods
which yet gives reasonable qualitative results is
the Cahn -- de Gennes approach~\cite{ch58,dg81}.
In this approach, it is possible to write down a continuum functional
which describes the contribution to the free energy of the
polymer chains in the solution. This procedure was introduced
by Edwards in the 60's ~\cite{e65}
and was applied to polymers at interfaces
by de Gennes~\cite{dg81}.
Below we present such a continuum version which can be studied
analytically. Another approach is a discrete one, where the monomers
and solvent molecules are put on a lattice. The latter approach
is quite useful in computer simulations and numerical self consistent field
(SCF) studies and is reviewed elsewhere~\cite{fleer}.

In the continuum approach and using a mean--field theory, the bulk
contribution to the adsorption free energy is written in terms of
the local monomer volume fraction $\phi(x)$, neglecting all kinds
of monomer-monomer correlations. The total reduction in the
surface tension $\Delta F$ (interfacial free energy per
unit area and in
units of $k_B T$) is then

\bea
\Delta F = -\gamma_1\phi_s + \int_{0}^{\infty} \dd x \Bigl[
G(\phi)\Bigl({d\phi\over{dx}}\Bigr)^2 + {F_b}(\phi)- {F_b}(\phi_b)
+\mu_p(\phi-\phi_b) \Bigr]~,
\label{d2}
\eea

\noindent
where $\gamma_1$ was defined in
Eq.~(\ref{d1}). The stiffness function $G(\phi)$ represents the
energy cost of local concentration fluctuations and its form is
specific to long polymer chains. For low polymer concentration it
can be written as~\cite{degennes}:

\bea
G(\phi)={1\over a^3}\Bigl({a^2\over 24\phi}\Bigr)~. \label{d3}
\eea

\noindent
 The other terms in Eq.~(\ref{d2}) come from the
Cahn-Hilliard free energy of mixing of the polymer solution,
$\mu_p$ being the polymer chemical potential, and~\cite{Flory1}

\bea
{F_b}(\phi)= {1\over a^3}\Bigl( {\phi\over N}\log \phi + \half
\tilde{v}_2\phi^2 +{1\over 6} \tilde{v}_3\phi^3+
\cdots \Bigr)~, \label{d4}
\eea

\noindent where $N$ is the polymerization index. In the following,
we neglect the first term in Eq.~(\ref{d4}) (translational
entropy), as can be justified in the long chain limit, $N\gg 1$.
The second and third dimensionless virial coefficients are
$\tilde{v}_2 = v_2/a^3$
and $\tilde{v}_3= v_3/a^6$, respectively.
Good, bad and theta solvent conditions
are achieved, respectively, for positive, negative or zero $\tilde{v}_2$.
We concentrate hereafter only on good solvent conditions,
$\tilde{v}_2>0$, in which case the higher order $\tilde{v}_3$-term can be safely
neglected. In addition, the local monomer density is assumed to
be small enough, in order to justify the omission of higher
virial coefficients. Note that for small molecules the
translational entropy always acts in favor of desorbing from the
surface. As was discussed in the Sec.~1, the vanishing small
translational entropy for polymers results in a stronger
adsorption (as compared with small solutes) and makes the polymer
adsorption much more of an irreversible process.

The key feature in obtaining Eq.~(\ref{d2}) is the so-called {\em
ground state dominance}, where for long enough chains $N\gg 1$,
only the lowest energy eigenstate (ground state) of a
diffusion-like equation is taken into account. This approximation
gives us the leading behavior in the $N\rightarrow \infty$
limit~\cite{dg69}. It is based on the fact that the weight of the
first excited eigenstate is smaller than that of the ground state
by an exponential factor: $\exp(-N\,\Delta E)$ where $\Delta
E=E_1-E_0>0$ is the difference in the eigenvalues between the two
eigenstates. Clearly, close to the surface more details on the
polymer conformations can be important. The adsorbing chains have
tails (end-sections of the chains that are connected to the
surface by only one end), loops (mid-sections of the chains that
are connected to the surface by both ends), and trains (sections
of the chains that are adsorbed on the surface), as depicted in
Fig.~\ref{fig3o}a. To some extent it is possible to get profiles of the
various chain segments even within mean--field theory, if the
ground state dominance condition is relaxed as is discussed further below.

Taking into account all those simplifying assumptions and
conditions, the mean--field theory for the interfacial free energy
can be written as:

\bea
\Delta F = -\gamma_1\phi_s+{1\over a^3} \int_{0}^{\infty}\dd
x\,\,\Bigl[{a^2\over 24\phi} {\Bigl({\dd\phi\over \dd x}\Bigr)}^2
+ \half \tilde{v}_2 (\phi(x)-\phi_b)^2\Bigr]~, \label{d5}
\eea

\noindent where the monomer bulk chemical potential $\mu_p$ is given
by $\mu_p=\partial {f(\phi)}/\partial \phi|_b =\tilde{v}_2\phi_b$.

It is also useful to define
the total amount of monomers per unit
area which take part in the adsorption layer. This is the so-called
surface excess $\Gamma$; it is measured experimentally
using, \eg, ellipsometry, and is defined as

\bea
\Gamma={1\over a^3}\int_0^\infty \dd x\,\,[\phi(x)-\phi_b]~.
\label{d6}
\eea

\noindent
(A different quantity, not used in our review, is the so-called
adsorbed amount, which measures the total amount of polymers per
unit area that have at least one monomer in contact with the substrate.)
The next step is to minimize the free energy functional
(\ref{d5}) with respect to both $\phi(x)$ and $\phi_s=\phi(0)$.
For the following algebraic manipulations,
it is more convenient to re-express Eq.~(\ref{d5}) in terms of
the square root of the monomer volume fraction,
$\psi(x)=\phi^{1/2}(x)$
and $\psi_s=\phi_s^{1/2}$

\bea
\Delta F = -\gamma_1\psi_s^2+{1\over a^3}
\int_{0}^{\infty}\dd x\,\,\Bigl[{a^2\over 6} {\Bigl({\dd\psi\over
\dd x}\Bigr)}^2 + \half \tilde{v}_2 (\psi^2(x)-\psi_b^2)^2\Bigr]~.
\label{d6a}
\eea

\noindent Minimization of Eq.~(\ref{d6a}) with respect to
$\psi(x)$ and $\psi_s$ leads to the following profile equation and
boundary condition

\bea
{a^2\over 6}{\dd^2\psi\over \dd x^2} &=&
\tilde{v}_2\psi(\psi^2-\psi_b^2)
\nonumber\\
& & \nonumber \\
{{1\over \psi_s}{\dd \psi\over \dd x}}\Bigl|_s &=&
-{6a}\gamma_1=-{1\over 2D}~. \label{d7}
\eea

\noindent The second equation sets a boundary condition on the
logarithmic derivative of the monomer volume fraction, $\dd
\log\phi/\dd x|_s =2\psi^{-1}\dd \psi/\dd x|_s=-1/D$, where the
strength of the surface interaction $\gamma_1$ can be expressed in
terms of a length $D\equiv 1/(12 a\gamma_1)$. Note that exactly
the same scaling of $D$ on $\gamma_1/T$ is obtained  in
Eq.~(\ref{Dnonideal}) for the single chain behavior if one sets
$\nu = \varphi = 1/2$ (ideal chain exponents). This  is strictly
valid at the upper critical dimension ($d=4$) and is a very good
approximation in three dimensions.

The profile equation (\ref{d7}) can  be integrated
once, yielding

\be
{a^2\over 6}\left({\dd\psi\over \dd x}\right)^2
= \half \tilde{v}_2(\psi^2-\psi_b^2)^2~.
\label{d7a}
\ee

\noindent
The above differential equation can now be solved analytically for
adsorption ($\gamma_1>0$) and depletion ($\gamma_1<0$).

We first present the results in more detail for polymer adsorption
and then repeat the main findings for polymer depletion.

\subsubsection{The Adsorption Case}

Setting $\gamma_1>0$ as is applicable for the adsorption case, the
first-order differential equation (\ref{d7a}) can be integrated
and together with the boundary condition Eq.~(\ref{d7}) yields

\bea
\phi(x)&=&\phi_b\coth^2\Bigl({x+x_0\over\xi_b}\Bigr)~,
\label{d8ad}
\eea

\noindent where the length $\xi_b=a(3\tilde{v}_2 \phi_b)^{-1/2}$ is the
Edwards correlation length characterizing the exponential decay of
concentration fluctuations in the bulk~\cite{degennes,e65}.
(See also the discussion in Sec.~7.2).
The
length $x_0$ is not an independent length since it depends on $D$
and $\xi_b$, as can be seen from the boundary condition
Eq.~(\ref{d7})

\bea
x_0={\xi_b\over 2}{\rm arcsinh} \Bigl({4D\over \xi_b}\Bigr)=
\xi_b {\rm arccoth} (\sqrt{\phi_s/\phi_b})~.
\label{d9}
\eea

\noindent
Furthermore, $\phi_s$ can be directly related to the
surface interaction $\gamma_1$ and the bulk value $\phi_b$

\be \frac{\xi_b}{2D}=\frac{6a^2\gamma_1}{ \sqrt{3 \tilde{v}_2\phi_b}}=
\sqrt{\frac{\phi_b}{\phi_s}}\left(\frac{\phi_s}{\phi_b}-1\right)~.
\label{d10} \ee

In order to be consistent with the semi-dilute concentration
regime, the correlation length $\xi_b$ should be smaller than the
size of a single chain, $R=a N^{\nu}$, where $\nu= 3/5$ is the
Flory exponent in good solvent conditions. This sets a lower bound
on the polymer concentration in the bulk, $c_m>c_m^*$.

So far three length scales have been introduced: the Kuhn length
or monomer size $a$, the adsorbed-layer width $D$, and the bulk
correlation length  $\xi_b$. It is more convenient for the
discussion to consider the case where those three length scales
are quite separated: $a\ll D \ll \xi_b$. Two conditions must be
satisfied. On one hand, the adsorption parameter is not large, $12
a^2 \gamma_1\ll 1$ in order to have $D\gg a$. On the other hand,
the adsorption energy is large enough to satisfy $12 a^2
\gamma_1\gg  \sqrt{3 \tilde{v}_2 \phi_b}$ in order to have $D\ll \xi_b$.
The latter inequality can be regarded also as a condition for the
polymer bulk concentration. The bulk correlation length is large
enough if indeed the bulk concentration (assumed to be in the
semi-dilute concentration range) is not too large. Roughly, let
us  assume in a typical case that the three length scales are
well separated: $a$ is of the order of a few Angstroms, $D$ of the
order of a few dozens of Angstroms, and $\xi_b$ of the order of a
few hundred Angstroms.

When the above two inequalities are satisfied,
three spatial regions of adsorption can be differentiated:
the proximal, central, and distal regions,
 as is outlined below. In addition, as soon as $\xi_b \gg D$,
$x_0\simeq 2 D$, as follows  from Eq.~(\ref{d9}).

\begin{itemize}
\item
Close enough to the surface, $x \sim a$, the adsorption profile
depends on the details of the short range interactions between the
surface and monomers. Hence, this region is not universal. In the
proximal region, for $a\gg x\gg D$, corrections to the mean--field
theory analysis  (which assumes the concentration to be constant)
are presented below similarly to the treatment of the single chain
section. These corrections reveal a new scaling exponent
characterizing the concentration profile. They are of particular
importance close to the adsorption/desorption transition.

\item
In the distal region, $x\gg\xi_b$, the excess polymer
concentration decays exponentially to its bulk value

\be
\phi(x)-\phi_b \simeq 4 \phi_b \e^{-2x/\xi_b}~,
\label{d11}
\ee

\noindent as follows from Eq.~(\ref{d8ad}). This behavior is very
similar to the decay of fluctuations in the bulk with $\xi_b$
being the correlation length.

\item
Finally, in the central region (and with the assumption that
$\xi_b$ is the largest length scale in the problem), $ D \ll x \ll
\xi_b$, the profile is universal and from Eq.~(\ref{d8ad}) it can
be shown to decay with a power law

\bea
\phi(x)&=& {1\over 3\tilde{v}_2}{\Bigl({a\over x+2D}\Bigr)}^2 ~.
\label{d12ad}
\eea

\end{itemize}

\begin{figure}
 \epsfxsize=10cm
 \centerline{\vbox{\epsffile{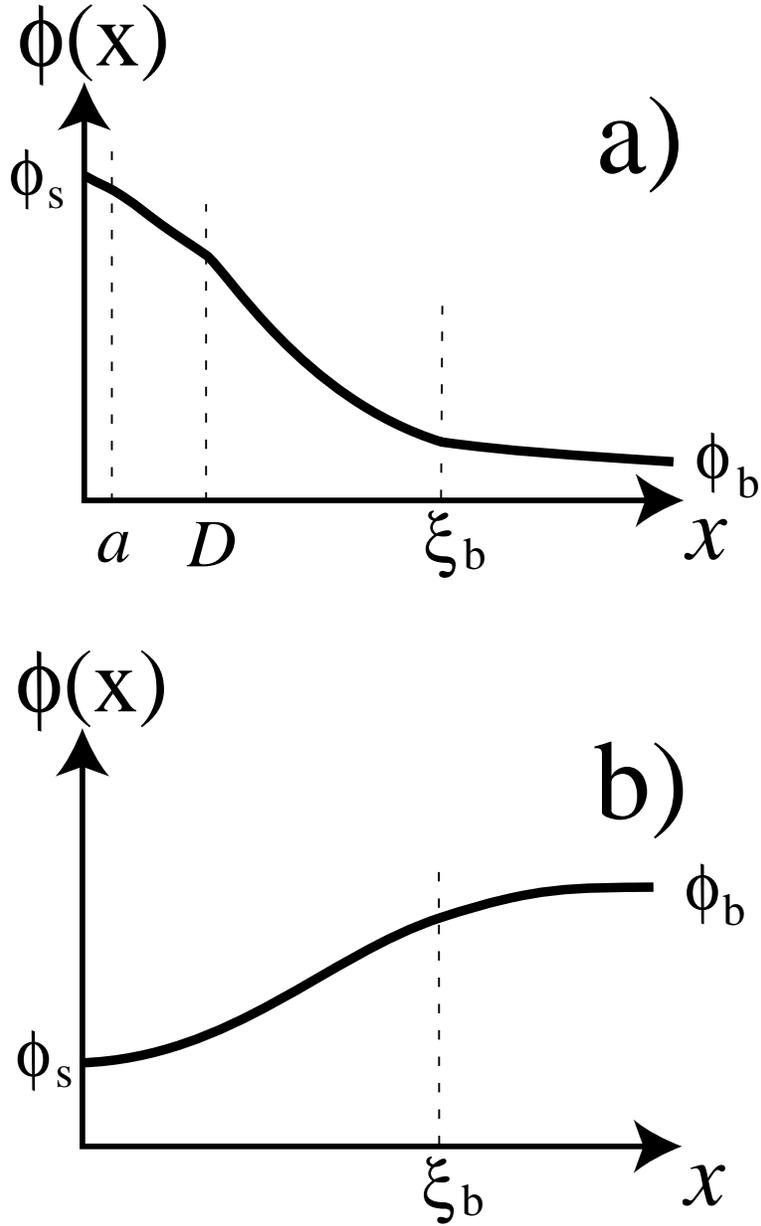} } }
\caption{ \label{fig8o} a) The schematic monomer volume fraction
profile for the case of adsorption from a semi-dilute solution; we
distinguish a layer of molecular thickness $x \sim a$ where the
polymer density depends on details of the interaction with the
substrate and the monomer size, the proximal region $a < x<  D$
where the decay of the density is governed by a universal power
law (which cannot be obtained within mean--field theory), the
central region for $D< x< \xi_b$ with a self-similar profile, and
the distal region for $\xi_b < x$, where the monomer volume
fraction relaxes exponentially towards its bulk value $\phi_b$. b)
The density profile for the case of depletion, where the
concentration close to the surface is $\phi_s$ and relaxes to its
bulk value, $\phi_b$, at a distance of the order of the bulk
correlation length $\xi_b$.}
\end{figure}

\noindent A sketch of the different scaling regions in the
adsorption profile is given in Fig.~\ref{fig8o}a. Included in this figure
are corrections in the proximal region, which is discussed further
below.

A special consideration should be given to the formal limit of
setting the bulk concentration to zero, $\phi_b \rightarrow 0$
(and equivalently $\xi_b \rightarrow \infty$), which denotes the limit
of an adsorbing layer in contact with a polymer
reservoir of vanishing concentration.
It should be emphasized that this limit  is not
consistent with the assumption of a semi-dilute polymer solution
in the bulk. Still,  some information on the polymer density
profile close to the adsorbing surface, where the polymer solution
is locally semi-dilute~\cite{bd87}, can be obtained.
Formally, we
take the limit $\xi_b \rightarrow \infty$ in Eq.~(\ref{d8ad}), and
the limiting expression, given by Eq.~(\ref{d12ad}), does not
depend on $\xi_b$. The profile in the central region decays
algebraically. In the case of zero polymer concentration in the
bulk, the natural cutoff is not $\xi_b$ but rather $R$, the coil
size of a single  polymer in solution. Hence, the distal region
loses its meaning and is replaced by a more complicated scaling
regime~\cite{g92}. The length $D$ can be regarded as the layer
thickness in the $\xi_b\rightarrow \infty$ limit in the sense that
a finite fraction of all the monomers are located in this layer of
thickness $D$ from the surface. Another observation is  that
$\phi(x) \sim 1/x^2$ for $x\gg D$. This power law is a result of
the mean--field theory  and its modification is discussed below.

It is now possible to calculate within the mean--field theory the
two physical quantities that are measured in many experiments: the
surface tension reduction $\Delta F$ and the surface excess
$\Gamma$.

The surface excess, defined in Eq.~(\ref{d6}), can be calculated
in a close form by inserting Eq.~(\ref{d8ad}) into Eq.~(\ref{d6}),

\bea
\Gamma={1\over \sqrt{3\tilde{v}_2}a^2}\Bigl(\phi_s^{1/2}-\phi_b^{1/2}\Bigr)
={\xi_b\phi_b\over a^3}\Bigl(\sqrt{\phi_s\over \phi_b}-1\Bigr) ~.
\label{d13}
\eea

\noindent For strong adsorption, we obtain from Eq.~(\ref{d10})
that  $\phi_s \simeq (a/2D)^2/3 \tilde{v}_2\gg \phi_b$, and Eq.~(\ref{d13})
reduces to

\bea
\Gamma ={1\over 3\tilde{v}_2 a^2}\Bigl({a\over D}\Bigr) \sim \gamma_1~,
\label{d14}
\eea

\noindent while the surface volume fraction scales as $\phi_s\sim
\gamma_1^2$. As can be seen from Eqs.~(\ref{d14}) and
(\ref{d12ad}), the surface excess as well as the entire profile
does not depend (to leading order) on the bulk concentration
$\phi_b$. We note again that the strong adsorption condition is
always satisfied in the $\phi_b \rightarrow 0$ limit. Hence,
Eq.~(\ref{d14}) can be obtained directly by integrating the
profile in the central region, Eq. (\ref{d12ad}).

Finally, let us calculate the reduction  in surface tension
for the adsorbing case.
Inserting the variational
equations (\ref{d7}) in Eq.~(\ref{d5})
yields

\bea \Delta F=- \gamma_1\phi_s+{ \sqrt{3\tilde{v}_2}\over
9a^2}\phi_s^{3/2} \Bigl[1-3\Bigl({\phi_b\over\phi_s}\Bigr)+
2\Bigl({\phi_b\over\phi_s}\Bigr)^{3/2}\Bigr] ~. \label{d16}
\eea

\noindent
The surface term in Eq.~(\ref{d16}) is negative
while the second term is positive.
For strong adsorption this reduction of $\Delta F$
does not depend on $\phi_b$
and reduces to

\bea \Delta F\sim -{\left({a^2\gamma_1}\right)}^3~ {1\over a^2} +
{\cal O}(\gamma_1^{4/3})~, \label{d17}
\eea

\noindent
where the leading term is just the contribution of
the surface monomers.

\subsubsection{The Depletion Case}
We highlight the main differences between  the polymer adsorption
and polymer depletion. Keeping in mind that $\gamma_1<0$ for
depletion, the solution of the same profile equation (\ref{d7a}),
with the appropriate boundary condition results in

\bea
\phi(x)&=&\phi_b\tanh^2\Bigl({x+x_0\over\xi_b}\Bigr)~,
\label{d8de}
\eea
%
which is schematically plotted in Fig.~\ref{fig8o}b.
The limit $\phi_b \rightarrow 0$ cannot be taken
in the depletion case since
depletion with respect to a null reservoir has no meaning.
However, we can, alternatively, look at the strong depletion limit,
defined by the condition $\phi_s \ll \phi_b$. Here
we find
\bea
\phi(x)&=& {3\tilde{v}_2\phi_b^2}{\Bigl({ x+2D\over a}\Bigr)}^2 ~.
\label{d12de}
\eea
%
In the same limit, we find for the surface volume fraction $\phi_s
\sim \phi_b^2\gamma_1^{-2}$, and the exact expression for the
surface excess Eq.~(\ref{d13}) reduces to

\bea
\Gamma  =-{1\over a^2}\sqrt{\phi_b\over 3\tilde{v}_2}
\simeq -{\phi_b\xi_b\over a^3}~.
\label{d15}
\eea

\noindent
The negative surface excess  can be
directly estimated from a profile
varying from $\phi_b$ to zero over a length scale of order $\xi_b$.

The dominating behavior for the surface tension can be calculated
from Eq.~(\ref{d5}) where both terms are now positive. For the
strong depletion case we get

\bea \Delta F
 \simeq {1 \over a^2}\Bigl({a\over \xi_b}\Bigr)^3
\sim \phi_b^{3/2}~.
\label{d18}
\eea


\subsection{Beyond Mean--Field Theory:
Scaling Arguments for Good Solvents}

One of the mean--field theory results that should be corrected is
the scaling of the correlation length with $\phi_b$. In the
semi-dilute regime, the correlation length can be regarded as the
average mesh size created by the overlapping chains. It can be
estimated using very simple scaling arguments~\cite{degennes}
similar to our derivation of the overlap concentration in
Eq.~(\ref{overlap}). The volume fraction of monomers inside a coil
formed by a subchain consisting of $g$  monomers embedded in $d$
dimensional space is $\phi_b \sim g^{1-d\nu}$ where $\nu$ is the
Flory exponent. The spatial scale of this subchain is given by
$\xi_b \sim a g^\nu$. Combining these two relations we obtain the
general scaling of the correlation length

\be
\xi_b \simeq a\phi_b^{\nu/(1-d\nu)} ~,
\label{d19}
\ee
%
and for good solvent condition and $d=3$
\be
\xi_b\simeq a \phi_b^{-3/4} ~.
\label{d20}
\ee
This relation corrects the  mean--field theory result
$\xi_b \sim \phi_b^{-1/2}$  which can be obtained from, \eg,
Eq.~(\ref{d5}), and also directly from Eq.~(\ref{d19}) by setting $d=4$ and
inserting the Gaussian exponent $\nu=1/2$.

\subsubsection{Scaling for Polymer Adsorption}

We repeat here an argument due to de Gennes~\cite{dg81}. The main
idea is to assume that the relation Eq.~(\ref{d19}) holds locally:
$\phi(x)=[\xi(x)/a]^{-4/3}$, where $\xi(x)$ is the local ``mesh
size'' of the semi-dilute polymer solution. Since there is no
other length scale in the problem beside the distance from the
surface, $x$, the correlation length $\xi(x)$ should scale as the
distance $x$ itself, $\xi(x)\simeq x$ leading to the profile

\be
\phi(x) \simeq \left(\frac{a}{x}\right)^{4/3}~.
\label{d22}
\ee

\noindent We note that this argument holds only in the central
region $D\ll x\ll \xi_b$. It has been confirmed experimentally
using neutron scattering~\cite{neutscatt} and neutron
reflectivity~\cite{neutrefl}. Equation (\ref{d22}) satisfies the
distal boundary condition: $x\rightarrow \xi_b$, $\phi(x)
\rightarrow \phi_b$, but for $x > \xi_b$ we expect the regular
exponential decay behavior of the distal region, Eq.~(\ref{d11}).
De Gennes also proposed (without a rigorous proof) a convenient
expression for $\phi(x)$, which has the correct crossover from the
central to the mean--field proximal region~\cite{dg81}

\be
\phi(x)=\phi_s\left(\frac{{4\over3}D}{x+{4\over3}D}\right)^{4/3}
\simeq \left(\frac{a}{x+{4\over 3}D}\right)^{4/3}~.
\label{d23}
\ee

\noindent
Note that the above equation reduces to Eq.~(\ref{d22})
for $x\gg D$. The extrapolation of Eq.~(\ref{d23}) also gives the
correct definition of $D$: $D^{-1}= -\dd \log \phi/\dd x|_s$. In
addition, $\phi_s$ is obtained from the extrapolation to $x=0$ and
scales as

\be
\phi_s= \phi(x=0)=\left(\frac{a}{D}\right)^{4/3}~.
\label{d24}
\ee

\noindent
For strong adsorption ($\phi_s\gg \phi_b$), we have

\bea
\phi_s &\simeq& {\left(\frac{a}{D}\right)}^{4/3} \sim \gamma_1^2~,
\nonumber \\ &&\nonumber\\
D &\simeq& a \left(\frac{1}{a^2\gamma_1}\right)^{3/2} \sim
\gamma_1^{-3/2}~,
\nonumber\\ &&\nonumber\\
\Gamma &\simeq & a^{-2}\,\left({a^2\gamma_1}\right)^{1/2} \sim
\gamma_1^{1/2} ~,
\nonumber \\ &&\nonumber \\
\Delta F &\simeq &-\frac{1}{a^2}\phi_s^{3/2} \sim -\gamma_1^3~.
\label{d25} \eea

\noindent It is interesting to note that although $D$ and $\Gamma$
have different scaling with the surface interaction $\gamma_1$ in
the mean--field theory and scaling approaches, $\phi_s$ and
$\Delta F$ have the same scaling using both approaches. This is a
result of the same scaling $\phi_s \sim \gamma_1^2$, which, in
turn, leads to $\Delta F\simeq \gamma_1\phi_s\sim \gamma_1^3$.

\subsubsection{Scaling for Polymer Depletion}

For polymer depletion similar arguments led de Gennes~\cite{dg81} to propose
the following scaling form for the central and mean--field proximal
regions, $a<x<\xi_b$,

\bea
\phi(x)=\phi_b\left( \frac{x+{5\over 3}D}{\xi_b} \right)^{5/3}~,
\label{d26}
\eea
%
where the depletion thickness is $\xi_b-D$
whereas in the strong depletion regime ($\phi_s\ll \phi_b$)

\bea
\phi_s &\simeq &\phi_b\left(\frac{D}{\xi_b}\right)^{5/3}\sim
\phi_b^{9/4}\gamma_1^{-5/2}~,
\nonumber \\ & &\nonumber \\
D&=&a\left({a^2\gamma_1}\right)^{-3/2} ~,
\nonumber \\ & &\nonumber \\
\Gamma &\simeq& -\phi_ba^{-3}(\xi_b-D)\sim \phi_b^{1/4} ~,
\nonumber \\ & &\nonumber \\
\Delta F &\simeq& \frac{1}{a^2} \phi_b^{3/2} ~. \label{d27} \eea

Note that the above scaling of the surface tension
with the bulk concentration, $\phi_b$ is the same as that obtained
by the mean--field theory approach in Sec.~7.1.2, Eq.~(\ref{d18}).

\subsection{Proximal Region Corrections}

So far we did not address any corrections in the proximal region:
$a<x<D$ for the many chain adsorption. In the mean--field theory
picture the profile in the proximal region is featureless and
saturates smoothly to its extrapolated surface value, $\phi_s>0$.
However, in relation to surface critical phenomena which is in
particular relevant close to the adsorption-desorption phase
transition (the so-called `special' transition),  the polymer
profile in the proximal region has a scaling form with another
exponent $m$.

\be
\phi(x)\simeq \phi_s\left(\frac{a}{x}\right)^m
\label{d28}
\ee
\noindent where $m=(\varphi+\nu-1)/\nu$ is the proximal exponent,
Eq.~(\ref{proxm}). This is similar to the single chain treatment
in Sec.~5.

For good solvents, one has $m\simeq 1/3$, as was derived  using
analogies with surface critical phenomena, exact enumeration of
polymer configurations, and Monte-Carlo simulations~\cite{ekb82}.
It is different from the exponent 4/3 of the central region.

With the proximal region correction, the polymer profile
can be written as~\cite{dgp83}

\bea
\phi(x) \simeq
     \left\{
     \begin{array}{llll}
     & \phi_s & {\rm for} &  0<x<a  \\
     &&& \\
     &   \phi_s\left(\frac{a}{x}\right)^{1/3} & {\rm for} &  a<x<D  \\
     &&& \\
     & \phi_s\left(\frac{a}{x}\right)^{1/3}\left(\frac{D}{x+D}\right)
                                    & {\rm for} & D<x<\xi_b  \\
                \end{array} \right.
\label{d29}
\eea

\bigskip \noindent where

\medskip
\be
\phi_s=\frac{a}{D}
\label{d30}
\ee

\medskip
\noindent
The complete adsorption profile is shown schematically in Fig.~\ref{fig8o}a.
By minimization of the free energy with respect to the layer
thickness $D$ it is possible to show that $D$
is proportional to  $1/\gamma_1$

\be
D\sim \gamma_1^{-1}~,
\label{d31}
\ee

\noindent in accord with the exact field-theoretic results for a
single chain as discussed in Sec.~5.

The surface concentration, surface excess and surface tension have
the following scaling \cite{dgp83}:

\bea
\phi_s &\simeq& \frac{a} {D} \sim \gamma_1
\nonumber\\ &&\nonumber\\
\Gamma &\simeq& a^{-3}D \left(\frac{a}{D}\right)^{4/3} \sim \gamma_1^{1/3}
\nonumber\\ &&\nonumber\\
\Delta F &\simeq& -{\gamma_1^2 a^2} \sim \gamma_1^2 ~.\label{d32}
\eea

Note the differences in the scaling of the surface tension
and surface excess in Eq.~(\ref{d32}) as compared with the
results obtained with no proximity exponent ($m=0$) in the previous
section, Eq.~(\ref{d25}).

At the end of our discussion of polymer adsorption from solutions,
we would like to add that for the case of adsorption from dilute
solutions, there is an intricate crossover from the single-chain
adsorption behavior, as discussed in Sec.~5, to the adsorption
from semi-dilute polymer solutions, as discussed in this
section~\cite{bd87}. Since the two-dimensional adsorbed layer has
a higher local polymer concentration than the bulk, it is possible
that the adsorbed layer forms a two-dimensional semi-dilute state,
while the bulk is a truly dilute polymer solution. Only for
extremely low bulk concentration or for very weak adsorption
energies  the adsorbed layer has a single-chain structure with no
chain crossings between different polymer chains.

\subsection{Loops  and Tails}
It has been realized quite some time ago that the so-called
central region of an  adsorbed polymer layer is characterized  by
a rather broad distribution of loop and tail
sizes~\cite{fleer,taleoftails,johner93}. A loop is defined as a
chain region located between two points of contact with the
adsorbing surface, and a tail is defined as the chain region
between the free end and the closest contact point to the surface,
while a train denotes a chain section which is tightly bound
to the substrate (see Fig.~\ref{fig3o}a).
The relative statistical weight of loops and tails
in the adsorbed layer is clearly of importance to applications.
For example, it is expected that polymer loops which are bound at
both ends to the substrate are more prone to entanglements with
free polymers than tails and, thus, lead to enhanced friction
effects. It was found in detailed numerical mean--field theory
calculations that the external part of the adsorbed layer is
dominated by dangling tails, while the inner part  is mostly built
up by loops~\cite{fleer,taleoftails}.

Recently, an analytical theory was formulated which correctly
takes into account the separate contributions of loops and tails
and which thus goes beyond the {\em ground state dominance}
assumption made in ordinary mean--field theories.  The theory
predicts  that a crossover between tail-dominated and
loop-dominated regions occurs at some distance $x^*\simeq a
N^{1/(d-1)}$~\cite{semenov95} from the surface, where $d$ is the
dimension of the embedding space. It is well known that mean--field
theory behavior can formally be obtained by setting the embedding
dimensionality equal to the upper critical dimension, which is for
self-avoiding polymers given by $d=4$ ~\cite{Cloizeaux}. Hence,
the above expression predicts a crossover in the adsorption
behavior at a distance $x^*\simeq a N^{1/3}$. For good-solvent
conditions in three dimensions ($d=3$), $x^*\simeq a N^{1/2}$. In
both cases, the crossover occurs at a separation much smaller than
the size of a free polymer $R\sim a N^{\nu}$ where, according to
the classical Flory argument \cite{Flory1}, $\nu = 3/(d+2)$.

A further rather subtle result of these improved mean--field
theories is the occurrence of a depletion hole, \ie, a region at a
certain separation from the adsorbing  surface where the monomer
concentration is smaller than the bulk concentration
\cite{semenov95}. This depletion hole results from an interplay
between the depletion of free polymers from the adsorbed layer and
the slowly decaying density profile due to dangling tails. It
occurs at a distance from the surface comparable with the radius
of gyration of a free polymer, but also shows some dependence on
the bulk polymer concentration.
In a different formulation, the interfacial free energy of an
adsorbed layer consisting of finite-length polymers has been
calculated for the full concentration range from
dilute to dense solutions \cite{Manghi}.
These and other effects, related
to the occurrence of loops and tails in the adsorbed layer, have
been recently reviewed \cite{semrev}.

\section{Adsorption of Polyelectrolytes -- Mean Field}
\label{sectionpe}
\setcounter{equation}{0}

In Sec.~6 we have been reviewing the behavior of single PE chains
close to a charged wall (or surface). This will be now extended
to include adsorption of PE from bulk (semi-dilute) solutions
having a bulk concentration $c_m^b$. As before the chains are
assumed to have a fraction $f$ of charged monomers, each carrying
a charge $e$ resulting in a linear charge density, $\tau=f/b$.
The interesting case of {\it polyampholytes} having negative and positive
charges is not considered in this section.
The
solution can also contain salt (small ions) of concentration
$c_\s$ which is directly related to the Debye-H\"uckel screening
length, $\kappa^{-1}$. For simplicity, the salt is assumed
throughout this section to be monovalent ($z=1$).

We will consider  adsorption only onto a single flat and charged
surface. Clearly the most important quantity is the profile of
the polymer concentration, $c_m(x)=\phi(x)/a^3$, as function of
$x$, the distance from the surface. Another useful quantity
mentioned already in Sec.~7 is
the polymer surface excess (per unit area)

\begin{equation}
\Gamma= \int_0^\infty [c_m(x)-c_m^b] dx
=\frac{1}{a^3}\int_0^\infty[\phi(x)-\phi_b] dx~.
\end{equation}
Related to the surface excess $\Gamma$ is the amount of charges
(in units of $e$) carried by the adsorbing PE chains, $f\Gamma$.
In some cases the adsorbed polymer layer carries a higher charge (per unit area)
than the charged surface itself, $f\Gamma> \sigma$, and the
surface charge is overcompensated by the PE as we will see later.
This does not violate global charge neutrality in the system because of
the presence of counterions in solution.

In many experiments, the total amount of polymer  surface excess
$\Gamma$ is measured as a function of the bulk polymer
concentration, pH and/or ionic strength of the bulk solution
~\cite{Peyser}-\cite{Hoogeveen}.
(For reviews see, \eg\ Refs.~\cite{fleer,Cohen1,Cohen2,Norde}).
More recently, spectroscopy ~\cite{Meadows} and ellipsometry
~\cite{Shubin} have been used to measure the width of the
adsorbed PE layer. Other techniques  such as neutron scattering
can be employed to measure the entire profile $c_m(x)$ of the
adsorbed layer ~\cite{neutrefl,auroy1}.

Electrostatic interactions play a crucial role in the adsorption
of PE's \cite{fleer,Cohen1,Cohen2}. Besides the fraction
$f$ of charged monomers, the important parameters are the surface
charge density (or surface potential in case of conducting
surfaces), the amount of salt (ionic strength of low molecular
weight electrolyte) in solution and, in some cases, the solution
pH.

For PE's the electrostatic interactions between
the monomers themselves (same charges) are always repulsive,
leading to an effective stiffening of the
chain~\cite{barrat2,Netz2}. Hence, this interaction will {\em
favor} the adsorption of single polymer chains, because their
configurations are already quite extended~\cite{Netz4}, but it
will {\em oppose} the formation of dense adsorption layers close
to the surface \cite{fleer}.
If the PE chains and
the surface are oppositely charged, the electrostatic interactions
between them will {\em enhance} the adsorption.
In addition,
the presence of salt has a subtle  effect. It simultaneously
screens the monomer-monomer repulsive interactions as well as the
attractive interactions between the oppositely charged surface and
polymer. Presence of multivalent salt ions (not considered in this section)
makes the PE adsorption even more complex.

Two limiting adsorbing cases can be discussed separately: (i) a
non-charged surface on which the chains like to adsorb due to, {\it e.g.,}
van-der-Waals attraction. Here the
interaction between the surface and the chain does not have an
electrostatic component. However, as the salt screens the
monomer-monomer electrostatic repulsion, it leads to enhancement
of the adsorption. (ii) The surface is charged but does not
interact with the polymer besides the electrostatic interaction.
This is called the pure electro-sorption case. At low--salt
concentration, the polymer charge completely compensates the
surface charge. At high--salt concentration some of the
compensation is done by the salt, leading to a decrease in the
amount of adsorbed polymer. In some cases, over-compensation
of the surface charges by the polymer charges can
also occur (as is reviewed below in Sec.~8.5), where
the PE chains form a condensed layer and reverse the
sign of the total surface charge. This is used, \eg, to build a
multi-layered structure of cationic and anionic PE's
--- a process that can be continued for few dozen or even few
hundred times \cite{Decher,caruso}.
The phenomenon of over-compensation
is discussed in Refs.~\cite{Netz4,JFJ99,jpc99,Solis3,Nguyen} but is still not
well understood.

In practice, electrostatic and other types of interactions with
the surface can occur in parallel, making the analysis more
complex.
 In spite of the difficulties
 to treat  theoretically PE's in solution
because of the delicate interplay between the chain connectivity
and the long range nature of electrostatic interactions
~\cite{Oosawa,degennes,Odijk2,Dobrynin}, several simple
approaches treating adsorption exist.
One approach is a discrete {\em multi--Stern layer} model
~\cite{vanderSchee}-\cite{Linse},
 where
the system is placed on a lattice whose sites can be occupied by
a monomer, a solvent molecule or a small ion. The electrostatic
potential is determined self--consistently (mean-field theory)
together with the concentration profiles of the polymer and the
small ions.
   In another approach,  the electrostatic potential
and the PE concentration are treated as continuous functions
~\cite{Muthukumar,jpc99,Varoqui91,Varoqui93,epl95,Chatellier,epr98,mm98}.
These quantities are obtained from two coupled differential
equations derived from the total free energy of the system. In
some cases the salt concentration is considered explicitly
\cite{Varoqui93,epl95}, while in other works, (\eg, in
Ref.~\cite{Muthukumar,Netz4}) it induces a screened Coulombic
interaction between the monomers and the substrate. We will
review the main results of the continuum approach, presenting
numerical solutions of the mean field
equations and scaling arguments.

\subsection{Mean-Field Theory and Its Profile Equations}

The charge density on the polymer chains is assumed to be
continuous and uniformly distributed along the chains. Further
treatments of the polymer charge distribution
 (annealed and quenched models)
can be found in Refs.~\cite{epl95,epr98}.
   Within mean--field approximation, the free energy of the
system can be expressed in terms of the local electrostatic potential
 $U(\rr)$, the local monomer concentration $c_m(\rr)$
and the local concentration of positive and negative ions
 $c^{\pm}(\rr)$. The mean-field approximation means that the
 influence of the charged surface and the inter and intra-chain interactions
 can be expressed in term of an external potential which will
 determine the local concentration of the monomers, $c_m(\rr)$.
 This external potential depends both on the electrostatic
 potential and on the excluded volume interactions between the
 monomers and the solvent molecules.
The excess free energy with respect to the bulk can then be
calculated using another important approximation, the ground
state dominance. This approximation is used often for neutral
polymers ~\cite{degennes} (see Sec.~7) and is valid for very long polymer
chains, $N\gg 1$. As before, we introduce the
(dimensionless) polymer order parameter $\psi(\rr)$, where
$\psi(\rr)=\sqrt{\phi(\rr)}=\sqrt{a^3 c_m(\rr)}$, and express the
adsorption free energy ${\cal F}$ (in units of $k_B T$) in terms
of $\psi$ and $U$ ~\cite{Varoqui91,Varoqui93,epl95,epr98,mm98}
\begin{eqnarray}
   {\cal F} &=& \int \dr \left\{ F_{\rm pol}(\rr) + F_{\rm ions}(\rr)
                        + F_{\rm el}(\rr) \right\}~.
   \label{F}
\end{eqnarray}
  The polymer contribution is
\begin{eqnarray}
   F_{\rm pol}(\rr)  &=&   \asix|\nabla\psi|^2
                 + \half \tilde{v}_2 (\psi^4-\psi_b^4)
                 - \mu_p (\psi^2-\psi_b^2)~,
        \label{fpol}
\end{eqnarray}
where the first term is the polymer elastic energy. Throughout
this section we restrict ourselves to flexible chains described
by a Kuhn length $a$. The second term is the excluded volume
contribution where the dimensionless second virial coefficient
$\tilde{v}_2$ is positive and of order unity.
The case of negative virial coefficients (bad-solvent condition)
has been treated in Ref.~\cite{Dobrynin5}.
The last term couples the system to a
polymer reservoir via a chemical potential $\mu_p$, and
$\psi_b=\sqrt{\phi_b}$ is
 related to the bulk monomer concentration, $c_m^b=\phi_b/a^3$.

   The entropic contribution of the small (monovalent) ions is
\begin{eqnarray}
   F_{\rm ions}(\rr) &=& \sum_{i=\pm}
                   \left[ c^i\ln c^i - c^i
                        - c_\s\ln c_\s + c_\s \right]
                 - \mu^i (c^i-c_\s)~,
        \label{fent}
\end{eqnarray}
   where $c^i(\rr)$ and $\mu^i$ are, respectively,
the local concentration and the chemical potential of the $i=\pm$
ions, while $c_\s$ is the bulk concentration of salt.

Finally, the electrostatic contributions (per $k_B T$) are
\begin{eqnarray}
    F_{\rm el}(\rr)  &=& \left [f e\psi^2 U +e c^+U -e c^- U
                 - \frac{\varepsilon}{8\pi}|\nabla U|^2\right ]/k_B T~.
        \label{fel}
\end{eqnarray}
The first three terms are the
electrostatic energies of the monomers
(carrying $f$ fractional charge per monomer),
the positive ions and the negative ions, respectively.
  The last term is the self energy of the electric field, where
 $\varepsilon$ is the dielectric constant of the solution.
Note that the electrostatic contribution, Eq.~(\ref{fel}), is
equivalent to the well known result: $(\varepsilon/8\pi k_B T)
\int \dr |\nabla U|^2$ plus surface terms. This can be seen by
substituting  the Poisson--Boltzmann equation (as obtained below)
into Eq.~(\ref{fel}) and then integrating by parts.

Minimization of the free energy Eqs.~(\ref{F})-(\ref{fel}) with
respect to $c^\pm$, $\psi$ and $U$ yields a Boltzmann distribution
for the density of the small ions,
 $c^\pm(\rr)=c_\s \exp(\mp e U/k_B T)$, and two
coupled differential equations for $\psi$ and $U$:
\begin{eqnarray}
   \nabla^2U(\rr)
   &=& \frac{8\pi e}{\varepsilon}c_\s \sinh(eU/k_B T)
    -  \frac{4\pi e}{\varepsilon}
       \left( f\psi^2 - f\psi_b^2 \e^{eU/k_B T} \right)~,
   \label{PBs} \\
   \asix \nabla^2\psi(\rr)
   &=& \tilde{v}_2(\psi^3-\psi_b^2\psi) + f\psi  eU/k_B T~.
   \label{SCFs}
\end{eqnarray}
   Equation~(\ref{PBs}) is a generalized
Poisson--Boltzmann equation including the free ions as well as
the charged polymers. The first term represents the salt
contribution and the second term is due to the charged monomers
and their counterions.
 Equation~(\ref{SCFs}) is a generalization of the self--consistent
field equation of neutral polymers Eq.~(\ref{d7}) \cite{degennes}. In the bulk,
the above equations are satisfied by setting $U\to 0$ and $\psi\to
\psi_b$.

\subsection{Constant $U_{\rm s}$: the Low--Salt Limit}
\subsubsection{Numerical Solutions of Mean Field Equations}

  When the surface is taken as
ideal, {\it \ie}, flat and homogeneous, the solutions of the Mean-Field
equations
depend only on the distance $x$ from the surface. The surface
imposes boundary conditions on the polymer order parameter
$\psi(x)$ and the electrostatic potential $U(x)$. Due to global
electroneutrality,
all charge carriers in solution should exactly
balance the surface charges. The
Poisson--Boltzmann equation~(\ref{PBs}), the self--consistent
field equation~(\ref{SCFs}) and the boundary conditions uniquely
determine the polymer concentration profile and the electrostatic
potential. In all cases of interest, these two coupled non--linear equations
can only be solved  numerically.

We present now numerical profiles obtained for surfaces with a
constant potential $U_{\rm s}$:
\begin{eqnarray}
  \left. U \right|_{x=0} = U_{\rm s} ~.
  \label{BCpsi0}
\end{eqnarray}
The boundary conditions for $\psi(x)$ depend on the nature of the
short range non-electrostatic interaction of the monomers and the
surface. For simplicity, we  take a non--adsorbing surface and
require that the monomer concentration will  vanish there:
\begin{eqnarray}
  \left. \psi \right|_{x=0} = 0 ~.
  \label{BCphi0}
\end{eqnarray}
We note that the boundary conditions chosen in
Eqs.~(\ref{BCpsi0})-(\ref{BCphi0}) model the particular situation
of electrostatic attraction at constant surface potential
in competition with a
steric (short range) repulsion of non-electrostatic origin.
Possible variations of these boundary conditions
include surfaces with a constant surface charge (discussed below)
and surfaces with a non-electrostatic short range attractive (or
repulsive) interaction with  the polymer ~\cite{JFJ99,DAJFJ00}.
 Far from the surface
($x\rightarrow\infty$) both $U $ and $\psi $ reach their bulk
values and their derivatives vanish:
 $ U'|_{x\rightarrow\infty} = 0$ and
 $\psi'|_{x\rightarrow\infty} = 0$.

The numerical solutions of the mean--field equations~(\ref{PBs}),
(\ref{SCFs}) together with the boundary conditions discussed
above are presented in Fig.~\ref{fig11}, for several different
physical parameters in the low--salt limit.
The polymer is positively charged and is attracted to
the non-adsorbing surface
held at a constant negative potential.
The aqueous solution contains a small amount of monovalent salt
($c_\s=0.1$\,mM). The reduced concentration profile
$\phi(x)/\phi_b$ is plotted as a function of the distance from
the surface $x$.
Different curves correspond to different values of
the reduced surface potential $u_{\rm s}\equiv e U_{\rm s}/k_B T$, the charge
fraction $f$  and the Kuhn length $a$.
   Although the spatial variation of the  profiles
differs in detail, they all have a single peak which can be
characterized by its height and width. This observation serves as
a motivation to using scaling arguments.

\subsubsection{Scaling Arguments}

The numerical profiles of the previous section indicate that it
may be possible to obtain simple analytical results for the PE
adsorption by assuming that  the adsorption is characterized by
one dominant length scale $D$.  Hence, we write the polymer order
parameter profile in the form
\begin{eqnarray}
   \psi(x)=\sqrt{\phi_M}h(x/D) ~,
   \label{scaling_prof}
\end{eqnarray}
where $h(x/D)$ is a dimensionless function
normalized to unity at its maximum and
 $\phi_M$ sets the scale of polymer adsorption, such that
 $\psi(D)=\sqrt{\phi_M}$.
The free energy can now be expressed  in terms of $D$ and $\phi_M$,
while the exact form of $h(x/D)$ affects only the numerical
prefactors.

 As discussed  below,  the scaling
form  Eq.~(\ref{scaling_prof}),
which describes the density profile as a function of a single
scaling variable $x/D$,
is only valid as long as
$\kappa^{-1}$ and $D$ are not of the same order of magnitude.
Otherwise, the scaling function
$h$ should be a function of both $\kappa x$ and $x/D$.
We concentrate now on the limiting low--salt regime,
$D\ll\kappa^{-1}$, where Eq.~(\ref{scaling_prof}) can be
justified. In the other extreme of high--salt, $D\gg\kappa^{-1}$,
the adsorption crosses over to a depletion, as is discussed below
(Secs. 8.3 and 8.4).
Note that the latter limit is in agreement with the single-chain
adsorption (Sec.~6), where in the high--salt limit and for weakly
charged chains, the PE desorbs from the wall.

In the low--salt regime the effect of the small ions can be
neglected and the free energy (per unit surface area),
Eqs.~(\ref{F})-(\ref{fel}), can be evaluated using the
scaling form Eq.~(\ref{scaling_prof}) and turns out to be given by
 (see also
Refs.~~\cite{Varoqui93,mm98})
\begin{eqnarray}
{F}  \simeq  {1\over 6a D}\phi_M -  f|u_{\rm s}|\phi_M {D\over a^3}
          + 4\pi  l_B f^2 \phi_M^2 {D^3\over a^6}
          + \half  \tilde{v}_2 \phi_M^2 {D\over a^3} ~.
  \label{scalingF}
\end{eqnarray}
In what follows we drop prefactors of
order unity from the various terms.
  The first term of Eq.~(\ref{scalingF}) is the elastic energy
characterizing the response of the polymer to concentration
inhomogeneities. The second term accounts for the
electrostatic attraction of the polymers to the charged surface.
 The third term represents the Coulomb repulsion
between adsorbed monomers. The last term represents the excluded
volume repulsion between adsorbed monomers, where we assume that
the monomer concentration near the surface is much larger than
the bulk concentration $\phi_M\gg\phi_b$ . (The opposite limit,
$\phi_M\le\phi_b$, is consistent
with depletion and will be discussed separately in the
high--salt regime).

\begin{figure}
 \epsfxsize=12cm
 \centerline{\vbox{\epsffile{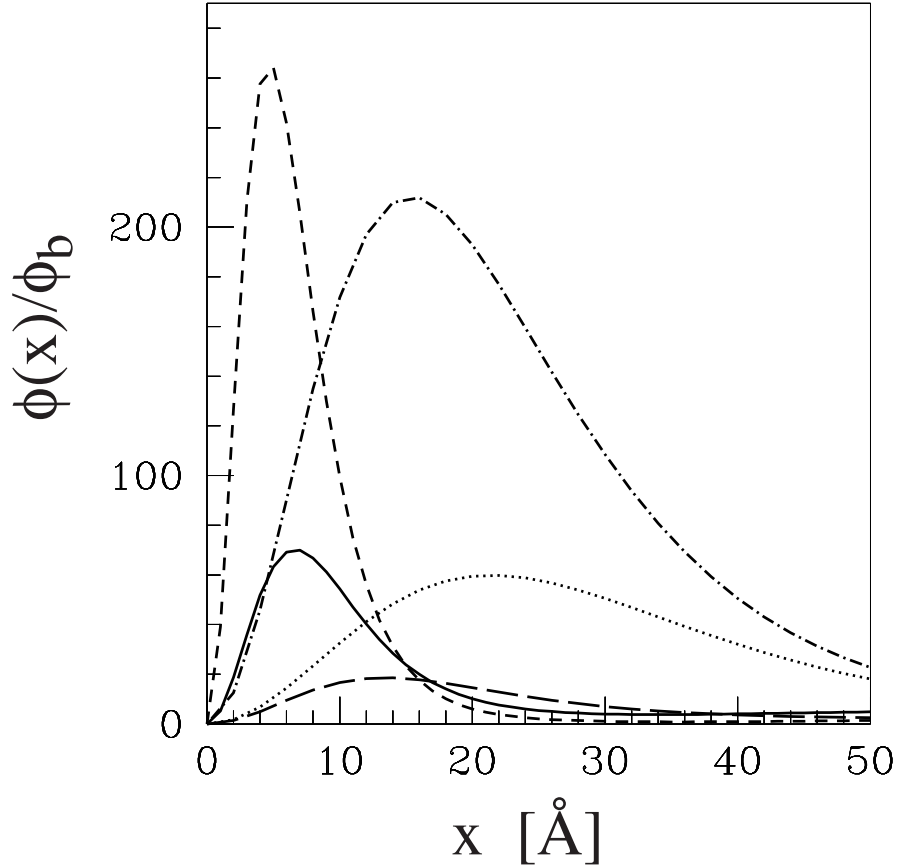} } }
\caption{  \label{fig11}
   Adsorption profiles obtained by numerical solutions of
Eqs.~(\ref{PBs}), (\ref{SCFs}) for several sets of physical
parameters in the low--salt limit. The monomer volume fraction
$\phi(x)$ scaled by its bulk value $\phi_b=\psi_b^2$ is plotted as
a function of the distance  $x$ from the surface.
  The different curves correspond to:
 $f=1$, $a=5$\,\AA\ and $u_s=e U_s /k_B T =-0.5$ (solid curve);
 $f=0.1$, $a=5$\,\AA\ and $u_s=-0.5$ (dots);
 $f=1$, $a=5$\,\AA\ and $u_s=-1.0$ (short dashes);
 $f=1$, $a=10$\,\AA\ and $u_s=-0.5$ (long dashes);
 and $f=0.1$, $a=5$\,\AA\ and $u_s=-1.0$ (dot--dash line).
For all cases  $c_m^b=\phi_b a^3=10^{-6}$\,\AA$^{-3}$,
$\tilde{v}_2=50$\,\AA$^3/a^3$,
 $\varepsilon=80$,
 $T=300$\,K and $c_\s=0.1$\,mM. Adapted from Ref.~\cite{mm98}.}
\end{figure}

   In the low--salt regime and for
strong enough PE's the electrostatic interactions are much
stronger than the excluded volume ones.
Neglecting the latter interactions and minimizing the free energy
with respect to $D$ and $\phi_M$ gives:
\begin{eqnarray}
  D^2 &\simeq&  {a^2 \over f|u_{\rm s}|}
      \sim {1\over f|U_{\rm s}|}
  \label{scalingD}
\end{eqnarray}
  and
\begin{eqnarray}
  \phi_M &\simeq&{a|u_{\rm s}|^2 \over 4\pi l_B }
      \sim |U_{\rm s}|^2 ~,
  \label{scalingcm}
\end{eqnarray}
recalling that $u_{\rm s}= e U_{\rm s}/k_B T$. These
expressions are valid as long as (i) $D\ll\kappa^{-1}$ and (ii)
the excluded volume term in Eq.~(\ref{scalingF}) is negligible.
 Condition (i) translates into
 $c_\s\ll f |u_{\rm s}|/(8\pi l_B a^2)$.
For
 $|u_{\rm s}|\simeq 1$, $a=5$\AA\ and $l_B=7$\AA\
this limits  the salt concentration to
 $c_\s/f \ll $\,0.4 M.
  Condition (ii) on the
magnitude of the excluded volume term
can be shown to be equivalent to
 $f\gg \tilde{v}_2 a|u_{\rm s}|/l_B$.
These requirements are consistent with the numerical data presented in
Fig.~\ref{fig11}.

We recall that the profiles presented in Fig.~\ref{fig11} were
obtained from the numerical solution of Eqs.~(\ref{PBs}) and
(\ref{SCFs}), including the effect of small ions and excluded
volume.
 The scaling relations
 are verified by plotting in
 Fig.~\ref{fig12} the same sets of data as in Fig.~\ref{fig11}~
using rescaled variables as defined in Eqs.~(\ref{scalingD}),
(\ref{scalingcm}). Namely, the rescaled electrostatic potential
 $U(x)/u_{\rm s}$ and polymer concentration
 $\phi(x)/\phi_M\sim \phi(x)/|u_{\rm s}|^2$ are plotted
as functions of the rescaled distance $x/D\sim x
f^{1/2}|u_{\rm s}|^{1/2}/a$. The different numerical data roughly collapse on
the same curve, which demonstrates that the scaling results
in Eqs.~(\ref{scalingD}),
(\ref{scalingcm}) are valid for a whole range of parameters in the low--salt regime.

  In many experiments the total amount of adsorbed polymer
per unit area (surface excess) $\Gamma$ is measured as function of the physical
characteristics of the system such as the charge fraction $f$,
the pH of the solution or
 the salt concentration $c_\s$ (see, \eg\ Refs.
~\cite{Peyser}-\cite{Hoogeveen}).
While in the next section we give general
predictions for a wide range of salt concentration, we comment here
on the low--salt limit, where
the scaling expressions, Eqs.~(\ref{scaling_prof}),(\ref{scalingD}) and
(\ref{scalingcm}), yield
\begin{eqnarray}
  \Gamma = {1\over a^3}\int_0^\infty [\phi(x)-\phi_b] dx
         \simeq {D\over a^3} \phi_M
         \simeq {|u_{\rm s}|^{3/2} \over l_B a f^{1/2}}
         \sim {|U_{\rm s}|^{3/2} \over f^{1/2}}~.
 \label{scalingGamma1}
\end{eqnarray}

   This scaling prediction for the  adsorbed amount $\Gamma(f)$
compares favorably with the numerical results shown in Fig.~\ref{fig13}a,
adapted from Ref.~\cite{san03},
for the low--salt limit (solid line corresponds to
$c_{\rm salt}=1.0$\,mM, and dashed line to 10\,mM).
As a consequence of Eq.~(\ref{scalingGamma1}),
$\Gamma$ decreases with increasing charge fraction $f$.
Similar behavior was also reported in experiments ~\cite{Denoyel}.
This effect is at first glance quite puzzling because as the
polymer charge increases, the chains are subject to a stronger
attraction to the surface. On the other hand, the
monomer--monomer repulsion is stronger and indeed, in this
regime, the  monomer--monomer Coulomb repulsion scales as $(f
\phi_M)^2$, and dominates over the adsorption energy that scales
as $f \phi_M$.

\subsection{Adsorption Behavior in the Presence of Finite Salt}

The full dependence of $\Gamma$ on $c_{\rm salt}$ and $f$, as obtained
from the numerical solutions of the mean-field equations with fixed $U_s$
boundary condition \cite{san03},
is presented in Fig.~\ref{fig13}. Our results
are in agreement with numerical solutions of discrete lattice
models  (the multi--Stern layer theory)
~\cite{fleer,Cohen1,Cohen2,vanderSchee,Papenhuijzen,Evers,vandeSteegSim,Linse}.
In
Fig.~\ref{fig13}a the dependence of $\Gamma$ on $f$ is shown for several
salt concentrations ranging from low--salt conditions, $c_{\rm
salt}$=1.0\,mM, all the way to high salt, $c_{\rm salt}$=0.5\,M.
For low enough $f$, $\Gamma<0$ indicates depletion (as is discussed below).
As $f$
increases, a crossover to the adsorption region, $\Gamma>0$, is
seen. In the adsorption region, a peak in $\Gamma(f)$ signals the
maximum adsorption amount at constant $c_{\rm salt}$. As $f$
increases further, beyond the peak, $\Gamma$ decreases as $f^{-1/2}$
for low--salt concentrations, in agreement with Eq.~(\ref{scalingGamma1}).
Looking at the variation of $\Gamma$ with salt, as $c_{\rm salt}$
increases, the peak in $\Gamma(f)$ decreases and shifts to higher
values of $f$. For very large amount of salt, {\it e.g.,} $c_{\rm
salt}=0.5$\,M, the peak occurs in  the limit $f\to 1$, and only an increase
in $\Gamma(f)$ is seen from the negative depletion values (small $f$)
towards the peak at $f\to 1$.

In Fig.~\ref{fig13}b,
we plot $\Gamma(c_{\rm salt})$ for several $f$ values: 0.03,
0.1, 0.3 and 1.0. For low enough salt condition, the surface
excess is almost independent of $c_{\rm salt}$. In this adsorption
regime, the surface excess is well characterized by the scaling
result of the previous section,  Eq.~(\ref{scalingGamma1}),
$\Gamma\sim f^{-1/2}$. As the
amount of salt increases above some threshold, the adsorption
regime crosses over to depletion quite sharply, signaling the
adsorption-depletion transition. The salt concentration at the
transition, $c_{\rm salt}^*$, increases with the charge fraction
$f$.

\subsection{Adsorption-Depletion Crossover in High--Salt
Conditions}

In the scaling discussion in Sec. 8.2.2, it was assumed implicitly that the PE chains
are adsorbing to the surface.
Namely, the electrostatic interaction with the surface is strong enough
so that it overcomes any compression and entropy loss of the polymers in the adsorbing layer.
This is not correct for highly screened systems (high salt) and weakly charged PE's.

The numerical PE profiles obtained from solving Eqs.~(\ref{PBs})-(\ref{SCFs}) \cite{san03}
demonstrating the adsorption-depletion
transition (which is not a sharp transition but rather
a crossover) are presented in Fig.~\ref{fig22}.
The profiles  were obtained by solving numerically the
differential equations for several values of $f$ in a range including the
adsorption--depletion transition. For salt concentration of about
$c_{\rm salt}^* \simeq 0.16 u_{\rm s} f/(l_{\rm B}a^2)$ (solid line in Fig.~\ref{fig22} with $f=0.09$),
the figure demonstrates the disappearance of the
peak in the concentration profile.
Our way of identifying this crossover is by looking at the surface excess, $\Gamma$.
The place where $\Gamma=0$ indicates an adsorption--depletion
transition, separating positive $\Gamma$ in the adsorption regime
from negative ones in the depletion regime.

The numerical phase diagrams displaying the
adsorption--depletion transition are presented in Fig.~\ref{fig23}, where
the line of vanishing surface excess, $\Gamma=0$, is located in
 the $(f,c_{\rm salt})$ plane while
fixing $u_{\rm s}$ (Fig.~\ref{fig23}a), and in the
$(|u_{\rm s}|,c_{\rm salt})$ plane while fixing $f$ (Fig.~\ref{fig23}b).
From
the figure it is apparent that  the adsorption--depletion  transition line fits
quite well a line of slope 1.0 in both Fig.~\ref{fig23}a and b plotted on a log-log
scale. Namely, $c_{\rm salt}^*\sim f$
for fixed $u_{\rm s}$, and $c_{\rm salt}^*\sim u_{\rm s}$ for fixed $f$.

These scaling forms of  $c_{\rm salt}^*$ at the adsorption-depletion transition can be
reproduced by using simplified scaling arguments, similar to the single-polyelectrolyte
adsorption situation in Sec. 6. There we found that the exact scaling of the
desorption transition is recovered by defining desorption to occur when the prediction
for the adsorption layer thickness $D$ reaches the screening length $\kappa^{-1}$.
The condition for adsorption is thus $\kappa D<1$. Using the scaling for $D$,
Eq.~(\ref{scalingD}), and the definition of $\kappa$, we find the
adsorption--depletion transition to occur at the salt concentration
\begin{equation}
c_{\rm{salt}}^* \simeq \frac{u_{\rm s} f}{l_{\rm B} a^2}.
 \label{depletion_condition_ys}
\end{equation}
in the case
of a fixed surface potential. This explains the numerical results
of Fig.~\ref{fig23}a and b.  We mention the analogous results for fixed surface charge
as well as the phenomenon
of overcompensation in the next subsection.

\subsection{Adsorption of PE's for Constant Surface Charge and its Overcompensation}

We turn now  to a different electrostatic boundary condition of
constant surface charge density and look at the interesting
phenomenon of charge overcompensation by the PE chains in relation to
experiments for PE adsorption on flat surfaces, as well as on
charged colloidal particles~\cite{Decher,donath,caruso}. What was
observed in experiments is that PE's  adsorbing on an oppositely
charged surface can overcompensate the original surface charge.
Because the PE's create a thin layer close to the surface, they
can act as an effective absorbing surface to a second layer of
PE's having an opposite charge compared to the first layer.
Repeating the adsorption of alternating positively and negatively
charged PE's, it is possible to create a multilayer structure of
PE's at the surface. Although many experiments and potential
applications for PE multilayers exist, the  theory of PE
overcompensation is only starting to be
developed~\cite{Netz4,JFJ99,jpc99,Nguyen,epr98,mm98,DAJFJ00}.

The scaling laws presented for constant $U_{\rm s}$ can be used also for
the case of constant surface charge. A surface held at a constant
potential $U_{\rm s}$ will  induce a surface charge density $\sigma$ (in units of $e$).
The two quantities are related by: $dU/dx=-4\pi \sigma
e/\varepsilon$ at $x=0$. We will now consider separately the two
limits: low salt $D \ll \kappa^{-1}$, and high salt  $D \ge
\kappa^{-1}$.

As will be explained in Sec. \ref{latcorr}, an alternative
mechanism for overcharging is produced by
lateral correlations between adsorbed PE's, which in conjunction
with screening by salt ions leads to strongly overcharged
surfaces ~\cite{Netz4,Nguyen}.

\subsubsection{Low--Salt Limit: $D\ll \kappa^{-1}$}

\begin{figure}
 \epsfxsize=9cm
 \centerline{\vbox{\epsffile{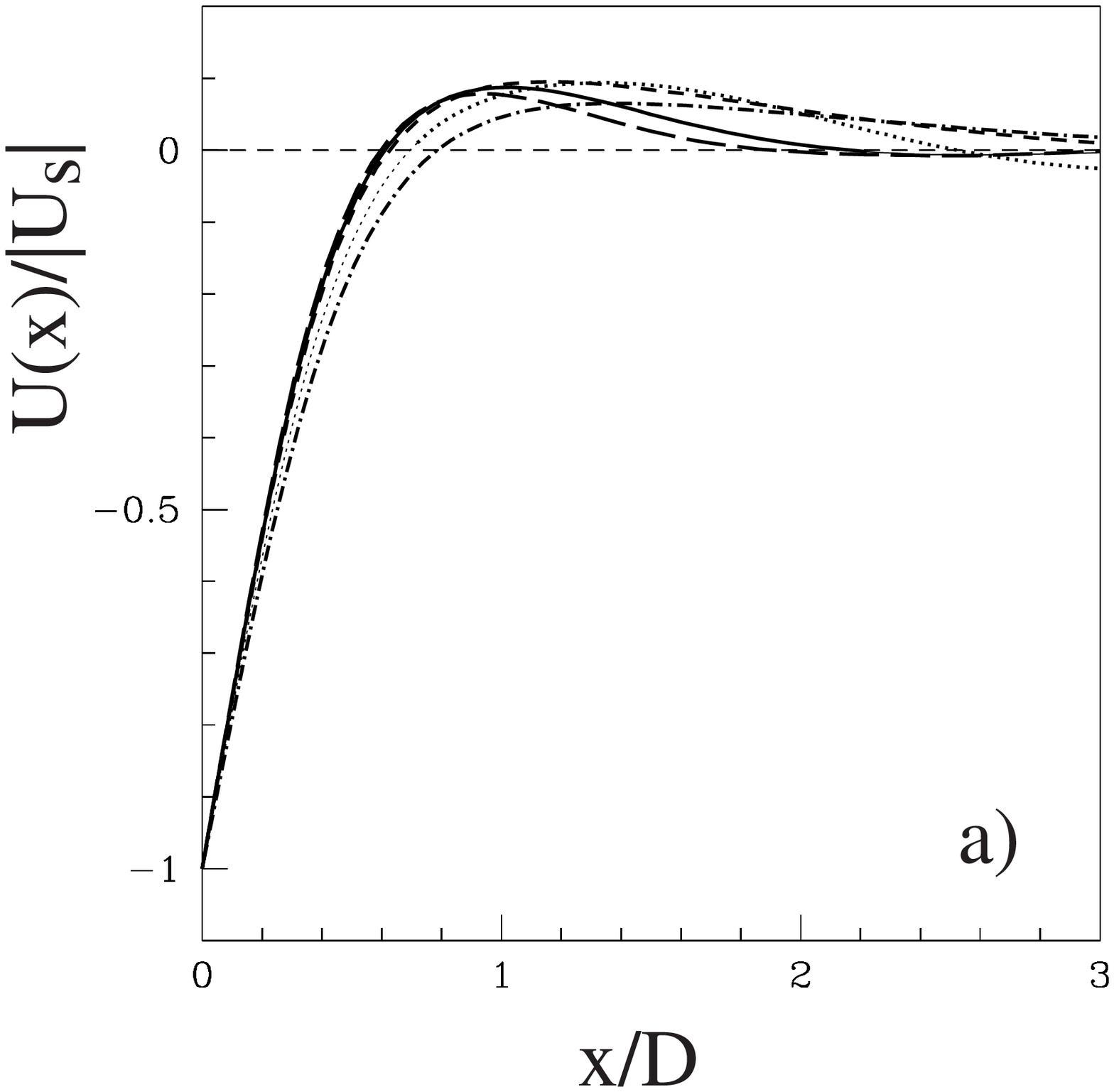} } }
 \epsfxsize=9cm
 \centerline{\vbox{\epsffile{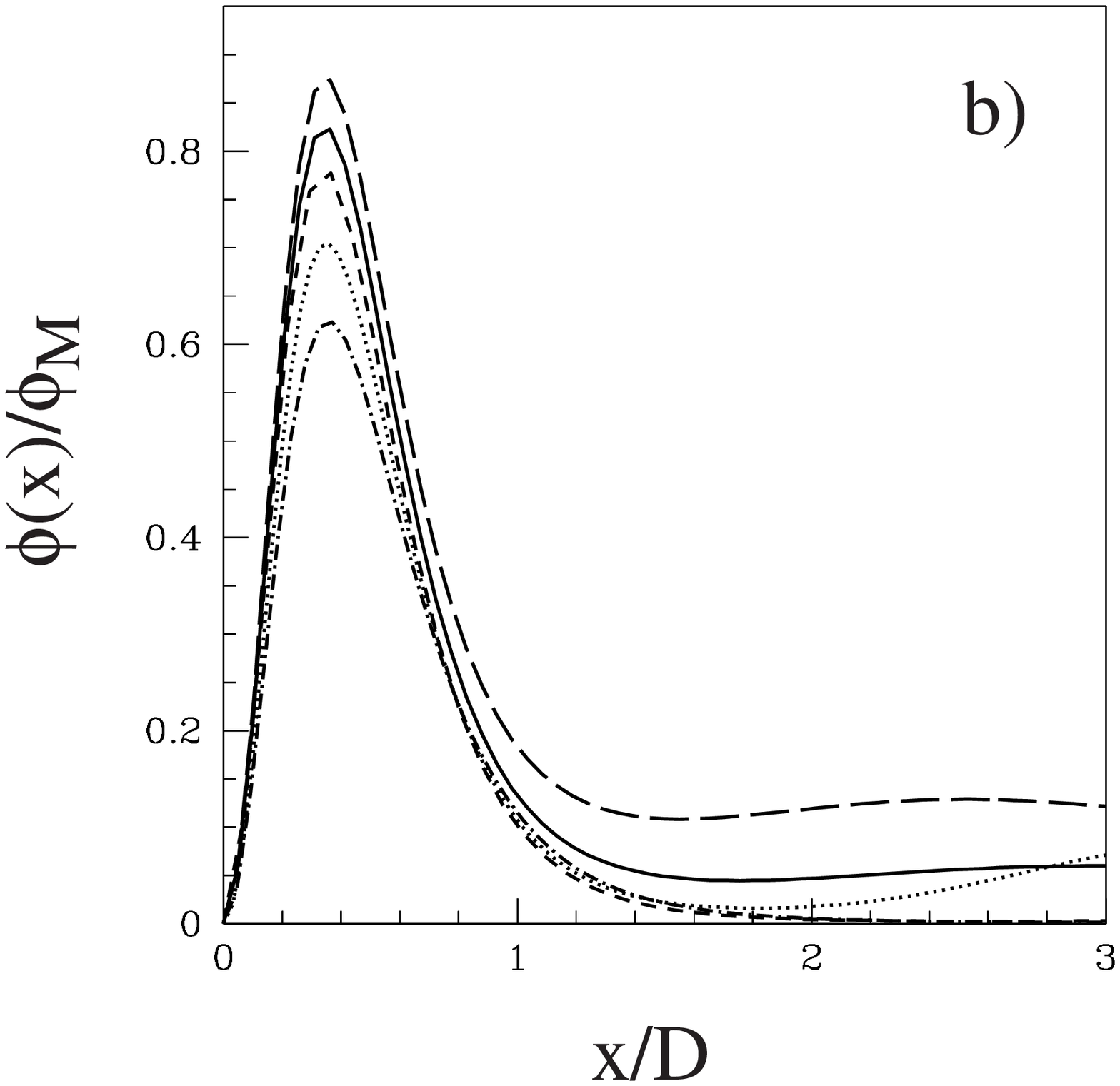} } }
\caption{  \label{fig12}
  Rescaled profiles of PE adsorption in the
low--salt regime confirming the scaling relations,
Eqs.~(\ref{scalingD}), (\ref{scalingcm}). (a) The
rescaled electrostatic potential $U(x)/|U_s|$ as a function of the
rescaled distance $x/D$. (b) The rescaled monomer volume fraction
 $\phi(x)/\phi_M$ as a function of the same rescaled distance.
The profiles are taken from Fig.~\ref{fig11} (with the same
notation). The numerical prefactors of a piecewise linear $h(x/D)$ profile
were used in the calculation of $D$ and $\phi_M$. Adapted from
Ref.~\cite{mm98}.}
\end{figure}

\begin{figure}
 \epsfxsize=17cm
\centerline{\vbox{\epsffile{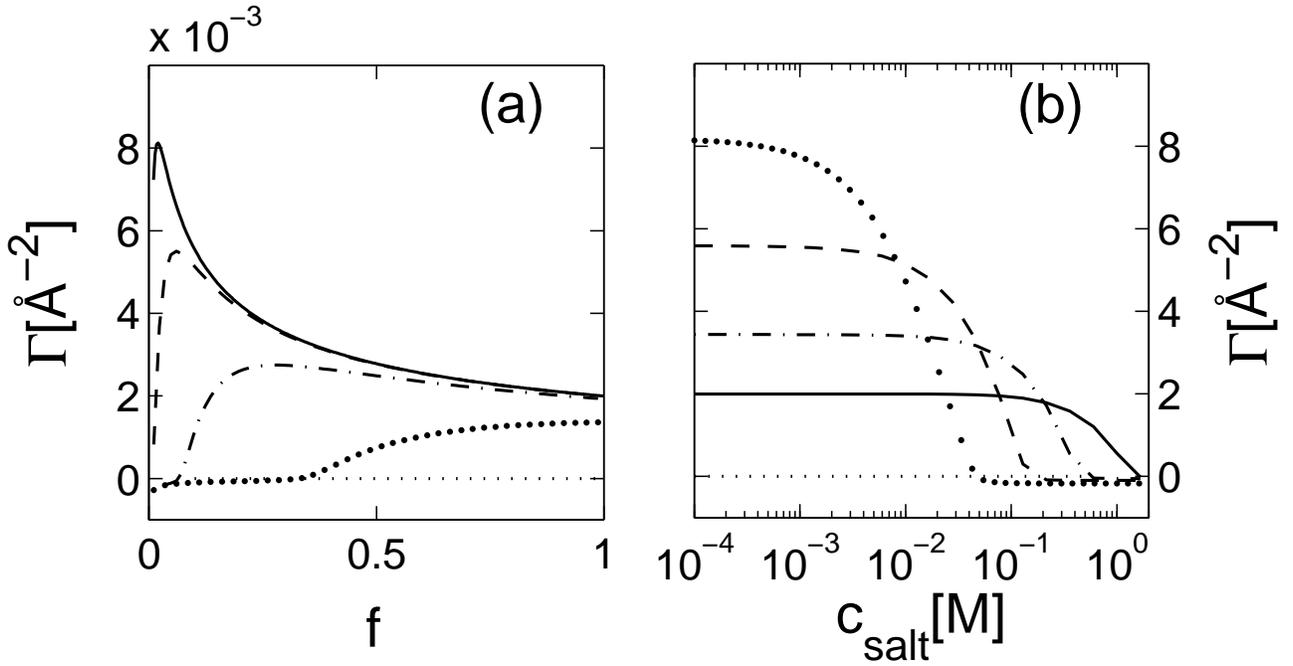} } } \caption{  \label{fig13}
(a) Surface excess of polyelectrolyte adsorption, $\Gamma$, as
function of the chain charged fraction $f$, for constant surface
potential  and for several $c_{\rm salt}$ concentrations: 1.0\,mM
(solid line), 10\,mM (dashed line), 0.1\,M (dash-dot line), 0.5\,M
(dots). As the salt concentration increases, the peak in $\Gamma$
shifts to higher $f$ values and disappears for $c_{\rm
salt}=0.5$\,M. The depletion-adsorption transition is defined to occur for
$\Gamma=0$. (b) Surface excess as function of salt concentration,
$c_{\rm salt}$, for constant surface potential and for several $f$
values: f=0.03 (dots), 0.1 (dashes), 0.3 (dot-dash), 1.0 (solid
line). $\Gamma$ is almost independent of $c_{\rm salt}$ for
low--salt concentrations in the adsorption region. It is then followed
by a strong descent into a depletion region.
The other parameters used here are: $u_{\rm s}=-1.0$,
$c_m^b=\phi_b a^3=10^{-6}$\,\AA$^{-3}$,
$v_2=\tilde{v}_2a^3=50$\,\AA$^3$, $a=5$\,\AA, $T=300$\,K,
$\varepsilon=80$.
Adapted from Ref.~\cite{san03}.}
\end{figure}

\begin{figure}
 \epsfxsize=10cm
\centerline{\vbox{\epsffile{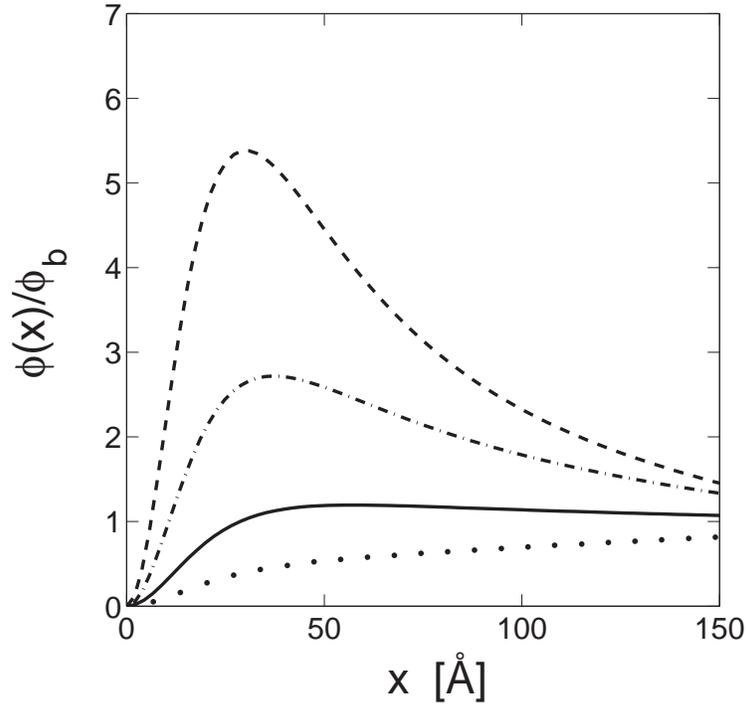} } }
\caption{\protect\label{fig22}
Numerical profiles exhibiting the transition from adsorption to
depletion. The dashed line corresponds to
$f=0.12$, the dot--dash line to $f=0.1$, the solid line to $f=0.09$,
and the dotted line to $f=0.08$.
From the condition $\Gamma=0$ the adsorption--depletion transition
is found to occur for $f\simeq 0.09$, corresponding to
$c_{\rm salt}^*\simeq 0.16{|u_{\rm s}| f}/{l_B a^2}$.
All profiles have $u_{\rm s}=-0.5$,
$c_{\rm salt}=70\,$mM, $c_m^b=\phi_b
a^3=10^{-6}$\,\AA$^{-3}$, $v_2=\tilde{v}_2a^3=50$\,\AA$^3$,\,
$a=5$\,\AA, $T=300$\,K,
$\varepsilon=80$. Adapted from Ref.~\cite{san03}.}
\end{figure}

\begin{figure}
 \epsfxsize=18cm
\centerline{\vbox{\epsffile{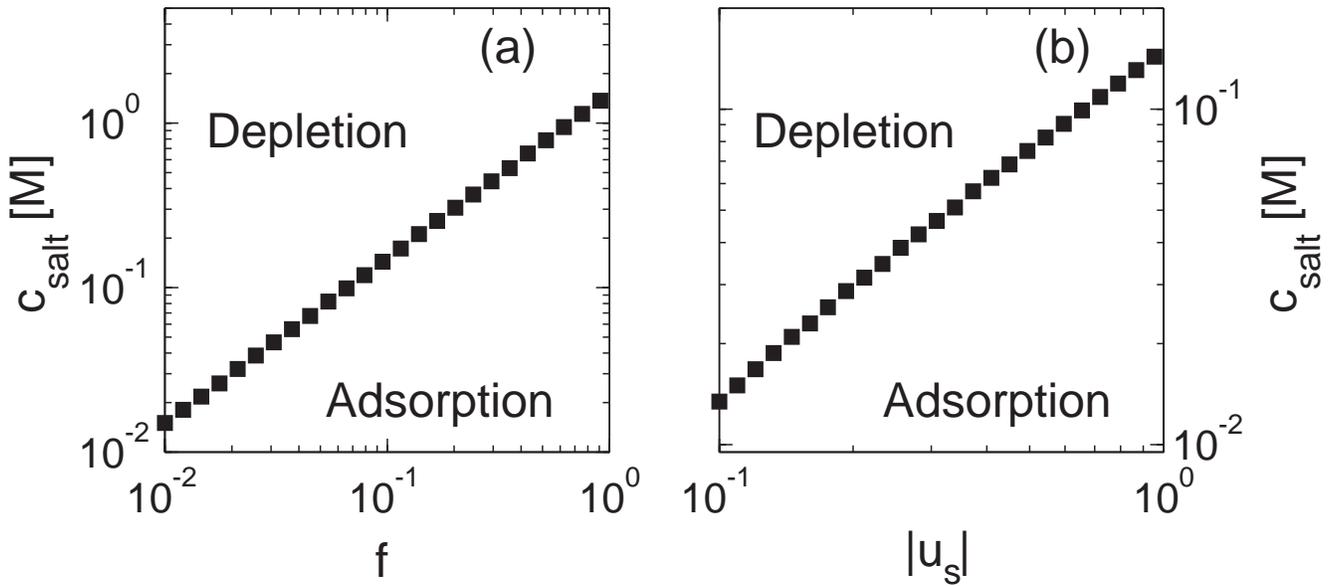} } }
\caption{\protect\label{fig23}
     Numerically calculated crossover diagram from adsorption to depletion
on
         a log-log scale for constant surface potential conditions. In
(a) the
         $(f,c_{\rm salt})$ parameter plane is presented for constant
$u_{\rm s}=-1.0$.
      The  least-mean-square fit
        has
         a slope of $1.00 \pm 0.02$, in excellent agreement with  the
scaling arguments,
     $c_{\rm salt}^*\sim f$. In (b)
         the $(u_{\rm s},c_{\rm salt})$ parameter plane is presented, for
constant $f=0.1$.
          The least-mean-square fit  has a slope of $1.04 \pm 0.02$,
     in good agreement with
     scaling arguments, $c_{\rm salt}^*\sim u_{\rm s}$. All profiles have
     $c_m^b=\phi_b
    a^3=10^{-6}$\,\AA$^{-3}$, $v_2=\tilde{v}_2a^3=50$\,\AA$^3$,\,
    $a=5$\,\AA, $T=300$\,K,
$\varepsilon=80$. Adapted from Ref.~\cite{san03}.}
     \end{figure}

Assuming that there is only one length scale characterizing the
potential behavior in the vicinity of the surface, as demonstrated
in Fig.~\ref{fig12}, the surface potential $U_{\rm s}$ and the
surface charge $\sigma$ are related by $U_{\rm s} \sim \sigma e D/\varepsilon$. In
the low--salt limit we find from  Eq.~(\ref{scalingD})

\begin{equation}
D \sim (f\sigma l_B)^{-1/3} ~ \label{Ds}
\end{equation}
in accord with the single-chain result Eq.~(\ref{delta2}).
Let us define two related concepts via the effective surface
charge density  defined as $\Delta \sigma \equiv f\Gamma - \sigma$,
which is sum of the adsorbed polymer charge density $f \Gamma$ and the
charge density $\sigma$ of the bare substrate. For $\Delta \sigma =0$ the
adsorbed polymer charge exactly {\it compensates} the substrate
charge. If $\Delta \sigma$ is positive the PE {\it
overcompensates} the substrate charge, more polymer adsorbs than
is needed to exactly cancel the substrate charge. If $\Delta
\sigma$ is positive and reaches the value $\Delta \sigma=\sigma$
it means that the PE charge is $f\Gamma=2\sigma$ and leads to an exact
{\it charge inversion} of the substrate charge.
In this case, the effective
surface charge consisting of the substrate charge plus the PE
layer has a charge density which is exactly opposite to the
original substrate charge density $\sigma$.

Do we obtain overcompensation or even charge inversion in the low--salt
limit within mean-field theory? Using scaling arguments this is
not clear since one finds that $\Delta \sigma\sim f\Gamma \sim
\sigma$. Namely each of the two terms in $\Delta \sigma$ scales
linearly with $\sigma$, and the occurrence of overcompensation or
charge inversion will depend on numerical prefactors (which are
difficult to obtain using scaling arguments) determining
the relative sign of the two opposing terms. However, if we look
on the numerical solution for the mean-field electrostatic
potential, Fig.~\ref{fig12}, we see indeed that all plotted profiles have
a maximum of $U(x)$ as function of $x$. An extremum in $U$ means
a zero local electric field. Or equivalently, using Gauss law,
this means that the integrated charge density from the surface to
this special extremum point (including surface charges) is
exactly zero. At this point the charges in solution exactly
compensate the surface charges.
For larger distances from the surface, the adsorption layer overcompensates
the substrate charge.

\subsubsection{High--Salt Limit: $D\ge \kappa^{-1}$ and Depletion}

 When we include salt in
the solution and look at the high--salt limit,
the only length characterizing the exponential
decay of $U$ close to the surface is the Debye-H\"uckel screening
length. Hence, using $dU/d x|_s \sim -\sigma e /\varepsilon$ yields
$U_{\rm s} \sim \sigma e / \kappa \varepsilon$ or $u_{\rm s} \sim \sigma \ell_B /\kappa$.
Inserting this relation into the adsorption threshold for constant
surface potential, Eq. (\ref{depletion_condition_ys}),
we obtain for the crossover between adsorption and
depletion
\begin{equation}
  c_{\rm{salt}}^* \simeq \sigma^{2/3} f^{2/3}l_{\rm B}^{-1/3}a^{-4/3} \sim \sigma^{2/3} f^{2/3},
 \label{depletion_condition_scharge}
\end{equation}
in accord with
Refs. \cite{Wiegel,Muthukumar,Dobrynin} and as confirmed by the numerical studies of
Eqs.~(\ref{PBs})-(\ref{SCFs}) with constant $\sigma$ boundary conditions.
More details can be found in Ref.~\cite{san03}.
We note that the same threshold is obtained by equating the adsorption
layer thickness in the constant-surface-charge ensemble, Eq. (\ref{Ds}),
with the screening length $\kappa^{-1}$.

We end this section with a short comment on the relation
between the semi-dilute and single-chain adsorption
behaviors. By construction of the scaling argument,
the desorption threshold obtained here in the semi-dilute regime
for fixed surface charge, Eq.~(\ref{depletion_condition_scharge}),
is the same as the single-chain desorption
transition, Eq.~(\ref{weak}). It is important to point out that
this equivalence is perfectly confirmed by our numerical solutions
of the full mean-field equations. Therefore, it follows that multi-chain
effects (within mean-field level) do not modify the location of the single-polyelectrolyte chain
adsorption transition.


\section{Lateral Correlation Effects in Polyelectrolyte Adsorption}
\label{latcorr}
\setcounter{equation}{0}

 In this section we go beyond
the mean-field approach by considering lateral correlation
effects (for recent reviews on related subjects
see \cite{Shklovrev,Levinrev}).
The mean-field theories discussed before average the
polymer profile in the lateral direction and only consider a
spatially varying profile in the direction perpendicular to the
substrate. Although mean-field equations can in principle be
formulated which take also lateral order into account, this would
be very involved and complicated. In this section we
generalize the discussion of the single PE chain
adsorption from Sec.~6 and consider the effect of interactions
between different adsorbed polymers on a simple scaling level. In
order to do so, we assume that the adsorption energy is  strong enough
such that the polymers essentially lie flat on the substrate.
Lateral correlations are large enough to locally induce the
polymers to form some type of ordered lattice. Due to the formation
of two-dimensionally ordered adsorbed layers, the local chain structure
becomes important and we therefore describe the PE chains as semi-flexible
polymers in this section.

\begin{figure}
 \epsfxsize=10cm
 \centerline{\vbox{\epsffile{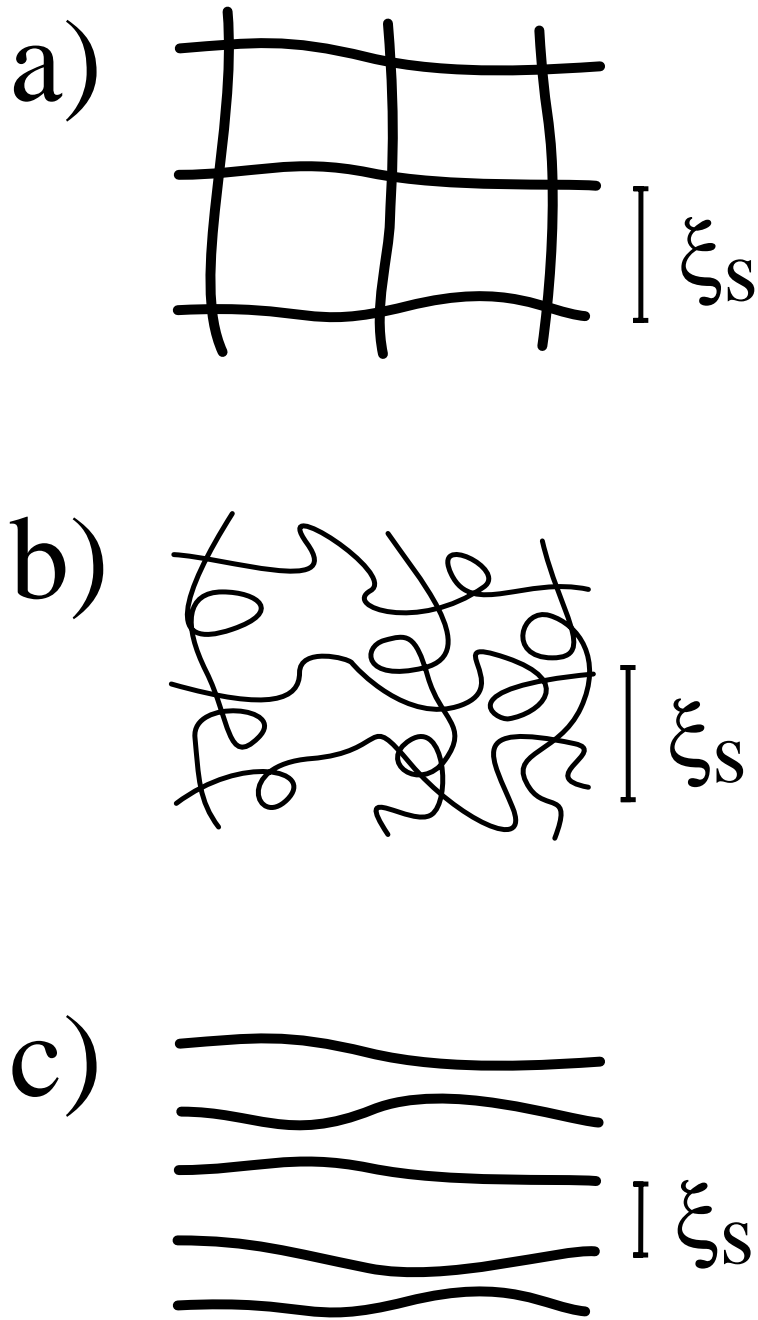} } }
\caption{ \protect \label{figschema}
Schematic top views  of the different adsorbed surface phases considered.
a) Disordered uncrumpled phase, with an average mesh size $\xi_s$ smaller than
the persistence length, exhibiting an average density of chain crossings
of $\sim 1/\xi_s^2$.
b) Disordered crumpled phase, with a mesh size $\xi_s$ larger than the
persistence length.
c) Lamellar phase, with a lamellar spacing $\xi_s$ smaller than the persistence
length.  }
\end{figure}

We follow here the
original ideas of Ref.~\cite{Netz4}, which were subsequently
elaborated by Nguyen et al.\cite{Nguyen}. To understand the idea,
consider Fig.~\ref{figschema}, where schematic top views of
different adsorbed phases are shown. A strongly adsorbed, flat
polymer phase can form a {\em disordered} surface pattern with
many chain crossings, characterized by a certain mesh size $\xi_s$
which corresponds to the average distance between chain
crossings. We distinguish two different cases: if the effective
persistence length $\ell_{\rm eff}$ is larger than the mesh size,
we obtain a disordered {\em uncrumpled} phase, as depicted
schematically in Fig.~\ref{figschema}a; if the effective
persistence length is smaller than the mesh size, we expect a
phase which is {\em crumpled}
 between consecutive chain crossings, as depicted in Fig.~\ref{figschema}b.
 We also anticipate a
{\em lamellar} phase  where different polymer strands are parallel
locally, characterized by an average lamellar spacing $\xi_s$, as
shown  in Fig.~\ref{figschema}c. The lamellar phase is stabilized
either by steric or by electrostatic repulsions between
neighboring polymers; we will in fact encounter both stabilization
mechanisms for different values of the parameters.

We now calculate the free energy and other characteristics of
these adsorbed phases. In all the following calculations, we
assume that we are inside the adsorbed regime of a single
polymer, as discussed in Sec.~6. We basically assume, later on,
that the desorption transitions obtained for the single-chain
case also apply to the case of  many-chain adsorption.
 As was shown in Ref.~\cite{Netz4}, to obtain the complete
phase diagram it is sufficient to consider the lamellar phase
depicted in Fig.~\ref{figschema}c, since the other phase
morphologies are metastable or degenerate. We assume that the
distance between neighboring polymer strands, $\xi_s$,  is much
smaller than the effective persistence length, $\xi_s < \ell_{\rm
eff}$ (this assumption is checked self-consistently at the end).
Since the possible conformations of the adsorbed polymers are
severely restricted in the lateral directions, we have to include,
in addition to the electrostatic interactions, a repulsive free
energy contribution coming from steric interactions between stiff
polymers~\cite{Odijk1}. This is the same type of entropic
repulsion that was used in Sec.~6 to estimate the pressure
inducing desorption from a substrate. The total free energy
density is given by
\begin{equation}
\label{freelam}
F_{\rm lam}  \simeq   -\frac{2 \pi \ell_B \sigma \tau}{\xi_s \kappa}
+ \frac{1}{\ell_{\rm eff}^{1/3} \xi_s^{5/3}} \ln
\left(\frac{\ell_{\rm eff}}{\xi_s} \right) + F_{\rm rep}~,
\end{equation}
where the first term comes from the electrostatic
attraction to the oppositely charged surface (which for consistency
is taken to be penetrable to ions), the second term is the
 Odijk entropic repulsion \cite{Odijk1} and
$F_{\rm rep}$ is the electrostatic repulsion of a lamellar array.

To obtain the electrostatic repulsive energy, we first note that the
reduced potential created by
 a charged line with line charge density $\tau = f/b$ at
a distance $\xi_s$  is within the Debye-H\"uckel approximation
 given by
\begin{equation}
\label{Vline}
V_{\rm line}(\xi_s)  = \tau
\int_{-\infty}^{\infty} {\rm d}s\;
v_{\rm DH}(\sqrt{\xi_s^2+s^2})  =
2 \ell_B \tau K_0[\kappa \xi_s]~,
\end{equation}
with the Debye-H\"uckel potential $v_{\rm DH}$ defined in
Eq.~(\ref{introDH}). $K_0$ denotes the modified Bessel function.
The repulsive electrostatic free energy density of an array of
parallel lines  with a nearest-neighbor distance of $\xi_s$ and
line charge density $\tau$ can thus be written as
\begin{equation}
\label{selfsum}
F_{\rm rep} = \frac{2 \ell_B \tau^2}{\xi_s} \sum_{j=1}^{\infty}
K_0[j \xi_s \kappa]~.
\end{equation}
This expression is also accurate for rods of finite radius $d$ as long as
$d \ll \xi_s$ holds.
In the limit $\xi_s \kappa \ll 1$, when the distance between strands is
much smaller than the screening length, the sum can be transformed
into an integral and we obtain
\begin{equation}
\label{selflam1}
F_{\rm rep} \simeq \frac{2 \ell_B \tau^2}{\xi_s}
\int_0^{\infty} {\rm d}s\; K_0[s \xi_s \kappa] =
\frac{ \pi \ell_B \tau^2}{\xi_s^2 \kappa} ~.
\end{equation}
This expression neglects effects due to the presence of a solid substrate.
For example, and as discussed in Ref.~\cite{Netz4},
for a low-dielectric substrate the electrostatic
interactions are enhanced by a factor of two close to the substrate
surface, a rather small effect which will be neglected
in the following.
Since the average adsorbed surface charge density is given
by $\sigma_{\rm ads} = \tau/\xi_s$, it follows that the self energy
Eq. (\ref{selflam1}) in the limit  $\xi_s \kappa \ll 1$  is
given by $F_{\rm rep}  \simeq  \pi \ell_B \sigma_{\rm ads}^2 \kappa^{-1}$ and
thus is identical to the self energy of a totally
smeared-out charge distribution~\cite{Netz4}.
In this case, lateral correlations therefore do not matter.

In the opposite limit, $\xi_s\kappa \gg 1$, when the polymers are
much farther apart than the screening length, the
sum in Eq. (\ref{selfsum}) is dominated by the first term and (using the
asymptotic expansion of the Bessel function) the free energy density (in units of $k_B T$)
becomes
\begin{equation}
\label{selflam2}
F_{\rm rep} \simeq \frac{ \sqrt{2 \pi} \ell_B \tau^2
{\rm e}^{-\xi_s \kappa}}{\xi_s^{3/2} \kappa^{1/2}} ~.
\end{equation}
In this limit, it is important to note that the smeared-out repulsive
energy Eq. (\ref{selflam1}) is much larger and thus
 considerably overestimates the
actual electrostatic repulsion between polymer strands.
Conversely, this reduction of the electrostatic repulsion between
polymers results in an enormous overcharging of the substrate,
as we will see shortly.

In order to determine the equilibrium distance between the polymer
strands, we balance the electrostatic attraction term, the first
term in Eq.~(\ref{freelam}), with the appropriate repulsion term.
There are three choices to do this. For $d< \kappa^{-1}  < \xi_s^*
<  \xi_s$ (with some crossover length $\xi_s^*$ to be determined
later on), the electrostatic repulsion between the polymers is
irrelevant (i.e. the last term in Eq.~(\ref{freelam}) can be
neglected), and the lamellar phase is {\em sterically} stabilized
in this case. The equilibrium  lamellar spacing  is given by
\begin{equation}
\label{stericB}
\xi_s \sim \left[ \frac{\kappa }
{\tau \sigma \ell_B \ell_{\rm eff}^{1/3}}
\ln \left( \frac{\tau \sigma \ell_B \ell_{\rm eff}}{\kappa}
\right) \right]^{3/2}~.
\end{equation}
In all what follows, we neglect the logarithmic cofactor.

For $ d< \kappa^{-1} < \xi_s < \xi_s^*  $, the steric repulsion
between the polymers is irrelevant (i.e. the second term in
Eq.~(\ref{freelam}) can be neglected). The free energy is
minimized by balancing the electrostatic adsorption term , the
first term in Eq.~(\ref{freelam}), with the electrostatic
repulsion term appropriate for the case $\xi_s \kappa > 1$, Eq.
(\ref{selflam2}), which leads to the {\em electrostatically }
stabilized lamellar spacing
\begin{equation}
\label{electB}
\xi_s \sim \kappa^{-1} \ln
\left[ \frac{\tau \kappa}{\sigma} \right] .
\end{equation}
The adsorbed charge density then follows from
$\sigma_{\rm ads}  \sim \tau /\xi_s$  as
\begin{equation} \label{chargereverse}
\sigma_{\rm ads}  \sim \sigma \frac{ \tau \kappa \sigma^{-1}  }{
\ln(\tau \kappa \sigma^{-1})} ~
\end{equation}
(note that in the previous section the adsorbed charge density was
obtained as the product of the surface amount $\Gamma$ and the
charged-monomer fraction $f$, $\sigma_{\rm ads} = f \Gamma$).
Therefore, the electrostatically stabilized lamellar phase shows
charge reversal as long as the spacing $\xi_s$ is larger than the
screening length. As we will see, this is always the case. The
crossover between the sterically stabilized lamellar phase,
described by Eq.~(\ref{stericB}), and the lamellar phase which  is
stabilized by electrostatic repulsion, Eq. (\ref{electB}), occurs
when the predictions for $\xi_s$ become simultaneously equal to
the crossover spacing $\xi_s^*$, leading to a crossover for a
surface charge density of (without logarithmic cofactors)
\begin{equation}
\label{st/el}
\sigma \sim
\frac{\kappa^{5/3} }{\tau \ell_{\rm eff}^{1/3}\ell_B } ~.
\end{equation}
For $\sigma $ larger than the crossover value in Eq.~(\ref{st/el})
the distance between neighboring polymer strands is smaller than
$\xi_s^*$ and the electrostatic stabilization mechanism is at
work, for  $\sigma$ smaller than the crossover value in
Eq.~(\ref{st/el}) the lamellar spacing $\xi_s$ is larger than the
characteristic crossover length $\xi_s^*$ and the Odijk repulsion
dominates. One notes that the  transition Eq.~(\ref{st/el}) is, on
the scaling level, the same as the adsorption threshold in Eq.
(\ref{strong}) and it is therefore not clear a priori whether the
sterically stabilized lamellar phase exits. However, we note that
additional non-electrostatic adsorption forces will stabilize the
sterically stabilized lamellar phase which should therefore occur
in a finite range of parameters\cite{Netz4}. The electrostatically
stabilized lamellar phase crosses over to the {\em
charge-compensated} phase when $\xi_s$ as given by Eq.
(\ref{electB}) becomes of the order of the screening length
$\kappa^{-1}$. In the charge-compensated phase, the lamellar
spacing is obtained by balancing the electrostatic adsorption
energy with the repulsion in the smeared-out limit
Eq.~(\ref{selflam1}) and  is given by
\begin{equation}
\label{compB}
\xi_s \simeq \frac{\tau}{\sigma} ~.
\end{equation}
In this case  the adsorbed surface charge density $\sigma_{\rm ads} = \tau/\xi_s$
exactly neutralizes the substrate charge density,
\begin{equation}
\sigma_{\rm ads} \sim \sigma ~.
\end{equation}
The crossover between the  charged-reversed phase and charge-compensated phase
is obtained by matching Eqs. (\ref{electB}) and (\ref{compB}), leading to
a threshold surface charge density of
\begin{equation}
\label{comp/el}
\sigma \sim
\tau \kappa ~.
\end{equation}

Finally, taking into account that the polymers have some width $d$,
there is an upper limit for
the amount of polymer that can be adsorbed in a single layer. Clearly, the
lateral distance between polymers in the {\em full} phase is given by
\begin{equation}
\label{fullB}
\xi_s \simeq d
\end{equation}
and thus the adsorbed surface charge density in the full phase reads
\begin{equation}
\sigma_{\rm ads} = \frac{\tau}{d}.
\end{equation}
The crossover between the full phase and the compensated phase is obtained by
comparing Eqs. (\ref{compB}) and (\ref{fullB}), leading to
\begin{equation}
\label{comp/full}
\sigma \sim
\tau /d .
\end{equation}

\begin{figure}
 \epsfxsize=14cm
 \centerline{\vbox{\epsffile{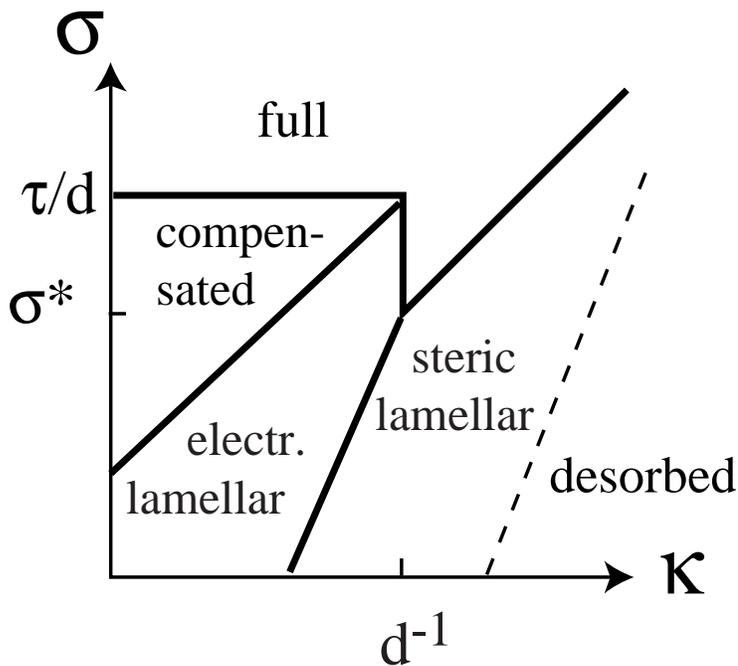} } }
\caption{ \label{figphasediag} Complete adsorption phase diagram
as a function of the substrate charge density $\sigma$ and the
inverse screening length $\kappa$. Note that we use logarithmic
scales on both axes. We find a desorbed regime, an
adsorbed lamellar phase stabilized by electrostatic repulsions
(which is strongly overcharged) and a lamellar phase
which is stabilized by steric repulsion between polymer strands,
an adsorbed charge-compensated phase,
and a full phase, where the substrate charge cannot be
compensated with  a single  adsorption layer because the layer is
close-packed.  }
\end{figure}

In Fig.~\ref{figphasediag} we show the adsorption diagram, for
strongly charged polymers, defined by $\tau \sqrt{\ell_B \ell_0}
> 1$, as a function of the substrate charge density $\sigma $ and
the inverse screening length $\kappa $. The electrostatically
stabilized lamellar phase shows strong charge
reversal as described by Eq. (\ref{chargereverse}). At slightly
larger surface charge densities we predict a charge-compensated phase
 which is not full (i.e. $\xi_s < d$) for a  range of surface
 charge densities  as determined
by Eqs. (\ref{comp/el}) and (\ref{comp/full}). At even larger
substrate charge density, the adsorbed polymer phase becomes close
packed, i.e. $\xi_s = d$. We note that since the full phase is not
charge reversed, the full phase can consist of a second adsorbed
layer (or even more layers). It should however be clear that close
to charge compensation the effective substrate charge density an
additional layer feels is so small that the condition for
adsorption is not met. At low substrate charge densities the
distance between adsorbed polymer strands becomes so large that
the entropic repulsion between polymers dominates the
electrostatic  repulsion, and finally, at even lower charge
densities, the polymers desorb. One notes that the transition
between the electrostatically and sterically stabilized adsorbed
phases, Eq.~(\ref{st/el}), has the same scaling form (disregarding
logarithmic factors) as the desorption transition of semi-flexible
polymers, Eq.~(\ref{strong}). We have shifted the desorption
transition to the right, though, because typically there are
attractive non-electrostatic interactions as well, which tend to
stabilize adsorbed phases. This is also motivated by the fact that
 the sterically stabilized phase
has been seen in experiments on DNA adsorption, as will be discussed below.
The critical charge density $\sigma^*$ where the full phase, the
electrostatically and the sterically stabilized phases meet at one point,
is given by $\sigma^* = 1/(d^{5/3} \ell_{\rm eff}^{1/3} \tau \ell_B)$.
In the phase diagram we have assumed that the charge density threshold for
 the full phase, $\sigma \sim \tau/d $,
 satisfies the inequality
$\tau/d > \sigma ^*$, which for a fully
charged PE at the Manning threshold, $\tau = 1/\ell_B$,
amounts to the condition $\ell_{\rm eff} > \ell_B^3 / d^2$,
which is true for a large class of PE's.

The most important result of our discussion
is that in the electrostatically stabilized
phase the substrate charge is strongly reversed by the adsorbed polymer layer.
This can give rise to a charge-oscillating multilayer formation if the adsorption
of oppositely charged polymer is done in a second step.
The general trend that emerges is that charge reversal is more
likely to occur for intermediate salt concentrations and rather
low substrate charge density. For too high--salt concentration
and too low substrate charge density, on the other hand, the
polymer does not adsorb at all.
In essence, the salt concentration and the substrate charge
density have to be tuned to intermediate values in order
to create charge multilayers.

In experiments on DNA  adsorbed on oppositely charged substrates
one typically observes a lamellar phase \cite{Fang,Raedler}. In
one experiment, the spacing between DNA strands was found to
increase with increasing salt concentration~\cite{Fang}. One
theoretical explanation invokes an effective interaction between
neighboring DNA strands mediated by elastic deformations of the
membrane, which forms the substrate in these experiments
\cite{Dan}. In the sterically stabilized regime, the distance
between adsorbed polymers increases  as $\xi_s \sim \kappa^{3/2}$
with the salt concentration, see Eq.~(\ref{stericB}), which offers
an alternative explanation for the experimental findings. It would
be interesting to redo DNA adsorption experiments on rigid
substrates, where the elastic coupling to the membrane is absent.
For high enough substrate charge densities and by varying the salt
concentration one should be able to see the crossover from the
electrostatically stabilized phase, Eq.~(\ref{electB}), where the
DNA spacing decreases with added salt, to the sterically
stabilized phase, Eq.~(\ref{stericB}), where the DNA spacing
increases with added salt.

Between the two limiting cases, diffusive mean-field adsorption profile
with no lateral correlations (as treated in Sec. 8),
and a flat, two-dimensional adsorption layer with short-ranged
lateral correlations (as discussed in this section),
there clearly exists a continuous crossover.

\section{Interaction between Two Adsorbed Layers}
\setcounter{equation}{0}

One of the many applications of polymers lies in their influence
on the interaction between colloidal particles suspended in a
solvent~\cite{napper}. Depending on the details of substrate-polymer
interactions and properties of polymers in solution, the effective
interaction between colloids in a polymer solution can be attractive
or repulsive, explaining why polymers are widely used as flocculants
and stabilizers in industrial processes~\cite{napper}.
The various regimes and effects obtained for the interaction
of polymer solutions between two surfaces have recently
been reviewed~\cite{kleinreview}. It transpires that force-microscope
experiments done on adsorbed polymer layers  form an ideal tool
for investigating the basic mechanisms of polymer adsorption,
colloidal stabilization and flocculation.

\subsection{Non Adsorbing Polymers}

Let us first discuss briefly the relatively simple case when the
polymers do not adsorb on the surface of the colloidal particles
but are repelled from it. For low concentration of polymer, i.e.
below the overlap concentration $c_m^*$, the depletion of
polymer around the colloidal particles induces a strong  attraction
between the colloidal particles. The range of this
attraction is about the same as the radius of an isolated polymer
and can lead to polymer-induced
flocculation~\cite{Asakura,Rudhardt}. The effects of polymer
excluded volume can be taken into account in analytical
theories~\cite{Hanke,Eisenriegler00}, while Monte-Carlo
simulations in the grand-canonical ensemble confirm the existence
and characteristics of these depletion-induced attractive
forces~\cite{Broukhno}. At polymer concentration higher than the
overlap concentration, the depletion zones around the particles
become of the order of the mesh-size in the solution. The
attraction in this case is predicted to set in at separations equal
to or smaller than the mesh-size~\cite{jldg79}.
The force apparatus was used to measure the interaction
between depletion layers~\cite{lk85}, as realized with polystyrene
in toluene, which is a good solvent for  polystyrene but does not
favor the adsorption of polystyrene on mica surfaces.
Surprisingly, the resultant
depletion force is too weak to be detected.

\subsection{Adsorbing Neutral Polymers}

{\it (i) Equilibrium Adsorption in Good Solvents:~~~}
The case when polymers adsorb on the colloidal surface is much
more complicated, and many cases have to be distinguished. If the
polymer concentration is rather high and under good-solvent
conditions, polymers show the experimentally well-known tendency
to stabilize colloids against flocculation, \ie, to
induce an effective repulsion between the colloidal particles and to
hinder them from coming close enough to each other so
that van-der-Waals
attractions will induce flocculation~\cite{napper}.  We should also
mention that in other applications, small
polymer concentrations and high-molecular weight
polymers are used in the opposite sense as flocculants, to induce
binding between unwanted sub-micron particles and, thereby,
removing them from solution. It follows that adsorbing
polymers can have different effects on the stability of colloidal
particles, depending on the detailed parameters.

Hereafter, we assume that the polymers
form an adsorbed layer around the colloidal particles, with a
typical thickness much smaller than the particle radius, and
curvature effects can be neglected. In that case, the effective
interaction between the colloidal particles with adsorbed polymer
layers can be traced back to the interaction energy between two
planar substrates covered with polymer adsorption layers.
In the case
when  the approach of the two particles is slow and the adsorbed
polymer chains are in {\em full equilibrium} with the chains in  solution,
the interaction between two opposing adsorbed layers is predominantly
attractive~\cite{dg82,scheut1}, mainly
because polymers form bridges between the two surfaces.
Recently, it has been shown that there is
a small repulsive component
to the interaction at large separations~\cite{avalos,semenov2}.
For the case of diblock copolymers, the force
between two surfaces depends in a subtle way on
the relative affinities of the blocks to the surfaces~\cite{Ennis}.

The typical equilibration times of polymers are extremely long.
This holds in particular for adsorption and desorption processes,
and is due to the slow diffusion of polymers and their rather high
adsorption energies. Note that the adsorption energy of a polymer
can be much higher than $k_B T$ even if  the adsorption energy of a
single monomer is small because many monomers of
a single chain can be attached to the surface. Therefore, for the typical
time scales of colloid contacts, the adsorbed polymers are not in
equilibrium with the polymer solution.

{\it (ii) Constrained Equilibrium:~~~}
This is also the case for most
of the experiments done with a surface-force apparatus, where two
polymer layers adsorbed on crossed mica cylinders are brought in
contact. In all these cases one has a {\em constrained
equilibrium} situation, where the polymer configurations and thus
the density profile can adjust only with the  constraint that the
total adsorbed polymer excess stays constant. This case has been
first considered by de Gennes~\cite{dg82} who found that two
fully saturated adsorbed layers will strongly repel each other if
the total adsorbed amount of polymer is not allowed to decrease.
The repulsion is mostly due to osmotic pressure and originates
from the steric interaction between the two opposing adsorption
layers. It was experimentally verified in a series of
force-microscope experiments on polyethylene-oxide layers in
water (which is a good solvent for PEO)~\cite{klein82}.

{\it (iii) Undersaturated Layers:~~~}
In other experiments, the formation of the adsorption  layer is
stopped before the layer is fully saturated. The resulting
adsorption layer is called {\em undersaturated}. If two of those
undersaturated adsorption layers approach each other, a strong
attraction develops, which only at smaller separation changes to
an osmotic repulsion~\cite{klein84}. The theory developed for such
non-equilibrium conditions predicts that any surface excess lower
than  the one corresponding to full equilibrium will lead to
attraction at large separations~\cite{rossi,mendez}. Similar mechanisms
are valid for colloidal suspensions, if the total surface
available for polymer adsorption is large compared to the total
amount of polymer in solution. In this case, the adsorption layers
are  undersaturated, and the resulting attraction is utilized
in applications of polymers as flocculation
agents~\cite{napper}.

{\it (iv) Bad Solvent Conditions:~~~}
Another distinct mechanism  leading to attractive forces between
adsorption layers was investigated in experiments with
dilute polymer solutions  in bad solvents. As an example we mention
polystyrene in
cyclohexane below the theta temperature~\cite{klein80}.
The subsequently developed theory~\cite{kleinpincus}
showed that the adsorption layers attract each other
since the local concentration in the outer part of the
adsorption layers is enhanced
over the dilute solution and lies in the unstable
two-phase region of the bulk phase diagram.
Similar experiments have been repeated at the theta temperature~\cite{ikp90}.

{\it (v) Dynamic Effects:~~~}
Additional effects that have been considered are the dynamical
approach between two surfaces bearing adsorbed polymer layers,
which is controlled by the flow of solvent through the polymer
network affixed to the surfaces~\cite{Fredrickson}.

\subsection{Adsorbing Charged Polymers}

More complicated effects are obtained for the interaction between
two charged surfaces in the presence of oppositely charged
PE's. Experimentally, this situation is encountered
when one tries to flocculate or stabilize charge-stabilized
dispersions by the addition of oppositely charged
PE's~\cite{napper}. In the absence of added
PE's, two similarly charged surfaces repel each other
over a range of the order of the screening length in the case of
added salt. This can be calculated on the mean-field level \cite{jpc99} and
agrees quantitatively with Monte-Carlo simulations and
experimental results for monovalent salt~\cite{Guldbrand,Moreira}.
For divalent or trivalent salt mean-field theory becomes
inaccurate and attractive forces are generated by ion-ion
correlations~\cite{Guldbrand,Moreira}. Attractive forces
between the surfaces can result, at some separation range,
even on a mean field level \cite{jpc99}
from a combination of electrostatic
interactions between all charged species and the adsorption energies of
PE chains on the surfaces.

In
simulations~\cite{Akesson,Granfeldt} and mean-field
theories~\cite{jpc99,Akesson,Bohmer,Podgornik} it has been found
that the predominant effect of added PE's is an
attraction between the surfaces, due to bridging between the surfaces
and screening of the surface repulsion. Like in the case of neutral
polymers between adsorbing surfaces, the force between the surfaces
depends on the adsorbed amount. Salt can be used to control
the amount of adsorbed polymers, and it also has an important effect
on the net force. Since the adsorbed amount for highly charged
PE's increases with added salt, the force becomes
less attractive in this case and, for large salt concentration,
is purely repulsive. For small salt concentrations, on the other
hand, the attraction is strongest. Clearly, in the case of
constrained equilibrium, i.e. when the amount of adsorbed polymer
is fixed as the plate separation changes, the force acquires an
additional repulsive component as the plates approach each other,
due to the force needed to compress the polymer layer. For larger
separations and for undersaturated polymer layers, on the other
hand, the forces are attractive. The precise crossover between
attraction due to undersaturation (at large separation) to
repulsion due to oversaturation (at small separations) depends on
the adsorbed amount. This can be experimentally controlled for
example by the total amount of added PE.

Measurements of the disjoining pressure in thin liquid films of
PE solutions as a function of film thickness
demonstrated an oscillatory
pressure~\cite{KlitzingPSS,oscill,oscill2,oscill3} with a period
of the oscillation of the order of the peak position in the bulk
structure factor (which was discussed in Sec.~3.6).
Theoretically, those oscillations have been seen in mean-field
calculations~\cite{Chatellier} as well as more elaborate
integral-equation calculations~\cite{Yethiraj}.

An effect which is missed by mean-field theories is
the so-called mosaic-binding of charged surfaces in the presence
of a very low concentration of oppositely charged
PE~\cite{napper}. In this case the adsorbed layers
of the separate surfaces are very undersaturated.
Individual polymer coils form isolated patches on the substrate,
where the local surface charge is reversed. The substrate shows a mosaic pattern
of oscillating charge patches. If two of those patterned surfaces
approach each other, the patterns will readjust in order to match
oppositely charged patches, resulting in a very strong,
irreversible binding~\cite{napper}.

\section{Polymer Adsorption on Heterogeneous Surfaces}
\setcounter{equation}{0}

Polymer adsorption can be coupled in a subtle way with lateral
changes in the chemical composition or density of the surface.
Such a surface undergoing  lateral rearrangements at
thermodynamical equilibrium is called an {\em annealed} surface
\cite{dg90,aj91}. A Langmuir monolayer of insoluble surfactant
monolayers at the air/water interface is an example of such an
annealed surface. As function of the temperature change, a Langmuir
monolayer can undergo a phase transition from a high-temperature
homogeneous state to a low-temperature demixed state, where domains
of dilute
and dense regions coexist. Alternatively, the transition from a
dilute phase to a dense one may be induced by compressing the
monolayer at constant temperature, in which case the adsorbed
polymer layer contributes to the pressure~\cite{aazpr94}. The
domain boundary between the dilute and dense phases can act as
nucleation site for adsorption of bulky molecules~\cite{nao96}.

The case where the insoluble surfactant monolayer interacts with a
semi-dilute polymer solution solubilized in the water subphase was
considered in some detail. The phase diagrams of the mixed
surfactant/polymer system were investigated within the framework
of mean--field theory~\cite{ca95}. The polymer enhances the
fluctuations of the monolayer and induces an upward shift of the
critical temperature. The critical concentration is increased if
the monomers are more attracted (or at least less repelled) by the
surfactant molecules than by the bare water/air interface. In the
case where the monomers are repelled by the bare interface but
attracted by the surfactant molecules (or vice versa), the phase
diagram may have a triple point. The location of the polymer
desorption transition line (\ie, where the substrate-polymer
interaction  changes from being repulsive to being attractive)
appears to have a big effect on the phase diagram of the
surfactant monolayer \cite{ca95}.

A similar effect is seen with DNA which adsorbs on a mixed lipid
bilayer consisting of cationic and neutral lipid molecules
\cite{Maier}. Experimentally, it is seen that the negatively
charged DNA  attracts the positively charged lipid molecules and
leads to a local demixing of the membrane \cite{Maier}.
Theoretically, this can be studied by formulating the
Poisson-Boltzmann theory for a single charged cylinder (which
models the rigid DNA molecule)  at some distance from a
surface with mobile charged lipids of a given density and size
\cite{Fleck}. For low--salt concentrations, the charged DNA leads
to a strong accumulation of cationic lipids in its vicinity.
Depending on the size of the lipid heads, this lipid
concentration profile can extend far away from the cylinder. For
high--salt concentrations on the other hand, this accumulation
effect is much weaker due to screening. Similar effects have been
studied for periodic arrays of adsorbed DNA cylinders
\cite{Bruinsma,Wagner} which describe experimental results for
bulk DNA-cationic lipid complexes \cite{Raedler}.

The adsorption of DNA on laterally structured substrates was
recently characterized by direct AFM visualization \cite{Clausen}.
Patches of positively charged lipids were embedded in a matrix of
negative surface potential, and the size of the cationic surface
patches was varied from the micrometer down to the nanometer scale.
DNA adsorption was found to depend both on the average surface
charge density and on the size of positively charged patches.
Similar phenomena were studied theoretically using
off-lattice Monte-Carlo simulations\cite{Ellis,Kong}.

\section{Polymer Adsorption on Curved and Fluctuating Interfaces }
\setcounter{equation}{0}

\subsection{Neutral Polymers}

The adsorption of polymers on rough substrates is of high interest to
applications. One  example is the reinforcement of rubbers by filler
particles such as carbon black or silica particles~\cite{vilgis}.
Theoretical models considered sinusoidal surfaces~\cite{hone},
rough and corrugated
substrates~\cite{blunt,marquesfractal}. In all cases,
enhanced adsorption was found and rationalized in terms of
the excess surface available for adsorption.

The adsorption on macroscopically  curved  bodies, such as spheres
and cylinders, leads to modified adsorption profiles
\cite{curved1}. Of considerable interest is the effective
interaction between two colloidal particles covered by adsorption
layers \cite{colloidads}. Another application is obtained for the
adsorption of polymers on flexible interfaces or membranes
\cite{dg90,flexads1,flexads2}. Here one interesting aspect concerns
the polymer-induced contribution to the elastic bending moduli of
the flexible surface. The elastic energy of a liquid-like
membrane can be expressed in terms of two bending moduli,
$\kappa$ and $\kappa_G$. The elastic energy (per unit area) is

\be
\frac{\kappa}{2} (c_1+c_2 -2c_0)^2 +\kappa_G c_1 c_2~,
\ee

\noindent where $\kappa$ and $\kappa_G$ are the elastic
and Gaussian bending moduli, respectively. The
principle curvatures of the surface are
given by $c_1$ and $c_2$, and $c_0$ is the spontaneous
curvature.  Quite generally, in presence of adsorbing polymers
 $\kappa_G$ turns out
to be positive and thus favors the formation of surfaces with
negative Gaussian curvature. An `egg-carton'
structure is an example to such a
multi-saddle surface. On the other hand, the
effective $\kappa$ is reduced, leading to a more deformable and
flexible surface due to the adsorbed polymer
layer~\cite{dg90,elasticity1,elasticity2}. The spontaneous
curvature $c_0$ is only non-zero if the  adsorption profile is
different on both sides of the membrane \cite{flexads1}. This can
be achieved, for example, by incubating vesicle solutions with
polymers, so that the vesicle interior is devoid of polymers
(neglecting polymer translocation through the membrane which is
indeed a rather slow process).
If the polymers do not adsorb on the membrane,
the spontaneous curvature is such that the membrane bends towards
the polymer solution \cite{Hanke,Eisenriegler00}. If, on the other
hand, the polymers do adsorb on the membrane, the membrane bends
away from the polymer solution with a continuous crossover
between the two cases as the adsorption strength is varied
\cite{elasticity3}.

\subsection{Charged Polymers}

Of particular interest is the adsorption of strongly charged
polymers on oppositely charged cylinders\cite{curved2,Park,Kunze2}
and spheres~\cite{Goeler}-\cite{Netz5}, because these are
geometries encountered in many colloidal science applications
and in bio-cellular processes.
When the curvature of the small
colloidal particles is large enough, it can lead to a much more
pronounced effect for PE adsorption as compared with
neutral polymer. This is mainly due to the fact that the
electrostatic energy of the adsorbed PE layer depends sensitively
on curvature~\cite{linse,mateescu,Netz5,Shklovsphere}.
Bending a charged
polymer around a small sphere costs a large amount of
electrostatic energy, which will disfavor adsorption of long,
strongly charged PE at too low--salt concentration.

\begin{figure}
 \epsfxsize=12cm
 \centerline{\vbox{\epsffile{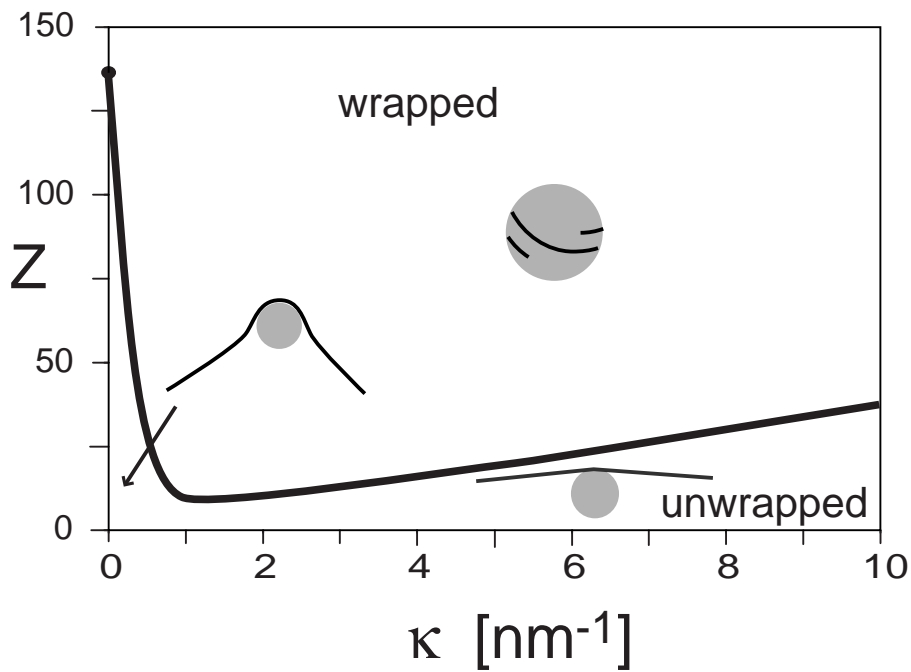} } }
\caption{  \label{fig10} Numerically determined adsorption
diagram for a negatively charged semi-flexible polymer of length
$L = 50$\,nm, linear charge density $\tau = 6$\,nm$^{-1}$,
persistence length $\ell_0 = 30$\,nm, interacting with a
oppositely charged sphere of radius $R_{sp}=5$\,nm. Shown is the main
transition from the unwrapped configuration (at the bottom) to
the wrapped configuration (at the top) as a function of sphere
charge $Z$ and inverse Debye-H\"uckel screening length $\kappa$. Wrapping is
favored at intermediate salt concentrations. The parameters are
chosen for the problem of DNA-histone complexation. Adapted from
Ref.~\cite{Kunze}.}
\end{figure}

In Fig.~\ref{fig10} we show the adsorption phase diagram of a
single stiff PE  of finite length which interacts with an
oppositely charged sphere of charge $Z$ (in units of $e$). The
specific parameters were chosen as appropriate for the
complexation of DNA (a negatively charged, relatively stiff
biopolymer) with positively charged histone proteins,
corresponding to a DNA length of $L = 50$\,nm, a chain persistence
length of $\ell_0 = 30$\,nm, and a sphere radius of
$R_{sp}=5$\,nm. The phase diagram was obtained by minimization of
the total energy including bending energy of the DNA,
electrostatic attraction between the sphere and the DNA, and
electrostatic repulsion between the DNA segments~\cite{Kunze}. All
interactions are represented by screened Debye-H\"uckel potentials
of the form of Eq.~(\ref{introDH}). Fluctuations of the DNA shape
are unimportant for such stiff polymers. Therefore, the
ground-state analysis performed is an acceptable approximation.

We show in Fig.~\ref{fig10} the main transition between an
unwrapped state, at low sphere charge $Z$, and the wrapped state,
at large sphere charge $Z$. It is seen that at values of the
sphere charge between $Z=10$ and $Z=130$ the wrapping only occurs
at intermediate values of the inverse screening length $\kappa\sim
c_\s^{1/2}$. At low--salt concentrations, (lower left corner in
the phase diagram), the self-repulsion between DNA segments
prevents wrapping, while at large salt concentrations, (lower
right corner in the diagram), the electrostatic attraction is not
strong enough to overcome the mechanical bending energy of the
DNA molecule. These results are in good agreement with
experiments on DNA/histone complexes~\cite{Yager}. Interestingly,
the optimal salt concentration, where a minimal sphere charge is
needed to wrap the DNA, occurs at physiological salt
concentration, for $\kappa^{-1} \approx 1$\,nm. For colloidal
particles of larger size and for flexible synthetic polymers,
configurational fluctuations become important. They have been
treated using a mean-field description in terms of the average
monomer density profile around the sphere~\cite{Goeler,sens}.


\section{Grafted Polymer Chains} \label{sectionbrush}
\setcounter{equation}{0}

The discussion so far assumed that all monomers of a polymer are
alike, showing the same tendency to adsorb  to the
substrate surface. For industrial and technological applications,
one is often interested  in {\em end-functionalized polymers}.
These are polymers which attach with one end only  to the
substrate, as is depicted in Fig.~\ref{fig3o}b, while the rest of the
polymer is not particularly attracted to (or even repelled from)
the grafting surface. Hence, it attains a random-coil structure in
the vicinity of the surface. Another possibility of block
copolymer grafting, as shown in Fig.~\ref{fig3o}c,
will be briefly discussed below as well.

The motivation to study such terminally attached polymers lies in
their enhanced power to stabilize particles  and surfaces against
flocculation. Attaching a polymer by its end  to the surface
opens up a much more effective route to stable surfaces. Bridging
and creation of polymer loops on the same surface, as encountered
in the case of homopolymer adsorption (and leading to attraction
between two particle surfaces and destabilization, see
Sec.~10), do not occur if the main-polymer section is chosen such
that it does not adsorb to the surface.

Experimentally, the end-adsorbed polymer layer can be built in
several different ways, depending on the application in mind.
First, one of the polymer ends can be {\em
chemically} bound to the grafting surface, leading to a tight and
irreversible attachment~\cite{auroy1} shown schematically in
Fig.~\ref{fig3o}b. The second possibility consists of  {\em physical}
adsorption of a specialized end-group which favors interaction
with the substrate. For example, polystyrene chains have been used
which contain a zwitterionic end group that adsorbs strongly on
mica sheets~\cite{taunton}.

Physical grafting is also possible with a suitably chosen diblock
copolymer (Fig.~\ref{fig3o}c), \eg, a polystyrene -- polyvinylpiridine
(PS-PVP) diblock in the solvent toluene
at a quartz substrate~\cite{field}. Toluene is a {\em selective
solvent} for this diblock. The PVP (polyvinylpyridine)
block is strongly adsorbed to the quartz substrate and forms a
collapsed anchor, while the PS (polystyrene) block is under
good-solvent conditions. It does not adsorb  to the substrate and
remains solubilized in the solvent. General adsorption scenarios for
diblock copolymers have been theoretically discussed, both for
selective and non-selective solvents, with special
consideration to the case when the asymmetry of the
diblock copolymer, \ie, the length difference between the two
blocks, is large~\cite{marques}.

Another experimental realization  is
possible with  diblock copolymers which are anchored at the
liquid-air~\cite{kent} or at a liquid-liquid interface of two
immiscible liquids~\cite{teppner}. This scenario offers the
advantage that the surface pressure can be directly measured. A
well studied example is that of a diblock copolymer of
polystyrene -- polyethylene oxide (PS-PEO). The PS block is shorter and
functions as an anchor at the air/water interface because it is
immiscible in water. The PEO block is miscible in water but because
of attractive interaction with the air/water interface it forms a
quasi-two dimensional layer at very low surface coverage. As the
surface pressure  increases and the area per polymer decreases, the PEO
block is expelled from the surface and forms a quasi polymer
`brush'.

\begin{figure}
 \epsfxsize=12cm
 \centerline{\vbox{\epsffile{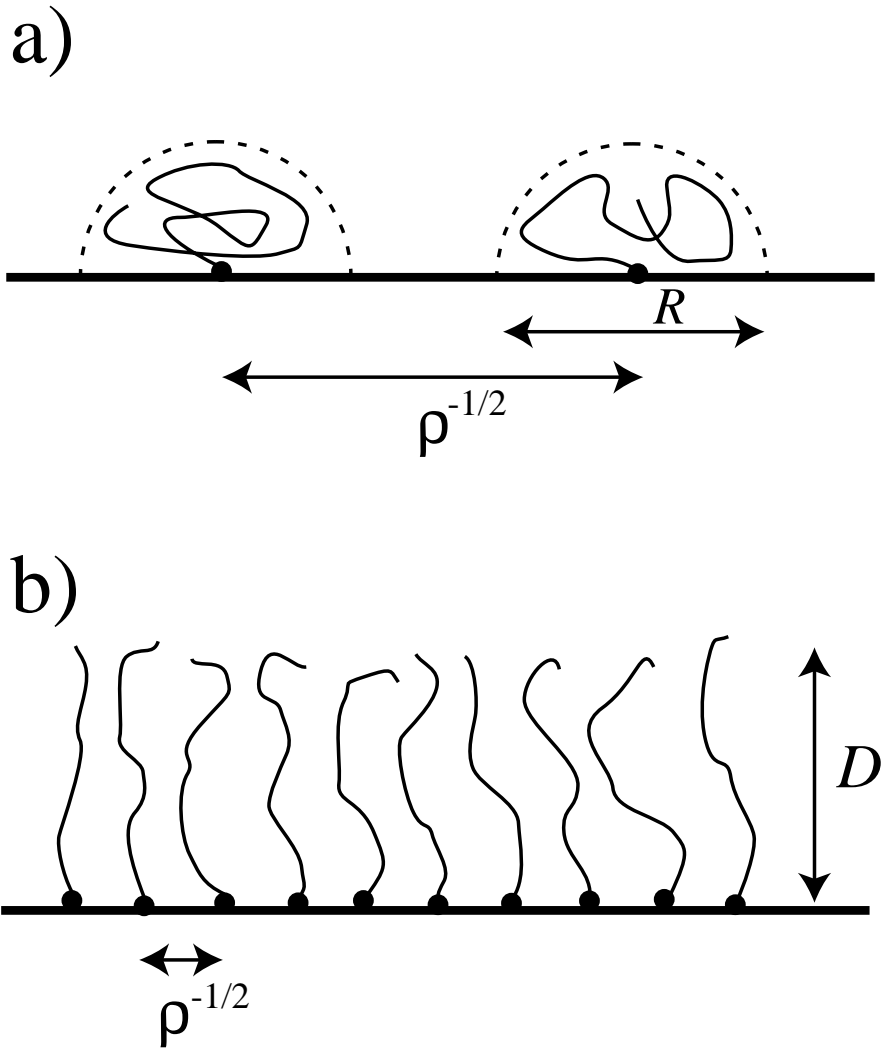} } }
\caption{ \label{fig9o}
For grafted chains, one distinguishes between: a) the mushroom regime,
where the distance between chains, $\rho ^{-1/2}$, is larger
than the size of a polymer coil; and, b) the brush regime, where
the distance between chains is smaller than the unperturbed coil
size. Here, the chains are stretched away from the surface due to
repulsive interactions between monomers. The brush height $D$
scales linearly with the polymerization index, $D \sim N$, and
thus is larger than the unperturbed coil radius $R \sim a N^\nu$.}
\end{figure}

In the following we simplify the discussion by assuming  that the
polymers are irreversibly grafted at one of their chain ends to the substrate.
We limit the discussion to good solvent conditions and
absence of any attractive interactions between the polymer chains and the surface.
The important new system parameter
 is the grafting density (or area per
chain) $\rho$, which is the inverse of the average area
available for each polymer at the surface. For small
grafting densities, $\rho < \rho^*$, the polymer chains will be far
apart from each other and hardly interact, as schematically shown
in Fig.~\ref{fig9o}a. The overlap grafting density for chains
in good solvent conditions (swollen chains) is
$\rho^* \sim a^{-2}N^{-6/5}$, where $N$ is the polymerization
index~\cite{gennes}.

For large grafting densities, $\rho > \rho^*$, the chains
begin to overlap. Since we assume the solvent to be good, monomers repel
each other. The lateral separation between the polymer coils is
fixed by the grafting density, so that the polymers extend away
from the grafting surface in order to avoid each other, as
depicted in Fig.~\ref{fig9o}b. The resulting structure is called a polymer
`brush', with a vertical height $D$ which greatly exceeds the
unperturbed coil radius~\cite{gennes,alex}. Similar stretched
structures occur in many other situations, such as diblock
copolymer melts in the strong segregation regime, or star polymers
under good solvent conditions~\cite{halperin92}. The universal
occurrence of stretched polymer configurations in several seemingly
unconnected situations warrants a  detailed discussion. Below, this discussion
is separated for neutral and charged grafted chains.

\subsection{Neutral Grafted Polymers}

The scaling behavior of the brush height $D$ can be analyzed
using a Flory-like mean--field theory, which is a simplified
version of the original Alexander theory~\cite{alex} for polymer brushes. The
stretching of the chain leads to an entropic free energy loss of
$D^2/(a^2 N)$ per chain, and the repulsive energy density due to
unfavorable monomer-monomer contacts is proportional to the
squared monomer density times the excluded-volume
parameter $v_2$ (introduced in Sec.~2.2). The free energy per
chain (and in units of $k_B T$) is then

\be
\label{flory}
{\cal F} \simeq \frac{D^2}{a^2 N} +
v_2 \left(\frac{\rho N}{D}
\right)^2 \frac{D}{\rho}~.
\ee
The equilibrium height is obtained by minimizing  Eq.~(\ref{flory})
with respect to $D$, and the result is~

\be \label{Floryh} D_{eq} = N \left( 2 v_2 a^2 \rho /3
\right)^{1/3}~ \ee where the numerical constants have been added
for numerical convenience in the following considerations. The
vertical size of the brush scales linearly with the polymerization
index $N$, a clear signature of the strong stretching of the
polymer chains, as was originally obtained by Alexander
\cite{alex}. At the overlap threshold, $\rho^* \sim a^{-2}
N^{-6/5}$, the height scales as $D_{eq} \sim N^{3/5}$, and thus
agrees with the scaling of an unperturbed chain radius in a good
solvent, Eq.~(\ref{swollenflex}), as it should. The simple scaling
calculation predicts the brush height $D$ correctly in the
asymptotic limit of long chains and strong overlap. It has been
confirmed by experiments~\cite{auroy1,taunton,field} and computer
simulations~\cite{cos87,murat}.

The above scaling result assumes that all chains are stretched to
exactly the same height, leading to a step-like shape of the
density profile. Monte-Carlo and numerical mean--field calculations
confirm the general scaling of the brush height, but exhibit a
more rounded monomer density profile which goes continuously to
zero at the outer perimeter~\cite{cos87}.
A big step towards a better understanding of stretched polymer
systems was made by Semenov~\cite{sem}, who recognized the
importance of {\em classical paths} for such systems.

The classical polymer path is defined as the path which minimizes
the free energy, for a given start and end positions, and thus
corresponds to the most likely path a polymer can take. The name
follows from the analogy with quantum mechanics, where the
classical motion of a particle is given by the quantum path with
maximal probability. Since for strongly stretched polymers the
fluctuations around the classical path are weak, it is expected
that a theory that takes into account only classical paths, is a
good approximation in the strong-stretching limit. To quantify the
stretching of the brush, let us introduce the (dimensionless)
stretching parameter $\beta$, defined as

\be
\beta \equiv N\left(\frac{3 v_2^2 \rho^2 }{2 a^2 }\right)^{1/3}
=\frac{3}{2} \left( \frac{D_{eq}}{a N^{1/2}} \right)^2~,
\ee

\noindent
where $D_{eq}$ is the
brush height according to Alexander's theory, compare
Eq.~(\ref{Floryh}). The parameter $\beta$ is proportional to the
square of the ratio of the Alexander prediction for the brush
height, $D_{eq}$, and the unperturbed Gaussian chain radius $R \sim a N^{1/2}$,
and, therefore, is a measure of the stretching of the brush.
Constructing a classical theory in the infinite-stretching limit,
defined as the limit $\beta \rightarrow \infty$, it was shown
independently by Milner et al.~\cite{mil} and Skvortsov et
al. \cite{skvor}
that the resulting monomer volume-fraction  profile
depends only on the vertical distance from the grafting surface
and  has in fact a {\em parabolic} profile. Normalized
to unity, the density profile is given by
\be \label{paraprofile}
\phi(x) =  \left(\frac{3 \pi}{4}\right)^{2/3} -
\left(\frac{\pi x }{2 D_{eq}}\right)^2~.
\ee

\noindent The brush height, \ie, the value of $x$ for which the
monomer density becomes zero, is given by  $x^* = (6/\pi^2)^{1/3}
D_{eq}$ and is thus proportional to the scaling prediction for the
brush height, Eq.~(\ref{Floryh}). The parabolic brush profile has
subsequently been confirmed in computer
simulations~\cite{cos87,murat} and experiments~\cite{auroy1} as
the limiting density profile in the strong-stretching limit, and
constitutes one of the cornerstones in this field. Intimately
connected with the density profile is the distribution of {\em
polymer end points}, which is non-zero everywhere inside the
brush, in contrast with the original scaling description leading
to Eq.~(\ref{Floryh}).

However, deviations from the parabolic profile become
progressively important as the length of the polymers $N$ or the
grafting density $\rho$ decreases. In a systematic derivation of
the mean--field theory for  Gaussian brushes~\cite{netzbrush} it
was shown that the mean--field theory is  characterized  by a
single parameter, namely the stretching parameter $\beta$. In the
limit $\beta \rightarrow \infty$, the difference between the
classical approximation and the mean--field theory vanishes, and
one obtains the parabolic density profile. For finite $\beta$ the
full mean--field theory and the classical approximation lead to
different results and  both show deviations from the parabolic
profile.

\begin{figure}
 \epsfxsize=12cm
 \centerline{\vbox{\epsffile{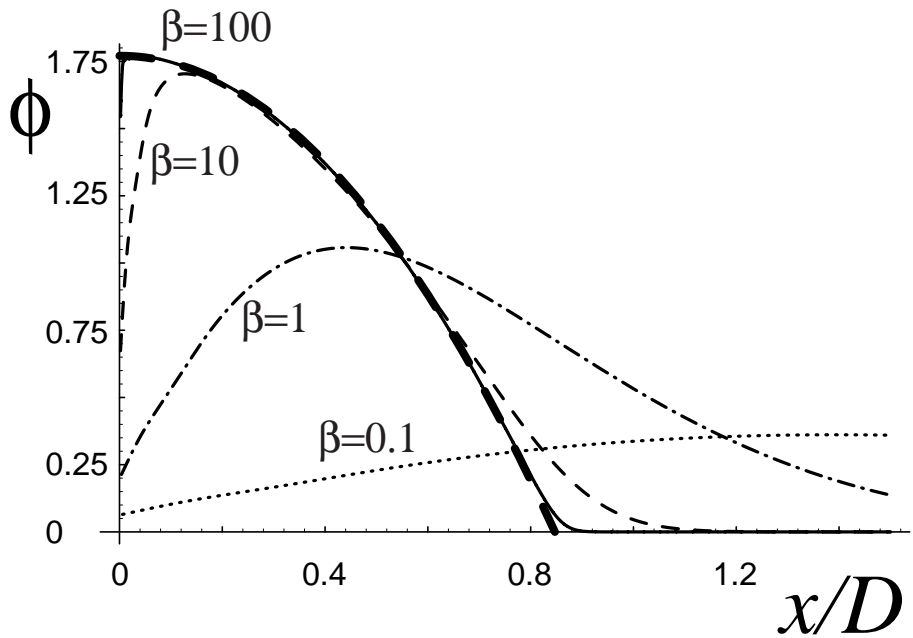} } }
\caption{ \label{fig10o} \protect
Results for the density profile (normalized to unity)
of a strongly compressed brush, as
obtained within  mean--field theory. As the
compression increases, described by the stretching parameter
$\beta$, which varies from 0.1 (dots) to 1 (dash-dots), 10
(dashes), and 100 (solid line), the density profile approaches the
parabolic profile (shown as a thick, dashed line) obtained within
a classical-path analysis (adapted from Ref.~\cite{netzbrush}).}
\end{figure}

In Fig.~\ref{fig10o} we show the density profiles
(normalized to unity)
for four different values
of $\beta$, obtained with the full mean--field
theory~\cite{netzbrush}. The parameter values used are $\beta =
100$ (solid line), $\beta=10$ (thin dashed line), $\beta = 1$ (
dotted-dashed line), and $\beta = 0.1$ (dotted line). For
comparison, we also show the asymptotic result according to
Eq.~(\ref{paraprofile}) as a thick dashed line. In contrast to
earlier numerical implementations~\cite{fleer}, the
self-consistent mean--field equations were solved in the continuum
limit, where the results depend only on the single
parameter  $\beta$ and direct comparison with other continuum
theories becomes possible. Already for $\beta = 100$ the
density profile obtained within mean--field theory is almost
indistinguishable from the parabolic profile denoted by a thick
dashed line.

Experimentally, the highest achievable $\beta$ values are in
the range of $\beta \simeq 20$. Namely, deviations from the
asymptotic parabolic profile are important. For moderately large
values of $\beta >10$, the classical approximation (not shown
here), derived from the mean--field theory by taking into account
only one polymer path per end-point position, is still a good
approximation, as judged by comparing density profiles obtained
from both theories~\cite{netzbrush}, except very close to the
surface. Unlike mean-field theory, the classical theory
misses completely the depletion
effects at the substrate. Depletion effects at the substrate lead to a
pronounced density depression close to the  grafting surface, as
is clearly visible in Fig.~\ref{fig10o}.

A further interesting question concerns the behavior of individual
polymer paths. As was already discussed for the infinite-stretching
theories ($\beta \rightarrow \infty$),  polymers paths
can end at any distance from the surface. Analyzing the polymer paths
which end at a common distance from the surface, two rather
unexpected features are obtained: i) free polymer ends,  in
general, are stretched; and, ii) the end-points lying close to the
substrate are pointing towards the surface
 (such that the polymer path first
turns away from the grafting surface before moving back
towards it). In contrast, end-points lying
 {\it beyond} a certain distance from the
substrate, point away from the surface
(such that the paths move monotonously towards the
surface).
We should point out that these two features have been recently
confirmed in molecular-dynamics simulations~\cite{Seidel}.
They are not an artifact of the continuous self-consistent theory
used in Ref.~\cite{netzbrush} nor are they due to the neglect of
fluctuations. These are interesting results, especially since it
has been long assumed that free polymer ends are
unstretched, based on the assumption that no
forces act on free polymer ends.

Let us now turn to the thermodynamic behavior of a polymer brush.
Using the Alexander description, we can calculate the
free energy per chain
by putting the result for the optimal brush height, Eq.~(\ref{Floryh}),
into the free-energy expression, Eq.~(\ref{flory}):

\be
{\cal F} \sim N \left( v_2 \rho / a \right)^{2/3}~.
\ee

\noindent In the presence of excluded-volume correlations, \ie,
when the chain overlap is rather moderate, the brush height $D$ is
still correctly predicted by the Alexander calculation, but the
prediction for the free energy is in error. Including correlations
\cite{alex}, the free energy is predicted to scale as ${\cal
F} \sim N \rho^{5/6}$. The osmotic surface pressure $\Pi$
is related to the free energy per chain by

\be
\Pi = \rho^2 \frac{\partial {\cal F}}{\partial \rho}~,
\ee
and should thus scale as $\Pi \sim \rho^{5/3}$ in the
absence of correlations, and as $\Pi \sim \rho^{11/6}$
in the presence of correlations. However,
these theoretical predictions do not compare well
with experimental results for the surface
pressure of a compressed brush~\cite{kent}.
At current, there is no explanation for this discrepancy.
An alternative theoretical method to study tethered chains is the so-called
single-chain mean--field method~\cite{carignano}, where the statistical
mechanics of a single chain is treated exactly, and the interactions
with the other chains are  taken into account on a mean-field level.
This method is especially useful for short chains, where
fluctuation effects are important, and for dense systems,
where excluded volume interactions play a role. The calculated
profiles and brush heights agree very well with experiments and
computer simulations. Moreover, these calculations explain the pressure isotherms
measured experimentally~\cite{kent} and in molecular-dynamics
simulations~\cite{grest}.

As we described earlier, the main interest in end-adsorbed or
grafted polymer  layers stems from their ability to stabilize
surfaces against van-der-Waals attraction. The force between
colloids with grafted polymers is repulsive if the polymers do not
adsorb on the grafting substrates~\cite{collgraft}. This is in
accord with our discussion of the interaction between adsorption
layers, where attraction was found to be caused mainly by bridging
and creation of polymer loops, which of course are absent for
non-adsorbing brushes. A stringent test of brush theories was
possible with accurate experimental measurements of the repulsive
interaction between two opposing grafted polymer layers using a
surface force apparatus~\cite{taunton}. The resultant force could
be fitted very nicely by the infinite-stretching theory of Milner
et al.~\cite{milner88b}. It was also shown that polydispersity
effects, as appear in experiments, have to be taken
into account theoretically in order to obtain a good fit of the
data~\cite{milner89}.

\subsection{Solvent and  Substrate Effects on
Polymer Grafting}

So far we assumed that the polymer grafted layer is in contact
with a good solvent. In this case, the grafted polymers try to
minimize their mutual contacts by stretching out into the solvent. If the
solvent is bad, the monomers try to avoid the solvent by forming a
collapsed brush, the height of which is considerably reduced with
respect to the good-solvent case. It turns out that the collapse
transition, which leads to phase separation in the bulk, is
smeared out for the grafted layer and does not correspond to a
true phase transition~\cite{halperin88}. The height of the
collapsed layer scales linearly in $\rho N$, which reflects the
constant density within the brush, in agreement with
experiments~\cite{auroy2}. Some interesting effects have been
described theoretically~\cite{marko93} and
experimentally~\cite{auroy2} for brushes in mixtures of good and
bad solvent, which can be rationalized in terms of a partial
solvent demixing.

For a theta solvent ($v_2=0$) the relevant interaction is
described by the third-virial coefficient; using a simple
Alexander approach similar to the one leading to
Eq.~(\ref{Floryh}), the brush height is predicted to vary with the
grafting density as $ D \sim \rho^{1/2}$, in agreement with
computer simulations~\cite{theta}.

Up to now we discussed planar grafting layers. It is of much
interest to consider the case where
polymers are grafted to {\it curved} surfaces. The first study taking
into account curvature effects of stretched and tethered polymers
was done in the context of star polymers~\cite{daoudcotton}. It
was found that chain tethering in the spherical geometry leads to
a universal density profile, showing a densely packed core, an
intermediate region where correlation effects are negligible and
the density decays as $\phi(r) \sim 1/r$, and an outside region
where correlations are important and the density decays as $\phi
\sim r^{-4/3}$. These considerations were  extended using  the
infinite-stretching theory of Milner et al.~\cite{ball},
self-consistent mean--field theories~\cite{dan},  and
molecular-dynamics simulations~\cite{murat91}. Of particular
interest is the behavior of the bending rigidity of a polymer
brush, which can be calculated from the free energy of a
cylindrical and a spherical brush and forms a conceptually simple
model for the bending rigidity of a lipid
bilayer~\cite{milnerbend}.

A different scenario is obtained with special functionalized
lipids linked to the polymer chain. If such lipids are
incorporated  into lipid vesicles, the water-soluble polymers
(typically one uses PEG (poly-ethylene glycol) for its non-toxic
properties) form well-separated mushrooms, or, at higher
concentration of PEG lipid, a dense brush. These modified vesicles
are very interesting in the context of drug delivery, because they
show prolonged circulation times in vivo~\cite{allen}. This is
probably due to a steric serum-protein-binding inhibition by
the hydrophilic brush coat consisting of the PEG lipids. Since the
lipid bilayer is rather flexible and undergoes thermal bending
fluctuations, there is an interesting coupling  between the
polymer density distribution and the membrane
shape~\cite{lipopol1,Hristova1,Hristova2}.
For non-adsorbing, anchored polymers, the membrane will bend away
from the polymer due to steric repulsion\cite{lipopol2,lipopol3,bickel},
 but for adsorbing
anchored polymer the membrane will bend towards the anchored
polymer~\cite{lipopol4,lipopol5}.

The behavior of a polymer brush in contact with a solvent, which
is by itself also a polymer, consisting of chemically identical
but somewhat shorter chains than the brush, had been first
considered by de Gennes~\cite{gennes}. A complete scaling
description has been given only recently~\cite{aubouy}. One
distinguishes different regimes where the polymer solvent is
expelled to various degrees from the brush. A somewhat related
question concerns the behavior of two opposing brushes brought
closely together, and
separated by a
solvent  consisting of a polymer solution~\cite{gast}. Here one
distinguishes a regime where the polymer solution leads to a
strong attraction between the surfaces via the ordinary depletion
interaction (compare to Ref.~\cite{jldg79}), but also a high
polymer concentration regime where the attraction is not  strong
enough to induce colloidal flocculation. This phenomenon is called
colloidal restabilization~\cite{gast}.

Considering a mixed brush made of
mutually incompatible grafted chains, a novel transition to a
brush characterized by a lateral composition modulation was
found~\cite{marko91}. Even more complicated spatial structures are
obtained with grafted diblock copolymers~\cite{brown}. Finally, we
would like to mention in passing that  these static brush
phenomena have interesting consequences on dynamic properties of
polymer brushes~\cite{halperin88b}.

\subsection{Charged Grafted Polymers}

Brushes can also be formed by
charged polymers which are densely end-grafted to a surface;
they are called polyelectrolyte  or charged brushes. They have
been the focus of numerous theoretical
~\cite{MIK88}-\cite{ZHU97} and experimental
~\cite{MIR95,GUE95,AHR97} studies. In addition to the basic
interest, charged brushes are considered for their applications
because they serve as an efficient mean for preventing colloids
in polar media (such as aqueous solutions) from flocculating and
precipitating out of solution~\cite{napper}. This stabilization
arises from steric (entropic) as well as electrostatic repulsion.
A strongly charged brush is able to trap its own counterions and
generates a layer of locally enhanced salt
concentration~\cite{PIN91}. It is thus less sensitive to the
salinity of the surrounding aqueous medium than a stabilization
mechanism based on pure electrostatics (\ie\  without polymers).
Little is known from experiments on the
scaling behavior of PE brushes, as compared to neutral brushes.
The thickness of the brush layer
has been calculated from neutron-scattering experiments on
end-grafted polymers ~\cite{MIR95} and charged
diblock-copolymers at the air-water interface ~\cite{AHR97}.

Theoretical work on PE brushes was initiated by the works of
Miklavic and Marcelja ~\cite{MIK88} and Misra et al.
~\cite{MIS89}. In 1991, Pincus ~\cite{PIN91} and Borisov,
Birshtein and Zhulina ~\cite{BOR91} presented scaling theories for
charged brushes in the so-called osmotic regime, where the brush
height results from the balance between the chain elasticity
(which tends to decrease the brush height) and the repulsive
osmotic counterion pressure (which tends to increase the brush
height). In later studies, these works have been generalized to
poor solvents ~\cite{ROS92} and to the regime where excluded
volume effects become important, the so-called quasi-neutral or
Alexander regime~\cite{BOR94}.

\begin{figure}
 \epsfxsize=10cm
 \centerline{\vbox{\epsffile{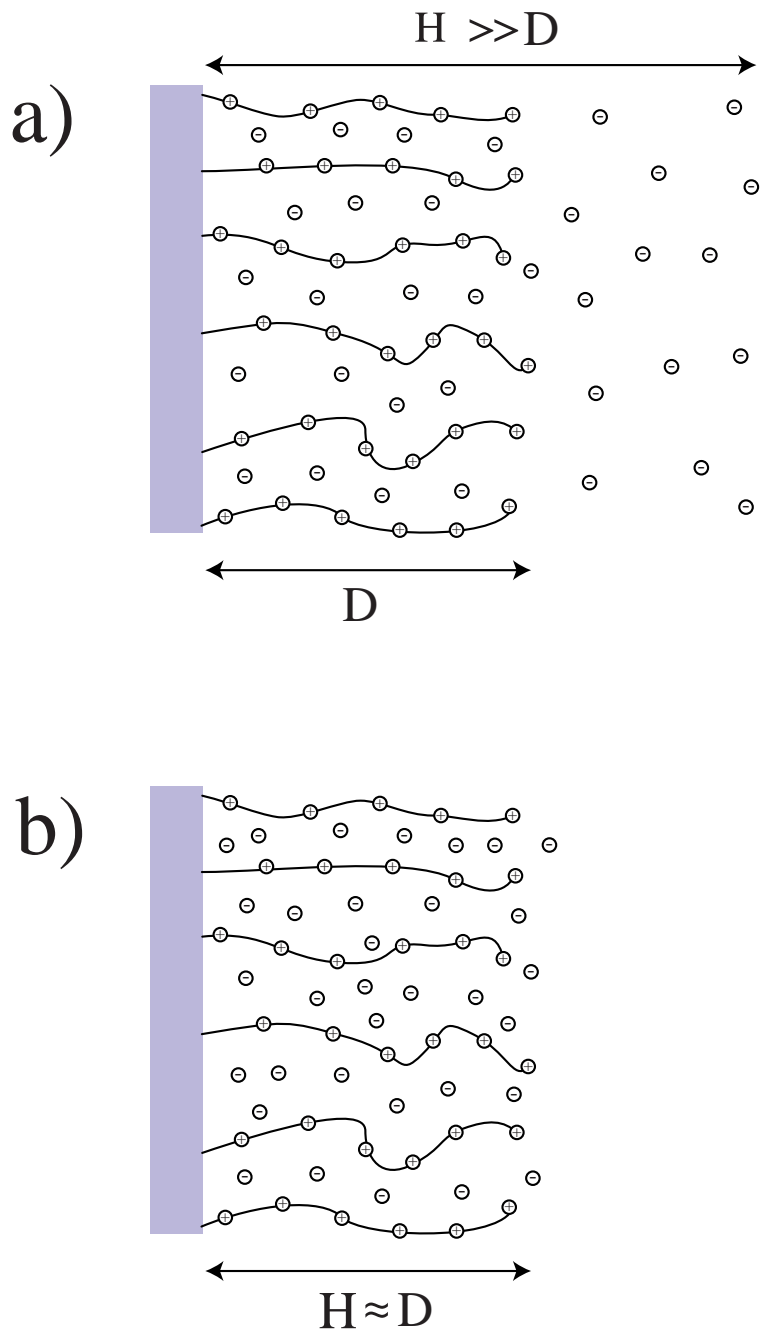} } }
\caption{ \label{fig16}
 Schematic PE brush structure. In a) we show the
weak-charge limit where the counterion cloud has a thickness $H$
larger than the thickness of the brush layer, $D$. In  b) we show
the opposite case of the strong-charge limit, where all
counterions are contained inside the brush and a single length
scale $D \approx H$  exists.}
\end{figure}

In what follows we assume that the charged brush is characterized
by two length scales: the average vertical extension of polymer
chains from the surface $D$, and the typical extent of the
counterion cloud, denoted by $H$. We neglect the presence of
additional salt, which has been discussed extensively in the
original literature, and only consider screening effects due to
the counterions of the charged brush. Two different scenarios
emerge, as is schematically presented in Fig.~\ref{fig16}. The
counterions can either extend outside the brush, $H\gg D$, as
shown in a), or be confined inside the brush, $H\approx D$ as
shown in b). As we show now, case b) is indicative of strongly
charged brushes, while case a) is typical for weakly charged
brushes.

The free energy density per unit area and in units of $k_B T$
contains several
contributions, which we now calculate one by one. We recall that
the grafting density of PE's is denoted by $\rho$, $z$ is the
counterion valency, $N$ the polymerization index of grafted
chains, and $f$ the charge fraction.
 The osmotic free energy, ${ F}_{\rm os}$,
associated with the ideal entropy cost of confining the
counterions to a layer of thickness $H$ is given by

\begin{equation}
    { F}_{\rm os}  \simeq
\frac{N f \rho}{z}
              \ln \left( \frac { N f \rho}{z H}\right)~.
\end{equation}

${ F}_{\rm v_2}$ is the second virial contribution to the free
energy, arising from steric repulsion between the monomers
(contributions due to counter ions are neglected). Throughout
this section, the polymers are assumed to be in a good solvent
(positive second virial coefficient $v_2 > 0$). The contribution
thus reads
\begin{equation}
  { F}_{\rm v_2}   \simeq  \frac{1}{2} \ D v_2
\left(
      \frac{N \rho}{D}\right)^2~.
\end{equation}
Finally, a direct electrostatic contribution ${ F}_{\rm el}$
occurs if the PE brush is not locally electro-neutral  throughout
the system, as for example is depicted in Fig.~\ref{fig16}a. This
energy is given by
\begin{equation}
{ F}_{\rm el} = \frac{2 \pi \ell_{\rm B}
(N f \rho)^2}{3}
                \frac{(H-D)^2}{H}~.
\end{equation}
This situation arises in the limit of low charge, when the
counterion density profile extends beyond the brush layer, \ie\
$H> D$.

\begin{figure}
 \epsfxsize=12cm
 \centerline{\vbox{\epsffile{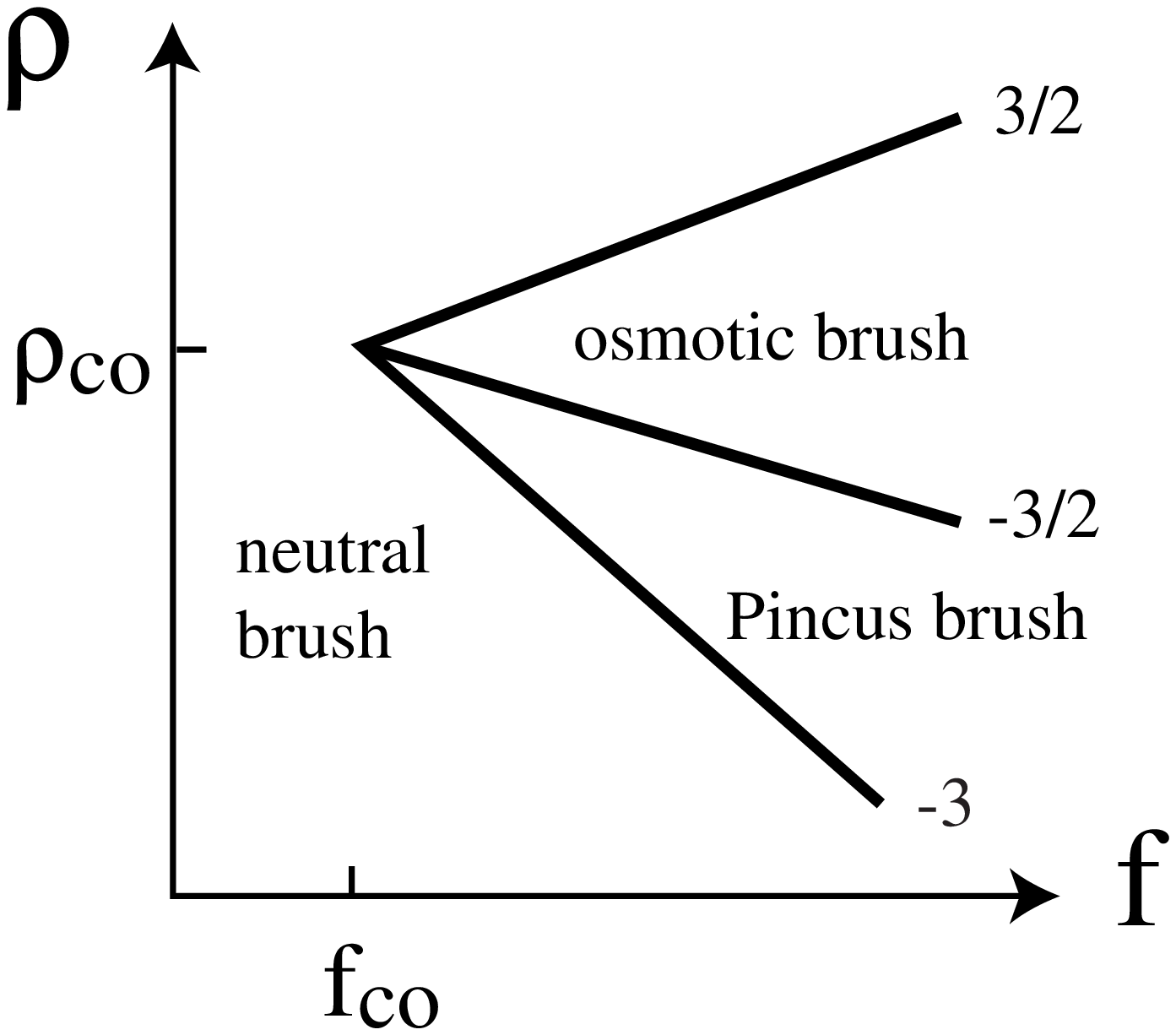} } }
\caption{  \label{fig17}
Scaling diagram for PE brushes on a
log-log plot as a function of the grafting density $\rho$ and the
fraction of charged monomers $f$. Featured are the Pincus-brush
regime, where the counterion layer thickness is much larger than
the brush thickness, the osmotic-brush regime, where all
counterions are inside the brush and the brush height is
determined by an equilibrium between the counterion osmotic
pressure and the PE stretching energy, and the neutral-brush
regime, where charge effects are not important and the brush
height results from a balance of PE stretching energy and
second-virial repulsion. The power law exponents of the various
lines are denoted by numbers.}
\end{figure}

The last contribution is the stretching energy of
the chains which is
\begin{equation}
{ F}_{\rm st}  = \frac{3 D^2}{2 N a^2}\, \rho~.
\end{equation}
Here, $a$ is the Kuhn length of the polymer,
implying that we neglect any chain stiffness for the brush
problem. The different free energy contributions lead, upon
minimization with respect to the two length scales $H$ and $D$,
to different behaviors. Let us first consider the weak charging
limit, \ie\  the situation where the counterions leave the brush,
$H > D$.
In this case, minimization of ${ F}_{\rm os} + { F}_{\rm el}$
with respect to the counterion height $H$ leads to
 \be
  H \sim \frac{1}{z \ell_B N f \rho}
 \ee
which has the same scaling as
the Gouy-Chapman length for $z$-valent counterions at a
surface of surface charge density $\sigma = N f \rho$. Balancing
now the polymer stretching energy ${ F}_{\rm st}$ and the
electrostatic energy ${ F}_{\rm el}$ one obtains the so-called
Pincus brush height
\begin{equation}
         D \simeq N^3 \rho\, a^2\ell_B f^{2}~,
         \label{Pinc}
\end{equation}
which results from the electrostatic attraction between
the counterions and the charged monomers. One notes the peculiar
dependence on the polymerization index $N$.
In the limit of $H\approx D$ where $D$ given by Eq.~(\ref{Pinc}),
the PE brush can be considered as neutral and the electrostatic
energy vanishes. There are two ways of balancing the remaining
free energy contributions. The first is obtained by comparing the
osmotic energy of counterion confinement, ${ F}_{\rm os}$, with the
polymer stretching term, ${ F}_{\rm st}$, leading to the height
 \be
D \sim \frac{N a f^{1/2}}{z^{1/2}}~,
 \ee
constituting the so-called osmotic brush regime. Finally
comparing the second-virial free energy, ${ F}_{\rm v_2}$, with the
polymer stretching energy, ${ F}_{\rm st}$, one obtains
 \be
  D \sim N a \left( v_2 \rho /a \right)^{1/3}~,
 \ee
and the PE brush is found to have the same scaling behavior as the
neutral brush~\cite{alex,gennes}, compare Eq.~(\ref{Floryh}).
Comparing the brush heights in all three regimes we arrive at the
phase diagram shown in Fig.~\ref{fig17}. The three scaling regimes
coexist at the characteristic charge fraction
\be
f_{\rm co} \sim \left( \frac{z v_2}{N^2 a^2 \ell_B} \right)^{1/3}~,
\ee
and the characteristic grafting density
\be
\rho_{\rm co} \sim \frac{1}{N \ell_B^{1/2} v_2^{1/2}}~.
\ee
For large values of the charge fraction $f$ and the grafting
density $\rho$ it has been found numerically that the brush height
does not follow any of the scaling laws discussed
here~\cite{Csajka0}. This has been recently rationalized in terms
of another scaling regime, the collapsed regime. In this regime
one finds that correlation and fluctuation effects, which are
neglected in the discussion in this section, lead to a net
attraction between charged monomers and
counterions~\cite{Csajka}. Similarly, two charged surfaces, one
decorated with a charged brush, the other one neutralized by
counter ions, attract each other at large enough grafting
densities \cite{Ennis2}.

Another way of creating a charged brush is to dissolve a diblock
copolymer consisting of a hydrophobic and a charged block in
water. The diblocks associate to form a hydrophobic core, thereby
minimizing the unfavorable interaction with water, while the
charged blocks form a highly charged corona or brush
\cite{Eisenberg}. The micelle morphology depends on different
parameters. Most importantly, it can be shown that salt acts as a
morphology switch, giving rise to the sequence spherical,
cylindrical, to planar micellar morphology as the salt
concentration is increased \cite{Eisenberg}. Theoretically, this
can be explained by the entropy cost of counterion confinement
in the charged corona \cite{Netzmicelles}. The charged corona can
be studied by neutron scattering \cite{Guenoun2} or atomic-force
microscopy \cite{Foerster2} and gives information on the behavior
of highly curved charged brushes.

\section{Concluding Remarks}
\label{conclusion} \setcounter{equation}{0}

We reviewed simple physical concepts underlying the main theories
which deal with equilibrium and static properties of
neutral and charged polymers
adsorbed or grafted to substrates. Most of the review dealt with
somewhat ideal situations: smooth and flat surfaces which are
chemically homogeneous; long and linear homopolymer chains where
chemical properties can be averaged on; simple phenomenological
type of interactions between the monomers and the substrate as
well as between the monomers and the solvent.

Even with all the simplifying assumptions, the emerging physical
picture is quite rich and robust. Adsorption of polymers from dilute
solutions can be understood in terms of
single-chain adsorption on the substrate. Mean--field theory is quite
successful but in some
cases fluctuations in the local monomer concentration play an
important role. Adsorption from more concentrated solutions results
in even more complex density profiles, with several regimes
(proximal, central, distal). Each regime is characterized by a
different physical behavior. We reviewed the principle theories
used to model the polymer behavior. We also mentioned briefly more
recent ideas about the statistics of polymer loops and tails.
For charged polymers, the structure of the adsorbed layer
is in part controlled by the counterion distribution which
is coupled to the polymer layer.

The second part of this review is about neutral and charged
polymers which are
terminally grafted on one end to the surface and are called
polymer brushes. The theories here are quite different since the
statistics of the grafted layer depends crucially on the fact that
the chain is not attracted to the surface but is forced to be in
contact to the surface since one of its ends is chemically or
physically bonded to the surface. Here as well we review the
classical mean--field theory and more advanced theories giving the
concentration profiles of the entire polymer layer as well as that
of the polymer free ends.

In general, the theory for neutral polymers is more advanced than
the one for charged polymers, partly because charged polymers
became the target for theoretical modelling fairly recently. In
addition, due to the long-range interactions between charged
monomers, and due to a number of additional relevant parameters
(such as salt concentration, pH), the resultant behavior for
charged polymers is more complex. We have introduced some of the
basic concepts of charged polymers, such as  the Manning
condensation of counterions and the electrostatic chain
stiffening. Due to this increased stiffness of polyelectrolytes,
their chain statistics is  described by semi-flexible models. We
have, therefore, introduced such models in some detail and also
demonstrated some effects specific to semi-flexible charged
polymers.

At present,
studies of polyelectrolytes
 in solutions and at surfaces is shifting
more towards biological systems. We mentioned in this
review the complexation of DNA and histones. This is
only one of many examples of interest where charged biopolymers,
receptors, proteins and DNA molecules interact with each other or
with other cellular components. The challenge for future
fundamental research will be to try to understand the role of
electrostatic interactions combined with specific biological
(lock-key) mechanisms and to infer on biological functionality of
such interactions.

In this review,
we also discussed additional factors that have an effect on the
polymer adsorption and grafted layers: the quality of the solvent,
undulating and flexible substrates such as fluid/fluid interfaces
or lipid membranes; adsorption and grafting on curved
surfaces such as spherical colloidal particles.

Although our main aim was to review the theoretical progress in
this field, we mentioned many relevant experiments. In this active
field several advanced experimental techniques are used to probe
adsorbed or grafted polymer layers: neutron scattering and
high-resolution x-ray reflectivity, light scattering using
fluorescent probes, ellipsometry, surface isotherms as well as
the surface force apparatus for the force measurement between two
surfaces.

This paper should be viewed as a general introduction to adsorption
phenomena  involving charged and neutral chains
and  can serve as a starting point
to understand more complex systems as encountered in applications
and current experiments.

\vspace{3cm}
\newlength{\tmp}
\setlength{\tmp}{\parindent}
\setlength{\parindent}{0pt}
{\em Acknowledgments}
\setlength{\parindent}{\tmp}

It is a pleasure to thank our collaborators G. Ariel, I. Borukhov,
M. Breidenich, Y. Burak, H. Diamant, J.-F. Joanny, K. Kunze, L.
Leibler, R. Lipowsky, A. Moreira, H. Orland, M. Schick, C. Seidel
and A. Shafir with whom we have been working on polymers and
polyelectrolytes. One of us (DA) would like to acknowledge partial
support from the Israel Science Foundation, Centers of Excellence
Program and under grant no. 210/02, the Israel--US Binational
Science Foundation (BSF) under grant no. 98-00429, and the
Alexander von Humboldt Foundation for a research award.

\newpage



\clearpage

\clearpage

\clearpage

\clearpage

\clearpage

\clearpage

\clearpage

\clearpage

\clearpage

\clearpage

\clearpage

\clearpage

\clearpage

\clearpage

\clearpage

\clearpage

\clearpage

\clearpage

\clearpage

\clearpage

\clearpage

\clearpage

\clearpage

\clearpage

\clearpage

\clearpage

\clearpage

\clearpage

\clearpage

\clearpage

\begin{thebibliography}{99}




\bibitem{fleer} G.J. Fleer, M.A. Cohen Stuart, J.M.H.M. Scheutjens, T. Cosgrove,
B. Vincent, Polymers at Interfaces,  Chapman \& Hall, London,
1993.

\bibitem{erich}
E. Eisenriegler, Polymers near Surfaces, World
Scientific, Singapore, 1993.

\bibitem{cscv86} M.A. Cohen Stuart, T. Cosgrove,  B. Vincent, Adv. Colloid
Interface Sci. 24 (1986) 143.

\bibitem{dg87} P.G. de Gennes, Adv. Colloid Interface Sci. 27 (1987) 189.

\bibitem{s96}I. Szleifer, Curr. Opin. Colloid Interface Sci. 1 (1996) 416.

\bibitem{fl97}G.J. Fleer, F.A.M. Leermakers, Curr. Opin. Colloid Interface
Sci. 2 (1997) 308.

\bibitem{Grosberg}
A.Yu. Grosberg, A.R. Khokhlov, Statistical Physics of Macromolecules,
AIP Press, New York, 1994.

\bibitem{Oosawa}
F. Oosawa,  Polyelectrolytes, Dekker, New York, 1971.

\bibitem{Dautzen}
H. Dautzenberg, W. Jaeger, B.P.J. K\"otz, C. Seidel, D.
Stscherbina, Polyelectrolytes: Formation, Characterization
and Application, Hanser Press, Munich, 1994.

\bibitem{Foerster}
S. F\"orster, M. Schmidt, { Adv. Polym. Sci.} 120 (1995)  50.

\bibitem{barrat1}
J.-L. Barrat, J.-F. Joanny, Adv. Chem. Phys.
94 (1996) 1.

\bibitem{Flory1}
P.J. Flory, Principles of Polymer Chemistry, Cornell
University, Ithaca, 1953.

\bibitem{Yamakawa} H. Yamakawa, Modern Theory of Polymer Solutions,
Harper \& Row, New York, 1971.

\bibitem{degennes}
P.G. de Gennes, Scaling Concepts in Polymer Physics, Cornell
University, Ithaca, 1979.

\bibitem{Cloizeaux}
J. des Cloizeaux, J. Jannink, Polymers in Solution,
Oxford University, Oxford, 1990.

\bibitem{tubulin}
F. Gittes, B. Mickey, J. Nettleton, J. Howard,
J. Cell Biol. 120 (1993) 923.

\bibitem{actin1}
J. K\"as, H. Strey, M. B\"armann,  E. Sackmann,
Europhys. Lett. 21 (1993) 865.

\bibitem{actin2}
A. Ott, M. Magnasco, A. Simon,  A. Libchaber,
Phys. Rev. E 48 (1993) R1642.

\bibitem{DNA}
C. Frontale, E. Dore, A. Ferrauto,  E. Gratton,
Biopolymers 18 (1979) 1353.


\bibitem{gpv}
P.G. de Gennes, P. A. Pincus, R.M. Velasco,  F. Brochard,
J. Phys. (France) 37 (1976) 1461.

\bibitem{Khokhlov}
A. R. Khokhlov,  K. A. Khachaturian, { Polymer} 23 (1982) 1742.

\bibitem{barrat2}
J.-L. Barrat,  J.-F. Joanny, { Europhys. Lett.} 24 (1993) 333.

\bibitem{Netz2}
R. R. Netz, H. Orland, { Eur. Phys. J. B} 8 (1999) 81.

\bibitem{Odijk0}
T. Odijk, { J. Polym. Sci. Part B: Polym. Phys.} 15 (1977) 477;
{ Polymer} 19 (1978) 989.

\bibitem{Skolnick}
J. Skolnick,  M. Fixman, Macromolecules 10 (1977) 944.


\bibitem{Sim1}
G.A. Christos,  S.L. Carnie, J. Chem. Phys. 92 (1990) 7661.

\bibitem{Sim2}
C. Seidel, H. Schlacken,  I. M\"uller,
Macromol. Theory Simul. 3 (1994) 333.

\bibitem{Sim3}
M. Ullner, B. J\"onsson, C. Peterson, O. Sommelius,
B. S\"oderberg, J. Chem. Phys. 107 (1997) 1279;
M. Ullner,  C.E. Woodward, Macromolecules 35 (2002) 1437.

\bibitem{Sim4}
U. Micka,  K. Kremer, Phys. Rev. E 54 (1996) 2653;
Europhys. Lett. 38 (1997) 279.

\bibitem{Ralf}
R. Everaers, A. Milchev,  V. Yamakov, Eur. Phys. J. E 8 (2002) 3.

\bibitem{Nguyen3}
T.T. Nguyen,  B.I. Shklovskii, Phys. Rev. E 66 (2002) 021801.



\bibitem{Li}
H. Li,  T. Witten, Macromolecules 28 (1995) 5921.

\bibitem{Ha2}
B.-Y. Ha,  D. Thirumalai, J. Chem. Phys. 110 (1999) 7533.

\bibitem{Ha1}
B.-Y. Ha,  D. Thirumalai, Macromolecules 28 (1995) 577.

\bibitem{Liverpool}
T.B. Liverpool,  M. Stapper, Europhys. Lett. 40 (1997) 485.

\bibitem{PEcoll0}
M. Olvera de la Cruz, L. Belloni, M. Delsanti, J.P. Dalbiez, O. Spalla,
M. Drifford, J. Chem. Phys. 103 (1995) 5781.

\bibitem{PEcoll1}
J. Wittmer, A. Johner,  J.F. Joanny,
J. Phys. II (France) 5 (1995) 635.

\bibitem{PEcoll2}
R.G. Winkler, M. Gold,  P. Reineker, Phys. Rev. Lett. 80 (1998) 3731.

\bibitem{PEcoll3}
N.V. Brilliantov, D.V. Kuznetsov,  R. Klein,
Phys. Rev. Lett. 81 (1998) 1433

\bibitem{PEcoll4}
M.O. Khan,  B. J\"onsson, Biopolymers 49 (1999) 121.

\bibitem{PEcoll5}
F.J. Solis,  M. Olvera de la Cruz, Eur. Phys. J. E 4 (2001) 143.

\bibitem{PEcoll6}
R. Golestanian, M. Kardar,  T.B. Liverpool,
Phys. Rev. Lett. 82 (1999) 4456.

\bibitem{PEcoll7}
P. L. Hansen, D. Svensek, V.A. Parsegian,  R. Podgornik,
Phys. Rev. E 60 (1999) 1956.

\bibitem{PEcoll8}
T.T. Nguyen, I. Rouzina,  B.I. Shklovskii,
Phys. Rev. E 60 (1999) 7032.

\bibitem{Ariel1}
G. Ariel,  D. Andelman, Europhys. Lett. 61 (2003) 67.

\bibitem{Ariel2}
G. Ariel,  D. Andelman, Phys. Rev. E 67 (2003) 0118xx .

\bibitem{cyl1}
N. Gr\o nbech-Jensen, R.J. Mashl, R.F. Bruinsma,  W.M. Gelbart,
Phys. Rev. Lett. 78 (1997) 2477.

\bibitem{cyl2}
B.-Y. Ha,  A.J. Liu, Phys. Rev. Lett. 79 (1997) 1289.

\bibitem{cyl3}
R. Podgornik,  V.A. Parsegian, Phys. Rev. Lett. 80 (1998) 1560.

\bibitem{cyl4}
J.J. Arenzon, J.F. Stilck,  Y. Levin, Eur. Phys. J. B 12 (1999) 79.

\bibitem{cyl5}
F.J. Solis,  M. Olvera de la Cruz, Phys. Rev. E 60 (1999) 4496.


\bibitem{Man1}
G.S. Manning,  J. Chem. Phys. {51} (1969) 924.

\bibitem{Man2}
G.S. Manning,  J. Chem. Phys. {51} (1969) 934.

\bibitem{Tracy}
C.A. Tracy,  H. Widom, Physica A 244 (1997) 402.

\bibitem{Man3}
G.S. Manning,  J. Ray, J. Biomol. Struct. Dyn.
16 (1998) 461.

\bibitem{Fixman2}
M. Fixman, J. Chem. Phys. 76 (1982) 6346.

\bibitem{LeBret}
M. Le Bret, J. Chem. Phys. 76 (1982) 6243.

\bibitem{Man4}
G.S. Manning,  U. Mohanty, Physica A 247 (1997) 196.

\bibitem{Deserno}
M. Deserno, C. Holm,  S. May, Macromolecules 33 (2000) 199.

\bibitem{Limbach}
H.J. Limbach,  C. Holm, J. Chem. Phys. 114 (2001) 9674.

\bibitem{Man5}
G.S. Manning, J. Chem. Phys. 89 (1988) 3772.

\bibitem{Kuhn}
P.S. Kuhn, Y. Levin,  M.C. Barbosa, Macromolecules 31 (1998) 8347.

\bibitem{Deshkovski}
A. Deshkovski, S. Obukhov,  M. Rubinstein,
Phys. Rev. Lett. 86 (2001) 2341.

\bibitem{Wandrey}
C. Wandrey, D. Hunkeler, U. Wendler, W. Jaeger, {
Macromolecules} {33} (2000) 7136.

\bibitem{Blaul}
J. Blaul, M. Wittemann, M. Ballauff,  M. Rehahn, { J. Phys.
Chem. B} {104} (2000) 7077.

\bibitem{Schiessel}
H. Schiessel,  P. A. Pincus,  { Macromolecules} {31} (1998) 7953;
H. Schiessel, { Macromolecules} {32} (1999) 5673.


\bibitem{houwaart}
T.O. Odijk,  A.C. Houwaart, J. Polym. Sci. 16 (1978) 627.

\bibitem{Fixman}
M. Fixman,  J. Skolnick, Macromolecules 11 (1978) 863.


\bibitem{Khokhlov2}
A.R. Khokhlov, J. Phys. A 13 (1980) 979.

\bibitem{Kantor}
Y. Kantor,  M. Kardar, Phys. Rev. E 51 (1995) 1299.

\bibitem{Dobrynin2}
A.V. Dobrynin, M. Rubinstein, S.P. Obukhov,
{ Macromolecules} 29 (1996) 2974.

\bibitem{Solis}
F.J. Solis,  M. Olvera de la Cruz,
Macromolecules 31 (1998) 5502.

\bibitem{Lyulin}
A.L.  Lyulin, B. D\"unweg, O.V. Borisov, A.A. Darinskii,
{ Macromolecules} 32 (1999) 3264.

\bibitem{Micka2}
U. Micka, C. Holm,  K. Kremer, Langmuir 15 (1999) 4033.

\bibitem{Waigh}
T.A. Waigh, R. Ober, C.E. Williams,  J.-C. Galin,
Macromolecules 34 (2001) 1973.


\bibitem{Nierlich1}
M. Nierlich, F. Boue, A. Lapp, R. Oberth\"ur, { Coll. Polym.
Sci.} 263 (1985) 955.

\bibitem{Nierlich2}
M. Nierlich, F. Boue, A. Lapp, R. Oberth\"ur, { J. Phys.
(France)}  {46} (1985) 649.

\bibitem{Spiteri}
M.N. Spiteri, F. Boue, A. Lapp,  J.P. Cotton,
Phys. Rev. Lett. 77 (1996) 5218.

\bibitem{Moussaid}
A. Moussaid, F. Schosseler, J.P. Munch,  S.J. Candau, { J.
Phys. II (France)} 3 (1993) 573.

\bibitem{Essafi1}
W. Essafi, F. Lafuma,  C.E. Williams, { J. Phys. II (France)}
5 (1995)  1269.

\bibitem{Nishida}
K. Nishida, K. Kaji,  T. Kanaya, { Macromolecules} 28 (1995) 2472.

\bibitem{Essafi2}
W. Essafi, F. Lafuma,  C.E. Williams, { Eur. Phys. J. B} 9 (1999) 261.

\bibitem{Borue}
V.Y. Borue,  I. Y. Erukhimovich, { Macromolecules} 21 (1988) 3240.

\bibitem{Joanny}
J.F. Joanny,  L. Leibler,  J. Phys. (France) 51 (1990) 545.

\bibitem{Yoshizaki}
T. Yoshizaki, H. Yamakawa,  Macromolecules 13 (1980) 1518.

\bibitem{Dymitrowska}
M. Dymitrowska,  L. Belloni, J. Chem. Phys. 109 (1998) 4659.

\bibitem{Yethi}
A. Yethiraj, J. Chem. Phys. 108 (1998) 1184.

\bibitem{NetzRPA}
R.R. Netz, to be published.

\bibitem{PEsemi1}
T.A. Witten,  P. A. Pincus, Europhys. Lett. 3 (1987) 315.

\bibitem{PEsemi2}
J.-L. Barrat,  J.-F. Joanny, J. Phys. II (France) 4 (1994) 1089.


\bibitem{Sim6}
M.J. Stevens,  K. Kremer, J. Chem. Phys. 103 (1995) 1669.

\bibitem{Sim7}
M.J. Stevens,  K. Kremer, J. Phys. II (France) 6 (1996) 1607.

\bibitem{Sim8}
H. Sch\"afer,   C. Seidel, Macromolecules 30 (1997) 6658.

\bibitem{Sim9}
M.J. Stevens,  S.J. Plimpton, Eur. Phys. J. B 2 (1998) 341.

\bibitem{KlitzingPSS}
R. von Klitzing, A. Espert, A. Asnacios, T. Hellweg, A. Colin,
D. Langevin, Colloids Surfaces A 149 (1999) 131.

\bibitem{lk85} P.F. Luckham,  J. Klein, Macromolecules 18 (1985) 721.

\bibitem{napper}D.H. Napper, Polymeric Stabilization of Colloidal Dispersions,
Academic Press, London, 1983.

\bibitem{lipopol1}
R. Lipowsky, Europhys. Lett. 30 (1995) 197.

\bibitem{star}
A. Halperin,  J.F. Joanny, J. Phys. II (France) 1 (1991) 623.

\bibitem{ring1}
B. van Lent, J. Scheutjens,  T. Cosgrove, Macromolecules 20 (1987) 366.

\bibitem{ring2}
G. Stratouras,  M. Kosmas, Macromolecules 25 (1992) 3307.

\bibitem{ran1}
C.M. Marques, J.F. Joanny, Macromolecules 23 (1990) 268.

\bibitem{ran2}
B. van Lent, J.M.H.M.
Scheutjens, J. Phys. Chem. 94 (1990) 5033.

\bibitem{ran3}
J.P. Donley, G.H. Fredrickson,  Macromolecules 27 (1994) 458.

\bibitem{ran4}
T. Garel, D.A. Huse, S. Leibler,  H. Orland,
Europhys. Lett. 8 (1989) 9.

\bibitem{ran5}
S.T. Milner,  G.H. Fredrickson, Macromolecules 28 (1995) 7953.

\bibitem{ran6}
S. Stepanow, J.-U. Sommer,  I.Y. Erukhimovich,
Phys. Rev. Lett. 81 (1998) 4412.

\bibitem{ran6b}
A. Maritan, M.P. Riva,  A. Trovato, J. Phys. A: Math Gen 32 (1999) L275.

\bibitem{ran7}
X. Chatellier,  J.-F. Joanny, Eur. Phys. J. E 1 (2000) 9.

\bibitem{fse53}
H.L. Frish, R. Simha, F.R. Eirish, J. Chem. Phys. 21, (1953) 365;
R. Simha, H.L. Frish, F.R. Eirish, J. Phys. Chem. 57 (1953) 584.

\bibitem{s62} A. Silberberg, J. Phys. Chem. 66 (1962) 1872; 66 (1962) 1884.

\bibitem{e65} S.F. Edwards, Proc. Phys. Soc. (London) 85 (1965) 613;
88 (1966) 255.

\bibitem{dm65} E.A. DiMarzio, J. Chem. Phys. 42 (1965) 2101;
E.A. DiMarzio,  F.L. McCrackin, J. Chem. Phys. 43 (1965) 539;
C. Hoeve, E.A. DiMarzio,  P. Peyser, J. Chem. Phys. 42 (1965) 2558.

\bibitem{r65} R.J. Rubin, J. Chem. Phys. 43 (1965) 2392.

\bibitem{jr77}I.S. Jones,  P. Richmond, J. Chem. Soc Faraday Trans 2 (1977) 73.

\bibitem{Israel}
J.N. Israelachvili, Intermolecular and Surface Forces,
Academic Press, London, 1992.

\bibitem{daoud77} M. Daoud,   P.G. de Gennes, J. Phys. (France) 38 (1977) 85.

\bibitem{dg69} P.G. de Gennes, Rep. Prog. Phys. 32 (1969) 187.

\bibitem{Flory2}
P.J. Flory, Statistical Mechanics of Chain Molecules, Hanser Press, Munich, 1988.

\bibitem{square}
J.M.J. van Leeuwen,  H.J. Hilhorst, Physica A 107 (1981) 319; T.W.
Burkhardt, J. Phys. A 14 (1981) L63; D.M. Kroll, Z. Phys. B 41 (1981) 345.

\bibitem{lip89}
R. Lipowsky,  A. Baumg\"artner, Phys. Rev. A 40 (1989) 2078;
R. Lipowsky, Physica Scripta T29 (1989) 259.

\bibitem{Borisov}
O. V. Borisov, E. B. Zhulina, T. M. Birshtein, { J. Phys. II
(France)} 4 (1994) 913.

\bibitem{netz95}
R.R. Netz, Phys. Rev. E 51 (1995) 2286.


\bibitem{pa1}
J.F. Joanny, J. Phys. II (France) 4 (1994) 1281.

\bibitem{pa2}
A.V. Dobrynin, M. Rubinstein,  J.F. Joanny, Macromolecules 30 (1997) 4332.

\bibitem{pa3}
R.R. Netz,  J.F. Joanny, Macromolecules 31 (1998) 5123.

\bibitem{pa4}
A.V. Dobrynin, S.P. Obukhov,  M. Rubinstein, Macromolecules 32 (1999) 5689.

\bibitem{pa5}
E. Zhulina, A.V. Dobrynin,  M. Rubinstein, Eur. Phys. J. E 5 (2001) 41.

\bibitem{pa6}
M.O. Khan, T. Akesson,  B. J\"onsson, Macromolecules 34 (2001) 4216.

\bibitem{Poland}
D. Poland,  H.A. Scheraga, J. Chem. Phys. 45 (1966) 1456;
45 (1966) 1464.

\bibitem{Fisher}
M.E. Fisher, J. Chem. Phys. 45 (1966) 1469.

\bibitem{gennes72}
P.G. de Gennes, Phys. Lett. A 38 (1972) 339.

\bibitem{ekb82}E. Eisenriegler, K. Kremer, K. Binder, J. Chem. Phys. 77 (1982) 6296;
E. Eisenriegler, J. Chem. Phys. 79 (1983) 1052.

\bibitem{dg76}P.G. de Gennes, J. Phys. (Paris) 37 (1976) 1445.

\bibitem{dgp83} P.G. de Gennes,  P. A. Pincus. J. Phys. Lett. (Paris)
44 (1983) L241.

\bibitem{Diehl}
H.W. Diehl,  M. Shpot, Phys. Rev. Lett. 73 (1994) 3431.

\bibitem{bd87} E. Bouchaud,  M. Daoud. J. Phys. (Paris) 48 (1987) 1991.

\bibitem{glassy}
J. Baschnagel, in Structure,  Properties of Glassy Polymers,
M.T. Tant,  A.J. Hill, eds., ACS Symposium Series 710.


\bibitem{Kerstin}
K. de Meijere, G. Brezesinski, H.  M\"ohwald, { Macromolecules}
30 (1997) 2337.

\bibitem{Kerstin2}
K. de Meijere, G. Brezesinski, K. Kjaer, H.  M\"ohwald, Langmuir
14 (1998) 4204.

\bibitem{Berlepsch}
H. von Berlepsch, C. Burger, H. Dautzenberg, { Phys. Rev. E}
58 (1998) 7549.

\bibitem{Stubenrauch}
C. Stubenrauch, P.-A. Albouy, R. von Klitzing,  D. Langevin,
Langmuir 16 (2000) 3206.

\bibitem{Asnacios2}
A. Asnacios, R. von Klitzing, D. Langevin, Colloids Surfaces A 167 (2000) 189.

\bibitem{Baltes}
H. Ahrens, H. Baltes, J. Schmitt, H. M\"ohwald,  C.A. Helm,
Macromolecules 34 (2001) 4504.

\bibitem{Fang}
Y. Fang, J.  Yang, { J. Phys. Chem. B} 101 (1997) 441.

\bibitem{Raedler}
J.O. R\"adler, I. Koltover, T. Salditt, C. R. Safinya, {
Science} 275 (1997) 810; T. Salditt, I. Koltover, J.
O.  R\"adler, C. R. Safinya, { Phys. Rev. Lett.} 79 (1997)
 2582.

\bibitem{Maier}
B. Maier, J. O. R\"adler, { Phys. Rev. Lett.} 82 (1999) 1911.


\bibitem{hong}
G. Decher, J.D. Hong, J.  Schmitt, { Thin Solid Films} 210/211
(1992) 831.

\bibitem{Klitzing}
R. von Klitzing, H. M\"ohwald, { Langmuir} 11 (1995)
3554; R. von Klitzing, H. M\"ohwald, { Macromolecules}
29 (1996) 6901.

\bibitem{Decher}
G. Decher, { Science} 277 (1997) 1232; M. L\"osche,
J. Schmitt, G. Decher, W.G. Bouwman, K. Kjaer, {
Macromolecules} 31 (1998) 8893.

\bibitem{caruso2}
F. Caruso, K. Niikura, D.N. Furlong, Y. Okahata, { Langmuir}
13 (1997) 3422.


\bibitem{donath}
E. Donath, G.B. Sukhorukov, F. Caruso,  S.A. Davis,
H. M\"ohwald,  Angew. Chem. Int. Ed. 16 (1998) 37;
G.B. Sukhorukov, E. Donath, S.A. Davis,
H. Lichtenfeld, F. Caruso,  V.I. Popov,
H. M\"ohwald,  Polym. Adv. Technol. 9 (1998) 759.

\bibitem{caruso}
F. Caruso, R.A. Caruso, H. M\"ohwald,
 Science 282 (1998) 1111; F. Caruso, Adv. Mater. 13 (2001) 11.


\bibitem{Wiegel}
F.W. Wiegel, J. Phys. A 10 (1977) 299.

\bibitem{Muthukumar}
  M. Muthukumar, { J. Chem. Phys.} 86 (1987) 7230.

\bibitem{xav98}
X. Chatellier, T.J. Senden, J.F. Joanny,  J.M. di Meglio, Europhys.
Lett. 41 (1998) 303.

\bibitem{Netz4}
R.R. Netz,  J.F. Joanny, Macromolecules 32 (1999) 9013.

\bibitem{Borisov2}
O.V. Borisov, F. Hakem, T.A. Vilgis, J.F. Joanny,  A. Johner,
Eur. Phys. J. E 6 (2001) 37.

\bibitem{Dobrynin4}
A.V. Dobrynin, A. Deshkovski,  M. Rubinstein,
 Macromolecules 34 (2001) 3421.

\bibitem{Odijk1}
T. Odijk, { Macromolecules} 16 (1983) 1340; 17
(1984) 502.

\bibitem{Maggs}
A. C. Maggs, D. A. Huse, S. Leibler, { Europhys. Lett.}
8 (1989) 615.

\bibitem{Gompper}
G. Gompper, T. W.  Burkhardt, { Phys. Rev. A} 40 (1989)
6124; G. Gompper, U. Seifert, { J. Phys. A} 23 (1990) L1161.

\bibitem{Bundschuh}
R. Bundschuh, M. L\"assig, R. Lipowsky, { Eur. Phys. J. E}
3 (2000) 295.

\bibitem{semi1}
D.V. Kuznetsov,  W. Sung, Macromolecules 31 (1998) 2679;
J. Chem. Phys. 107 (1997) 4729.

\bibitem{semi2}
S. Stepanow, J. Chem. Phys. 115 (2001) 1565.

\bibitem{semi3}
T. Sintres, K. Sumithra,  E. Straube, Macromolecules 34 (2001) 1352.

\bibitem{Yamakov}
V. Yamakov, A. Milchev, O. Borisov,  B. D\"unweg,
{ J. Phys.: Condens. Matter} 11 (1999) 9907.




\bibitem{ch58} J.W. Cahn,  J.E. Hilliard, J. Chem. Phys. 28 (1958) 258.

\bibitem{dg81}P.G. de Gennes, Macromolecules 14 (1981) 1637.

\bibitem{g92} O. Guiselin, Europhys. Lett. 17 (1992) 225.

\bibitem{neutscatt}
L. Auvray,  J.P. Cotton, Macromolecules 20 (1987) 202.

\bibitem{neutrefl}
L.T. Lee, O. Guiselin, B. Farnoux, A. Lapp, Macromolecules 24 (1991) 2518;
O. Guiselin, L.T. Lee, B. Farnoux, A. Lapp, J. Chem. Phys.
95 (1991) 4632; O. Guiselin, Europhys. Lett. 17 (1992) 57.



\bibitem{taleoftails}
J.M.H.M. Scheutjens, G.J. Fleer,  M.A. Cohen Stuart, Colloids
and Surfaces 21 (1986) 285; G.J. Fleer, J.M.H.M. Scheutjens, M.A. Cohen
Stuart, Colloids and Surfaces 31 (1988) 1.

\bibitem{johner93}
A. Johner, J.F. Joanny,  M. Rubinstein, Europhys. Lett. 22 (1993) 591.

\bibitem{semenov95}
A.N. Semenov,  J.F. Joanny, Europhys. Lett. 29 (1995) 279; A.N. Semenov,
J. Bonet-Avalos, A. Johner, J.F. Joanny, Macromolecules 29 (1996) 2179;
A. Johner, J. Bonet-Avalos, C.C. van der Linden, A.N. Semenov,
J.F. Joanny, Macromolecules 29 (1996) 3629.

\bibitem{Manghi}
M. Manghi,  M. Aubouy, Macromolecules 33 (2000) 5721.

\bibitem{semrev}
A.N. Semenov, J.F. Joanny, A. Johner. In: A. Grosberg, ed.,
Theoretical and Mathematical Models in Polymer Research,
Academic Press, Boston, 1998.


\bibitem{Peyser}
  P. Peyser, R.  Ullman,
  { J. Pol. Sci. A} 3 (1965)  3165.

\bibitem{Kawaguchi}
  M. Kawaguchi, H. Kawaguchi, A. Takahashi,
  { J. Coll. Interface Sci.} 124 (1988)  57.

\bibitem{Meadows}
  J. Meadows, P. A. Williams, M. J. Garvey,
  R. Harrop, G. O. Phillips,
  { J. Coll. Interface Sci.} 132 (1989) 319.

\bibitem{Denoyel}
  R. Denoyel, G. Durand, F. Lafuma, R. Audbert,
  { J. Coll. Interface Sci.} 139 (1990) 281.

\bibitem{Blaakmeer}
 J.  Blaakmeer, M. R. B\"ohmer, M. A. Cohen Stuart, G. J. Fleer.
  { Macromolecules} 23 (1990) 2301.

\bibitem{vandeSteegExp}
  H. G. A. van de Steeg,  A. de Keizer, M. A. Cohen Stuart,
 B. H.  Bijsterbosch,
  { Coll. Surf. A} 70 (1993) 91.

\bibitem{Shubin}
  V. Shubin, P. Linse,
  { J. Phys. Chem.} 99 (1995) 1285.

\bibitem{Hoogeveen}
  N. G. Hoogeveen, { Ph.D. Thesis},
  Wageningen Agricultural University,
  The Netherlands, 1996.

\bibitem{Cohen1}
M.A. Cohen Stuart, { J.  Phys. (France)} 49 (1988) 1001.

\bibitem{Cohen2}
   M. A. Cohen Stuart, G. J. Fleer, J. Lyklema,
   W. Norde, J. M. H. M. Scheutjens,
   { Adv. Coll. Interface Sci.} 34 (1991) 477.

\bibitem{Norde}
   C. A. Haynes, W.  Norde,
   { Coll. Surf. B} 2 (1994) 517.

\bibitem{auroy1}
P. Auroy, L. Auvray,  L. Leger, Phys. Rev. Lett. 66 (1991) 719;
Macromolecules 24 (1991) 2523; Macromolecules 24 (1991) 5158.

\bibitem{JFJ99}
J. F. Joanny, { Eur. Phys. J. B} 9 (1999) 117.

\bibitem{jpc99}
    I. Borukhov, D. Andelman, H. Orland,
    { J. Phys. Chem. B} 24 (1999) 5057.

\bibitem{Solis3}
F.J. Solis,  M. Olvera de la Cruz, J. Chem. Phys. 110 (1999) 11517.

\bibitem{Nguyen}
T.T. Nguyen, A.Y. Grosberg,  B.I. Shklovskii, J.
Chem. Phys. 113 (2000) 1110.

\bibitem{Odijk2}
  T. Odijk,
  { Macromolecules} 12 (1979) 688.

\bibitem{Dobrynin}
  A. V. Dobrynin, R. H.  Colby, M. Rubinstein,
  { Macromolecules} 28 (1995) 1859.


\bibitem{vanderSchee}
  H. A. van der Schee, J.  Lyklema,
  { J. Phys. Chem.} 88 (1984) 6661.

\bibitem{Papenhuijzen}
  J. Papenhuijzen, H. A.  van der Schee, G. J. Fleer,
  { J. Coll. Interface Sci.} 104 (1985) 540.

\bibitem{Evers}
  O. A. Evers, G. J. Fleer, J. M. H. M.  Scheutjens, J. M. H. M.,
  J. Lyklema,
  { J. Coll. Interface Sci.} 111 (1985) 446.

\bibitem{vandeSteegSim}
  H. G. M. van de Steeg, M. A.  Cohen Stuart, A. de Keizer,
   B. H. Bijsterbosch,
  { Langmuir} 8 (1992) 8.

\bibitem{Linse}
P.  Linse, { Macromolecules} 29 (1996) 326.

\bibitem{Varoqui91}
    R. Varoqui, A. Johner, A. Elaissari,
   { J. Chem. Phys.} 94 (1991) 6873.

\bibitem{Varoqui93}
    R. Varoqui, J. Phys. (France) II 3 (1993) 1097.

\bibitem{epl95}
   I. Borukhov,  D. Andelman, H. Orland,
   { Europhys. Lett.} 32 (1995) 499;
    I. Borukhov,
    in: {Short and Long Chains at Interfaces},
   J. Daillant, P. Guenoun, C. Marques, P. Muller,
    J. Tr\^an Thanh V\^an,  Eds.,
    Edition Frontieres, Gif-sur-Yvette, 1995, pp. 13-20.

\bibitem{Chatellier}
X. Chatellier, J.-F. Joanny,  J. Phys. II (France) 6 (1996) 1669.

\bibitem{epr98}
   I. Borukhov,  D. Andelman, H. Orland,
   { Eur. Phys. J. B}  5 (1998) 869.

\bibitem{mm98}
   I. Borukhov, D. Andelman, H. Orland,
    { Macromolecules} 31 (1998) 1665.

\bibitem{Dobrynin5}
A.V. Dobrynin, J. Chem. Phys. 114 (2001) 8145.

\bibitem{san03}
A. Shafir, D. Andelman, R.R. Netz, submitted to J. Chem. Phys. (2003).

\bibitem{DAJFJ00} D. Andelman, J. F. Joanny,  { C. R. Acad.
Sci. (Paris), Ser. IV} 1 (2000) 1153.

\bibitem{Shklovrev}
A.Y. Grosberg, T.T. Nguyen,  B.I. Shklovskii,
Rev. Mod. Phys. 74 (2002) 329.

\bibitem{Levinrev}
Y. Levin, Rep. Prog. Phys. 65 (2002) 1577.

\bibitem{Dan}
N. Dan,  { Biophys. J} 71 (1996) 1267;
73 (1997) 1842.



\bibitem{kleinreview}
J. Klein,  G. Rossi. Macromolecules 31 (1998) 1979.


\bibitem{Asakura}
S. Asakura,  F. Oosawa, J. Chem. Phys. 22 (1954) 1255.

\bibitem{Rudhardt}
D. Rudhardt, C. Bechinger,  P. Leiderer, Phys. Rev. Lett. 81 (1998) 1330.

\bibitem{Hanke}
A. Hanke, E. Eisenriegler,  S. Dietrich. Phys. Rev. E 59
(1999) 6853.

\bibitem{Eisenriegler00}
E. Eisenriegler, J. Phys.: Condens. Matter 12 (2000) A227.

\bibitem{Broukhno}
A. Broukhno, B. J\"onsson, T. Akesson, P.N. Vorontsov-Velyaminov,
J. Chem. Phys. 113 (2000) 5493.

\bibitem{jldg79} J.F. Joanny, L. Leibler, P.G. de Gennes. J. Polym Sci:
Polym Phys. Ed 17 (1979) 1073.

\bibitem{dg82}
P.G. de Gennes, Macromolecules 15 (1982) 492.

\bibitem{scheut1}
J.M.H.M. Scheutjens, G.J. Fleer, Macromolecules 18 (1985) 1882; G.J.
Fleer, J.M.H.M. Scheutjens, J. Coll. Interface Sci. 111 (1986) 504.

\bibitem{avalos}
J. Bonet-Avalos, J.F. Joanny, A. Johner, A.N. Semenov, Europhys.
Lett. 35 (1996) 97; J. Bonet-Avalos, A. Johner, J.F. Joanny, J. Chem.
Phys. 101 (1994) 9181.

\bibitem{semenov2}
A.N. Semenov, J. Phys. II (France) 6 (1996) 1759.

\bibitem{Ennis}
J. Ennis,  B. J\"onsson, J. Phys. Chem. B 103 (1999) 2248.

\bibitem{klein82} J. Klein,  P.F. Luckham,
Nature 300 (1982) 429; Macromolecules 17 (1984) 1041.

\bibitem{klein84} J. Klein,  P.F. Luckham,
Nature 308 (1984) 836; Y. Almog,  J. Klein, J. Colloid Interface Sci.
106 (1985) 33.

\bibitem{rossi} G. Rossi,  P.A. Pincus, Europhys. Lett.
5 (1988) 641;  Macromolecules 22 (1989) 276.

\bibitem{mendez}
J.M. Mendez-Alcaraz, A. Johner,  J.F. Johner, Macromolecules 31 (1998) 8297.

\bibitem{klein80}
J. Klein, Nature 288 (1980) 248; J. Klein,   P.F. Luckham,
Macromolecules 19 (1986) 2007.

\bibitem{kleinpincus}
J. Klein, P. A. Pincus, Macromolecules 15 (1982) 1129; K. Ingersent, J.
Klein,  P. A.  Pincus, Macromolecules 19 (1986) 1374.

\bibitem{ikp90}  K. Ingersent, J. Klein, P. A. Pincus, Macromolecules
23 (1990) 548.

\bibitem{Fredrickson}
G.H. Fredrickson,  P. A. Pincus, Langmuir 7 (1991) 786.

\bibitem{Guldbrand}
L. Guldbrand, B. J\"onsson, H. Wennerstr\"om,  P. Linse,
J. Chem. Phys. 80 (1984) 2221.

\bibitem{Moreira}
A.G. Moreira,  R.R. Netz, Phys. Rev. Lett. 87 (2001) 8301.

\bibitem{Akesson}
T. Akesson, C. Woodward, B. J\"onsson, J. Chem. Phys. 91 (1989) 2461.

\bibitem{Granfeldt}
M.K. Granfeldt, B. J\"onsson,  C.E. Woodward,
J. Phys. Chem. 95 (1991) 4819.

\bibitem{Bohmer}
M.R. B\"ohmer, O.A. Evers,  J.M.H.M. Scheutjens,
Macromolecules 23 (1990) 2223.

\bibitem{Podgornik}
R. Podgornik, J. Phys. Chem. 96 (1992) 884.


\bibitem{oscill}
A. Asnacios, A. Espert, A. Colin,  D. Langevin, Phys. Rev.
Lett. 78 (1997) 4974.

\bibitem{oscill2}
B. Kolaric, W. Jaeger,  R. von Klitzing, J. Phys. Chem. B 104 (2000) 5096.

\bibitem{oscill3}
R. von Klitzing, A. Espert, A. Colin,  D. Langevin,
Colloids Surfaces A 176 (2001) 109.

\bibitem{Yethiraj}
A. Yethiraj, J. Chem. Phys. 111 (1999) 1797.


\bibitem{dg90}P.G. de Gennes, J. Phys. Chem. 94 (1990) 8407.

\bibitem{aj91}
D. Andelman, J.F. Joanny, Macromolecules 24 (1991) 6040;
J.F. Joanny, D. Andelman. Makromol Chem. Macromol Symp 62 (1992) 35;
D. Andelman,   J.F. Joanny, J. Phys. II (France) 3 (1993) 121.

\bibitem{aazpr94}
V. Aharonson, D. Andelman, A. Zilman, P.A. Pincus,
E. Rapha\"el, Physica A 204 (1994) 1; 227 (1996) 158.

\bibitem{nao96}
R.R. Netz, D. Andelman, H. Orland, J. Phys. II (France) 6 (1996) 1023.


\bibitem{ca95}
X. Ch\^atellier,  D. Andelman, Europhys. Lett. 32 (1995) 567;
 X. Ch\^atellier,  D Andelman, J. Phys. Chem. 22 (1996) 9444.

\bibitem{Fleck}
C. Fleck, R.R. Netz,  H. von Gr\"unberg,
Biophysical J.  82 (2002) 76.


\bibitem{Bruinsma}
R. Bruinsma,  J. Mashl, Europhys. Lett. 41 (1998) 165.

\bibitem{Wagner}
D. Harries, S. May, W.M. Gelbart,  A. Ben-Shaul,
Biophys. J. 75 (1998) 159.

\bibitem{Clausen}
H. Clausen-Schaumann,  H.E. Gaub,
Langmuir 15 (1999) 8246.

\bibitem{Ellis}
M. Ellis, C.Y. Kong,  M. Muthukumar,
J. Chem. Phys. 112 (2000) 8723.

\bibitem{Kong}
J. McNamara, C.Y. Kong,  M. Muthukumar,
J. Chem. Phys. 117 (2002) 5354.

\bibitem{vilgis}
T.A. Vilgis, G. Heinrich, Macromolecules 27 (1994) 7846;
 G. Huber,  T.A. Vilgis, Eur. Phys. J. B 3 (1998) 217.

\bibitem{hone}
D. Hone, H. Ji, P.A. Pincus, Macromolecules 20 (1987) 2543;
 H. Ji,  D. Hone, Macromolecules 21 (1988) 2600.

\bibitem{blunt}
M. Blunt, W. Barford, R. Ball, Macromolecules 22 (1989) 1458.

\bibitem{marquesfractal}
C.M. Marques, J.F. Joanny, J. Phys. (France) 49 (1988) 1103.

\bibitem{curved1}
S. Alexander, J. Phys. (Paris) 38 (1977) 977.

\bibitem{colloidads}
P.A. Pincus, C.J. Sandroff,  T.A. Witten,
J. Phys. (France) 45 (1984) 725.

\bibitem{flexads1}
R. Podgornik, Europhys. Lett. 21 (1993) 245; Phys. Rev. E 51 (1995) 3368.

\bibitem{flexads2}
T. Garel, M. Kardar,  H. Orland, Europhys. Lett. 29 (1995) 303.

\bibitem{elasticity1}
J.T. Brooks, C.M. Marques,  M.E. Cates, Europhys. Lett. 14 (1991) 713; J.
Phys. II (France) 1 (1991) 673.

\bibitem{elasticity2}
 F. Clement,  J.F. Joanny, J. Phys. II
(France) 7 (1997) 973.

\bibitem{elasticity3}
M. Breidenich, R.R. Netz,  R. Lipowsky
(to be published).

\bibitem{curved2}
T. Odijk, Macromolecules 13 (1980) 1542.

\bibitem{Park}
S.Y. Park, R.F. Bruinsma,  W.M. Gelbart,
Europhys. Lett. 46 (1999) 454.

\bibitem{Kunze2}
K.K. Kunze,  R.R. Netz, Europhys. Lett.
58 (2002) 299.


\bibitem{Goeler}
F. von Goeler, M. Muthukumar, {J. Chem. Phys.} {100} (1994) 7796.

\bibitem{linse}
T. Wallin, P. Linse, {Langmuir} {12} (1996) 305; {J. Phys. Chem.}
100 (1996) 17873; {J. Phys. Chem. B} {101} (1997) 5506.

\bibitem{sens}
E. Gurovitch, P. Sens, {Phys. Rev. Lett.}  {82} (1999) 339.

\bibitem{Ramin}
R. Golestanian, Phys. Rev. Lett. 83 (1999) 2473.

\bibitem{mateescu}
E.M. Mateescu, C. Jeppesen,  P.A. Pincus. {Europhys. Lett} {46} (1999) 493.

\bibitem{Netz5}
R.R. Netz, J.F. Joanny, Macromolecules 32 (1999) 9026.

\bibitem{Shklovsphere}
T.T. Nguyen,  B.I. Shklovskii, Physica A 293 (2001) 324;
J. Chem. Phys. 114 (2001) 5905.

\bibitem{Kunze}
K. K. Kunze, R. R. Netz, { Phys. Rev. Lett.} 85 (2000) 4389.

\bibitem{Yager}
T.D. Yager, C.T. McMurray, K.E. van Holde, { Biochemistry}
28 (1989) 2271.



\bibitem{taunton}
H.J. Taunton, C. Toprakcioglu, L.J. Fetters,  J. Klein, Nature
332 (1988) 712; Macromolecules 23 (1990) 571.

\bibitem{field}
J.B. Field, C. Toprakcioglu, L. Dai, G. Hadziioannou, G. Smith,  W.
Hamilton, J. Phys. II (France) 2 (1992) 2221.

\bibitem{marques}
C.M. Marques, J.F. Joanny,  L. Leibler, Macromolecules 21 (1988) 1051;
%
C.M. Marques, J.F. Joanny, Macromolecules 22 (1989) 1454.

\bibitem{kent}
M.S. Kent, L.T. Lee, B. Farnoux,  F. Rondelez, Macromolecules 25
(1992) 6240; M.S. Kent, L.T. Lee, B.J. Factor, F. Rondelez,  G.S.
Smith, J. Chem. Phys. 103 (1995) 2320; H.D. Bijsterbosch, V.O. de
Haan, A.W. de Graaf, M. Mellema, F.A.M. Leermakers, M.A. Cohen
Stuart,  A.A. van Well, Langmuir 11 (1995) 4467; M.C. Faur\'e, P.
Bassereau, M.A. Carignano, I. Szleifer, Y. Gallot,  D. Andelman,
Eur. Phys. J. B 3 (1998) 365.

\bibitem{teppner}
R. Teppner, M. Harke,  H. Motschmann, Rev. Sci. Inst. 68 (1997) 4177;
R. Teppner,   H. Motschmann, Macromolecules 31 (1998) 7467.

\bibitem{gennes}
P.G. de Gennes, Macromolecules 13 (1980) 1069.

\bibitem{alex}
S. Alexander, J. Phys. (France) 38 (1977) 983.

\bibitem{halperin92}
A. Halperin, M. Tirell,  T.P. Lodge, Adv. Pol. Sc. 100 (1992) 31.

\bibitem{cos87}
T. Cosgrove, T. Heath, B. van Lent, F.A.M. Leermakers,  J. Scheutjens,
Macromolecules 20 (1987) 1692.

\bibitem{murat}
M. Murat, G.S. Grest, Macromolecules 22 (1989) 4054; A. Chakrabarti, R.
Toral, Macromolecules 23 (1990) 2016; P.Y. Lai, K. Binder, J.
Chem. Phys. 95 (1991) 9288.

\bibitem{sem}
A. N. Semenov, { Sov. Phys. JETP} 61 (1985) 733.

\bibitem{mil}
S. T. Milner, T. A. Witten, M. E. Cates, { Europhys. Lett.}
5 (1988) 413; { Macromolecules} 21 (1988) 610;
S. T. Milner,  { Science} 251 (1991) 905.

\bibitem{skvor}
A. M. Skvortsov, I. V. Pavlushkov, A. A. Gorbunov, Y. B. Zhulina,
O. V. Borisov, V. A. Pryamitsyn,  { Polym. Sci.} 30
(1988) 1706.

\bibitem{netzbrush}
R.R. Netz, M. Schick, Europhys. Lett. 38 (1997) 37; Macromolecules
31 (1998) 5105.

\bibitem{Seidel}
C. Seidel, R. R. Netz, { Macromolecules} 33 (2000) 634.

\bibitem{carignano}
M.A. Carignano, I. Szleifer, J. Chem. Phys. 98 (1993) 5006; J. Chem.
Phys. 100 (1994) 3210; Macromolecules 28 (1995) 3197; J. Chem. Phys.
102 (1995) 8662. A detailed summary of tethered layers is given in:
I. Szleifer, M.A. Carignano, Adv. Chem. Phys. XCIV (1996) 165.

\bibitem{grest}
G.S. Grest, Macromolecules 27 (1994) 418.

\bibitem{collgraft}
T.A. Witten,  P.A. Pincus, Macromolecules 19 (1986) 2509;
E.B. Zhulina,
O.V. Borisov,   V.A. Priamitsyn, J. Coll. Surf. Sci. 137 (1990) 495.

\bibitem{milner88b}
S.T. Milner, Europhys. Lett. 7 (1988) 695.

\bibitem{milner89}
S.T. Milner, T.A. Witten,   M.E. Cates, Macromolecules 22 (1989) 853.

\bibitem{halperin88}
A. Halperin, J. Phys. (France) 49 (1988) 547;
Y.B. Zhulina, V.A. Pryamitsyn, O.V. Borisov, Polymer Science 31 (1989) 205;
E.B. Zhulina, O.V. Borisov, V.A. Pryamitsyn,  T.M. Birshtein,
 Macromolecules 24 (1991) 140;
 D.R.M. Williams, J. Phys. II (France) 3 (1993) 1313.

\bibitem{auroy2}
P. Auroy, L. Auvray, Macromolecules 25 (1992) 4134.

\bibitem{marko93}
J.F. Marko, Macromolecules 26 (1993) 313.

\bibitem{theta}
P.Y. Lai, K. Binder, J. Chem. Phys. 97 (1992) 586;
G.S. Grest, M. Murat, Macromolecules 26  (1993) 3108.

\bibitem{daoudcotton}
M. Daoud, J.P. Cotton, J. Phys. (France) 43 (1982) 531.

\bibitem{ball}
R.C. Ball, J.F. Marko, S.T. Milner, A.T. Witten, Macromolecules
24 (1991) 693; H. Li, T.A. Witten, Macromolecules 27 (1994) 449.

\bibitem{dan}
N. Dan, M. Tirrell, Macromolecules 25 (1992) 2890.

\bibitem{murat91}
M. Murat, G.S. Grest, Macromolecules 24 (1991) 704.

\bibitem{milnerbend}
S.T. Milner, T.A. Witten,  J. Phys. (France) 49 (1988) 1951.

\bibitem{allen}
T.M. Allen, C. Hansen, F. Martin, C. Redemann,  A. Yau-Young,
Biochim. Biophys. Acta 1006 (1991) 29;
M.J. Parr, S.M. Ansell, L.S. Choi, P.R. Cullis,
 Biochim. Biophys. Acta 1195 (1994) 21.

\bibitem{Hristova1}
K. Hristova,  D. Needham,
Macromolecules 28 (1995) 991.

\bibitem{Hristova2}
K. Hristova, A. Kenworthy,  T.J. McIntosh,
Macromolecules 28 (1995) 7693.

\bibitem{lipopol2}
C. Hiergeist, R. Lipowsky, J. Phys. II (France) 6 (1996) 1465.

\bibitem{lipopol3}
M. Breidenich, R.R. Netz, R. Lipowsky, Europhys. Lett. 49 (2000) 431.

\bibitem{bickel}
T. Bickel, C. Marques,  C. Jeppesen, Phys. Rev. E 62 (2000) 1124.

\bibitem{lipopol4}
C. Hiergeist, V.A. Indrani, R. Lipowsky, Europhys. Lett. 36 (1996) 491.

\bibitem{lipopol5}
M. Breidenich, R.R. Netz, R. Lipowsky, Eur. Phys. J. E 5 (2001) 189.

\bibitem{aubouy}
M. Aubouy, G.H. Fredrickson, P. A. Pincus,  E. Raphael, Macromolecules
28 (1995) 2979.

\bibitem{gast}
A.P. Gast,  L. Leibler, Macromolecules 19 (1986) 686.

\bibitem{marko91}
J.F. Marko, T.A. Witten, Phys. Rev. Lett. 66 (1991) 1541.

\bibitem{brown}
G. Brown, A. Chakrabarti, J.F. Marko, Macromolecules 28 (1995) 7817;
E.B. Zhulina, C. Singhm, A.C. Balazs, Macromolecules 29 (1996) 8254.

\bibitem{halperin88b}
A. Halperin, S. Alexander, Europhys. Lett. 6 (1988) 329;
A Johner, J.F. Joanny, Macromolecules 23 (1990) 5299;
C. Ligoure, L. Leibler, J. Phys. (France) 51 (1990) 1313;
S.T. Milner, Macromolecules 25 (1992) 5487;
A. Johner, J.F. Joanny, J. Chem. Phys. 98 (1993) 1647.

\bibitem{MIK88}
S. J. Miklavic, S. Marcelja, { J. Phys. Chem.} 92 (1988) 6718.

\bibitem{MIS89}
S. Misra, S. Varanasi, P.P. Varanasi, Macromolecules 22 (1989) 5173.

\bibitem{PIN91}
P. A. Pincus,  Macromolecules 24 (1991) 2912.

\bibitem{BOR91}
O.V. Borisov, T.M. Birstein, E.B. Zhulina,
 J. Phys. II (France) 1 (1991) 521.

\bibitem{ROS92}
R.S. Ross, P. A. Pincus, Macromolecules 25 (1992) 2177;
 E.B. Zhulina, T.M. Birstein, O.V. Borisov,
  J. Phys. II (France) 2 (1992) 63.

\bibitem{WIT93}
J. Wittmer, J.-F. Joanny, { Macromolecules}
26 (1993) 2691.

\bibitem{ISR94}
R. Israels, F.A.M. Leermakers, G.J. Fleer, E.B. Zhulina,
{ Macromolecules} 27 (1994) 3249.

\bibitem{BOR94}
O.V Borisov, E.B. Zhulina, T.M. Birstein,
   { Macromolecules} 27 (1994) 4795.

\bibitem{ZHU97}
E.B. Zhulina, O.V. Borisov, { J. Chem. Phys.}
107   (1997) 5952.

\bibitem{MIR95}
Y. Mir, P. Auvroy, L. Auvray, { Phys. Rev. Lett.}
75 (1995) 2863.

\bibitem{GUE95}
P. Guenoun, A. Schlachli, D. Sentenac, J.M. Mays,
J.J. Benattar, { Phys. Rev. Lett.} 74 (1995) 3628.

\bibitem{AHR97}
H. Ahrens, S. F\"orster, C.A. Helm, Macromolecules 30 (1997) 8447;
Phys. Rev. Lett. 81 (1998) 4172.

\bibitem{Csajka0}
F. Csajka, C. Seidel, { Macromolecules} 33 (2000) 2728.

\bibitem{Csajka}
F.S. Csajka, R.R. Netz, C. Seidel,  J.F. Joanny,
Eur. Phys. J. E 4 (2001) 505.

\bibitem{Ennis2}
J. Ennis, L. Sj\"ostr\"om, T. Akesson,  B. J\"onsson,
J. Phys. Chem. B 102 (1998) 2149.

\bibitem{Eisenberg}
L. Zhang,  A. Eisenberg, Science {268} (1995) 1728;
L. Zhang, K. Yu,  A. Eisenberg, Science {272} (1996) 1777;
H. Shen, L. Zhang,  A. Eisenberg, J. Am. Chem. Soc.
{121} (1999) 2728.


\bibitem{Netzmicelles}
R.R. Netz, Europhys. Lett. 47 (1999) 391.

\bibitem{Guenoun2}
P. Guenoun, F. Muller, M. Delsanti, L. Auvray, Y.J. Chen, J.W. Mays,
M. Tirrell, Phys. Rev. Lett. 81 (1998) 3872;
P. Guenoun, M. Delsanti, D. Gazeau, J.W. Mays, D.C. Cook,
M. Tirrell, L. Auvray,  Eur. Phys. J. B {1} (1998) 77.


\bibitem{Foerster2}
S. F\"orster, N. Hermsdorf, W. Leube, H. Schnablegger, M. Regenbrecht,
S. Akarai, P. Lindner,  C. B\"ottcher, J. Phys. Chem. B 103 (1999) 6657;
M. Regenbrecht, S. Akarai, S. F\"orster,  H. M\"ohwald,
J. Phys. Chem. B 103 (1999) 6669.




\end{thebibliography}
\end{document}